\pdfoutput=1
\newcommand*{\ATLASLATEXPATH}{}
\documentclass[PAPER, atlasdraft=true, cernpreprint, texlive=2016, USenglish]{\ATLASLATEXPATH atlasdoc}
 
\usepackage[subcaption]{\ATLASLATEXPATH atlaspackage}
\usepackage{\ATLASLATEXPATH atlasbiblatex}
 
\usepackage{\ATLASLATEXPATH atlasphysics}
\usepackage{amsmath}
 
\graphicspath{{logos/}{figures/}}
 
\newcommand\upstrut{\rule{0pt}{12pt}}
\newcommand\downstrut{\rule[-6pt]{0pt}{6pt}}
 
\usepackage{caption}
\usepackage{savesym}
\savesymbol{tablenum}
\usepackage{multirow}
\usepackage[maxfloats=50]{morefloats}
\DeclareUnicodeCharacter{2212}{-}
\restoresymbol{SIX}{tablenum}

\AtlasTitle{Search for heavy neutral Higgs bosons produced in association with $b$-quarks and decaying to $b$-quarks at $\sqrt{s}=13$~TeV with the ATLAS detector}

\usepackage{authblk}

\usepackage{amsmath}

\AtlasVersion{3.1}
 
\AtlasRefCode{HIGG-2016-32}

\PreprintIdNumber{CERN-EP-2019-092}

\AtlasJournalRef{Phys. Rev. D 102 (2020) 032004}
\AtlasDOI{10.1103/PhysRevD.102.032004}
 
\AtlasAbstract{
A search for heavy neutral Higgs bosons produced in association with one or two $b$-quarks
and decaying to $b$-quark pairs is presented
using 27.8 fb$^{-1}$ of $\sqrt{s}=13$~TeV proton-proton collision data recorded by the ATLAS detector at the Large Hadron Collider
during 2015 and 2016.
No evidence of a signal is found.
Upper limits on the heavy neutral Higgs boson production cross section times its branching ratio to $b\bar{b}$
are set, ranging from 4.0 to 0.6~pb at  95\% confidence level over a Higgs boson mass range of 450 to 1400~GeV.
Results are interpreted within the two-Higgs-doublet model and the minimal supersymmetric Standard Model.
}

\AtlasCoverSupportingNote{Main Analysis Note}{https://cds.cern.ch/record/2205087}
 
\AtlasCoverCommentsDeadline{17 May 2020}
 
\AtlasCoverAnalysisTeam{Tim Barklow, Petar Bokan, Valerio Dao, Yuji Enari, Pawel Malecki, Tatsuya Masubuchi, Nicolas Morange, Giacinto Piacquadio, Dong Su, Song-Ming Wang, Bartlomiej Zabinski, Lidija Zivkovic}
 
\AtlasCoverEdBoardMember{Fabrizio~Parodi~(chair), Trisha Farooque, Paul Thompsom}

\AtlasCoverEgroupEditors{atlas-higg-2016-32-editors@cern.ch}
 
\AtlasCoverEgroupEdBoard{atlas-higg-2016-32-editorial-board@cern.ch}

\title{Search for heavy neutral Higgs bosons produced in association with $b$-quarks and decaying into $b$-quarks at $\sqrt{s}=13$~\TeV\ with the ATLAS detector}
\hypersetup{pdftitle={Search for neutral Higgs ... detector},pdfauthor={The ATLAS Collaboration}}
 
\begin{document}
 
\maketitle
 
\tableofcontents
 
\clearpage
\def\ifmath#1{\relax\ifmmode #1\else $#1$\fi}
\def\lsub#1{\ifmath{_{\lower1.5pt\hbox{$\scriptstyle #1$}}}}
\def\phm{\phantom{-}}

\newcommand\gfive{\ensuremath{\gamma_{\lower1.5pt\hbox{$\scriptstyle 5$}}}}
 
\section{Introduction}\label{sec:intro}

The measured properties of the Higgs boson discovered at the Large Hadron Collider (LHC) by the ATLAS and CMS collaborations~\cite{HIGG-2012-27,CMS-HIG-12-028} with a mass of 125~\GeV\ are consistent with those of the scalar particle that emerges from the mechanism of electroweak symmetry breaking in the Standard Model (SM) with its
one doublet of complex scalar fields~\cite{Englert:1964et,Higgs:1964ia,Higgs:1964pj,Guralnik:1964eu}.
Alternative electroweak symmetry breaking models which contain a scalar particle with properties similar to
the SM Higgs boson remain viable, however.
A simple, well-studied and well-motivated extension of the mechanism of electroweak symmetry breaking in the SM
is the two-Higgs-doublet model (2HDM), which contains two doublets of complex scalar fields~\cite{Branco:2011iw,Gunion:1989we}.
In the 2HDM there are,
assuming negligible \textit{CP-}violating
effects, two \textit{CP-}even scalar bosons, $h$ and $H$ which satisfy the mass relation $m_h < m_H$,  one \textit{CP-}odd pseudoscalar boson, $A$,
and two electrically charged scalar bosons, $H^\pm$.
The most general renormalizable, electroweak gauge invariant 2HDM contains tree-level
Higgs-boson-mediated flavor-changing neutral currents~\cite{Gunion:1989we} that are in conflict with experimental limits.
When symmetries are imposed to naturally suppress flavor changing neutral currents, four
model types emerge, distinguished from one another by their Yukawa couplings, as summarized in Table~\ref{yukawa_tab} for $h$, $H$, and $A$.
 
The agreement of SM predictions with measurements of the 125~\GeV\ Higgs boson,
assumed in this paper to be the scalar boson  $h$ in the 2HDM, is reducing the 2HDM  parameter space
toward the alignment limit of $\cos(\beta-\alpha)\approx 0$, where $\tan\beta$ is
the ratio of the vacuum expectation values of the two scalar doublets and $\alpha$ is the mixing angle of the two \textit{CP-}even scalar bosons~\cite{Asner:2013psa}.
In the alignment limit, decays of the $H$ and $A$ bosons
into gauge boson pairs $W^+W^-$ and $ZZ$ are heavily suppressed, and the fermion
coupling pattern simplifies to that of Table~\ref{yukawa_tab}. The suppression of $H/A$ couplings to $W^+W^-$ and $ZZ$,
along with  ATLAS and CMS limits on new particle production, implies that searches for the heavy neutral Higgs bosons of the 2HDM
mainly rely on their couplings to third-generation fermions.

\begin{table}[hb!]
\caption{Tree-level fermion couplings of the 2HDM $h$, $H$, and $A$ bosons for model types I, II, X (or lepton-specific), and
Y (or flipped).
Here $U$, $D$, and $E$ refer to up-type quarks, down-type quarks, and charged leptons, respectively,
$t_\beta \equiv \tan\beta$ is the ratio of the vacuum expectation values of the two scalar doublets,
and $\epsilon = \cos{(\beta-\alpha)}$ where $\alpha$ is the mixing angle of the two \textit{CP-}even scalar bosons~\cite{Asner:2013psa}.
The couplings are normalized to the SM Higgs boson couplings $h_\mathrm{SM}\bar{U}U$, $h_\mathrm{SM}\bar{D}D$,
and $h_\mathrm{SM}\bar{E}E$
and are given in the alignment limit $\cos{(\beta-\alpha)\approx 0}$ where the couplings of the light scalar boson $h$ are close to SM expectations.}
\label{yukawa_tab}
\centering
{\renewcommand\arraystretch{1.5}
\begin{tabular}{lccccccccc}
\toprule
&$h\bar{U}U$&$h\bar{D}D$&$h\bar{E}E$
&$H\bar{U}U$&$H\bar{D}D$&$H\bar{E}E$&$iA\bar{U}\gamma_{_5} U$&$iA\bar{D}\gamma_{_5} D$
& $iA\bar{E}\gamma_{_5} E$\\
\midrule
I &$1+{\epsilon \over t_\beta}$&$1+{\epsilon \over t_\beta}$&$1+{\epsilon \over t_\beta}$
&$-({1\over t_\beta}-\epsilon)$&$-({1\over t_\beta}-\epsilon)$&$-({1\over t_\beta}-\epsilon)$
&$-{1\over t_\beta}$&$\phantom{-}{1\over t_\beta}$&$\phantom{-}{1\over t_\beta}$\\
II&$1+{\epsilon \over t_\beta}$&$1-\epsilon t_\beta$&$1-\epsilon t_\beta$
&$-({1\over t_\beta}-\epsilon)$&$t_\beta+\epsilon$&$t_\beta+\epsilon$
&$-{1\over t_\beta}$&$-t_\beta$&$-t_\beta$\\
X &$1+{\epsilon \over t_\beta}$&$1+{\epsilon \over t_\beta}$&$1-\epsilon t_\beta$
&$-({1\over t_\beta}-\epsilon)$&$-({1\over t_\beta}-\epsilon)$&$t_\beta+\epsilon$
&$-{1\over t_\beta}$&$\phantom{-}{1\over t_\beta}$&$-t_\beta$\\
Y &$1+{\epsilon \over t_\beta}$&$1-\epsilon t_\beta$&$1+{\epsilon \over t_\beta}$
&$-({1\over t_\beta}-\epsilon)$&$t_\beta+\epsilon$&$-({1\over t_\beta}-\epsilon)$
&$-{1\over t_\beta}$&$-t_\beta$&$\phantom{-}{1\over t_\beta}$\\
\bottomrule
\end{tabular}}
\end{table}
 
The Higgs sector of the minimal supersymmetric Standard Model (MSSM) is a Type~II 2HDM, which has motivated searches for
heavy neutral Higgs bosons at
LEP~\cite{Schael:2006cr} and the LHC~\cite{Aaboud:2017sjh,Sirunyan:2018zut}.
These searches use decays of heavy neutral Higgs bosons into $\tau^+\tau^-$, and are sensitive to Type~II and lepton-specific 2HDMs.
They are not sensitive to flipped 2HDMs at large $\tan\beta$, however, and they do not cover the entire MSSM Type~II 2HDM parameter space since radiative corrections can significantly increase
the ratio of the $b\bar{b}$ and $\tau^+\tau^-$ partial widths beyond the tree-level value of $3m_b^2/m_\tau^2$~\cite{Carena:2005ek}.
 
This paper presents a search for heavy neutral Higgs bosons produced in association with one or two $b$-quarks
and decaying into $b$-quark pairs
using 27.8 fb$^{-1}$ of $\sqrt{s}=13$~\TeV\ proton-proton collision data recorded by the ATLAS detector at the LHC
during 2015 and 2016.
The search is sensitive to the Type II and flipped scenarios of the 2HDM in the regime where $\tan\beta\gg 1$.
In the 5-flavor scheme (5FS)~\cite{Harlander:2011aa},
processes such as those shown in Figure~\ref{fig:fd} lead to the production of heavy neutral Higgs bosons in association
with one $b$-quark (Figures~\ref{fig:1a} and \ref{fig:1b})  or two $b$-quarks (Figures~\ref{fig:1c} and \ref{fig:1d}).  In practice, the optimal balance
between signal efficiency and background rejection is achieved by requiring that signal events contain at least three $b$-quark-initiated jets.
The search is performed for neutral Higgs bosons in the mass range 450--1400~\GeV. A similar
search was performed by the CMS Collaboration for the mass range 300--1300~\GeV~\cite{Sirunyan:2018taj}.
 
The kinematic distributions for the production and decay of  $H$ and $A$ bosons are nearly identical, and therefore this search is insensitive to the CP properties
of the two heavy neutral Higgs bosons of the 2HDM.
The $\phi$ boson is used in this paper to represent the \textit{CP-}even $H$ boson, the \textit{CP-}odd $A$ boson, or a Higgs boson mass eigenstate with an
arbitrary mixture of \textit{CP-}even and \textit{CP-}odd eigenstates.

\begin{figure}
\centering
\begin{subfigure}[b]{0.20\textwidth}
\includegraphics[width=\textwidth]{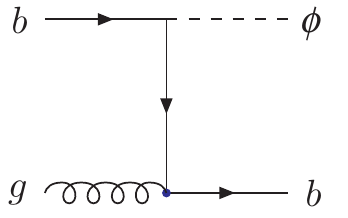}
\caption{\label{fig:1a} }
\end{subfigure}
\qquad
\begin{subfigure}[b]{0.20\textwidth}
\includegraphics[width=\textwidth]{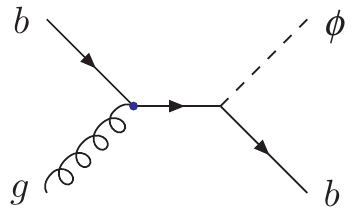}
\caption{\label{fig:1b} }
\end{subfigure}
\qquad
\begin{subfigure}[b]{0.20\textwidth}
\includegraphics[width=\textwidth]{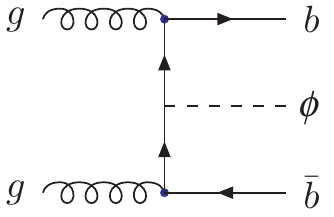}
\caption{\label{fig:1c} }
\end{subfigure}
\qquad
\begin{subfigure}[b]{0.20\textwidth}
\includegraphics[width=\textwidth]{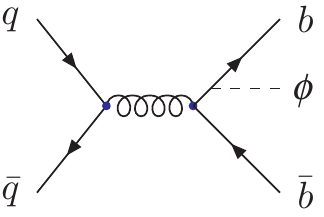}
\caption{ \label{fig:1d}}
\end{subfigure}
\caption{Feynman diagrams for some of the leading-order processes for the production of a heavy neutral Higgs boson (denoted here by $\phi$)
in association with one or two  $b$-quarks in the 5-flavor scheme. Diagrams (a) and (b) are unique to the 5-flavor scheme,
while diagrams (c) and (d) appear in both the 4- and 5-flavor schemes. \label{fig:fd}}
\end{figure}

\section{The ATLAS detector}
\label{sec:atlas}

\newcommand{\AtlasCoordFootnote}{
ATLAS uses a right-handed coordinate system with its origin at the nominal interaction point (IP) in the center of the detector and the $z$-axis coinciding with the axis of the beam pipe.  The $x$-axis points from the IP toward the center of the LHC ring, and the $y$-axis points upward. Cylindrical coordinates ($r$,$\phi$) are used in the transverse plane, $\phi$ being the azimuthal angle around the $z$-axis. The pseudorapidity is defined in terms of the polar angle $\theta$ as $\eta = - \ln \tan(\theta/2)$.  Angular distance is defined as $\Delta R \equiv \sqrt{(\Delta\eta)^{2} + (\Delta\phi)^{2}}$. Transverse momentum and energy are defined as $\pt=p\sin\theta$ and $\et=E\sin\theta$, respectively.}
 
The ATLAS experiment~\cite{PERF-2007-01} at the LHC is a multipurpose particle detector
with a forward-backward symmetric cylindrical geometry and a near $4\pi$ coverage in
solid angle.\footnote{\AtlasCoordFootnote}
It consists of an inner tracking detector surrounded by a thin superconducting solenoid
providing a \SI{2}{\tesla} axial magnetic field, electromagnetic and hadronic calorimeters, and a muon spectrometer.
The inner tracking detector covers the pseudorapidity range $|\eta| < 2.5$.
It consists of silicon pixel, silicon microstrip, and transition radiation tracking detectors.
The innermost pixel layer~\cite{Capeans:1291633,Abbott:2018ikt} was added before the start of collisions in 2015.
Lead/liquid-argon (LAr) sampling calorimeters provide electromagnetic (EM) energy measurements
with high granularity.
A hadronic steel/scintillator-tile calorimeter covers the central pseudorapidity range $|\eta| < 1.7$.
The endcap and forward regions are instrumented with LAr calorimeters
for both the EM and hadronic energy measurements up to $|\eta| = 4.9$.
The muon spectrometer surrounds the calorimeters and features
three large air-core toroid superconducting magnets with eight coils each.
The field integral of the toroids ranges from \num{2.0} to \SI{6.0}{\tesla$\cdot$\metre} across most of the detector.
It includes a system of precision tracking chambers and fast detectors for triggering.
A two-level trigger system~\cite{TRIG-2016-01} consisting of the level-1 (L1) trigger, implemented in hardware, and the software-based
high-level trigger (HLT),
selects interesting events. The L1 trigger uses a subset of detector information
to reduce the event rate to a design value of at most 100 kHz. The HLT, which can run offline reconstruction algorithms
and is used in this analysis for the triggering of $b$-quark-initiated jets, reduces the event rate to about 1 kHz.
 
\section{Data and simulated samples}\label{sec:samples}
 
\subsection{Data}
Proton-proton ($pp$) collision data recorded by the ATLAS detector at the LHC during 2015 and 2016
at a center-of-mass energy of $\sqrt{s}=13$~\TeV\ were used for the analysis described in this paper.
For this data, it is required that the LHC operate with stable beam conditions and that
all relevant detector systems be fully functional.
The data, corresponding to integrated luminosities
of $3.2\pm 0.1$~\ifb~and $24.5\pm 0.5$~\ifb~for 2015 and 2016, respectively,
were collected using a combination of HLT triggers, which  employ algorithms~\cite{TRIG-2016-01} to identify jets containing $b$-hadrons (resulting in `$b$-tagged  jets').
Maximum-likelihood algorithms were utilized in 2015, while the offline multivariate classifier MV2c20~\cite{PERF-2012-04,ATLAS:2016-012} was
used in 2016.
Events were recorded if they passed the L1 single-jet trigger with a transverse energy ($E_{\textrm{T}}$) threshold of $E_{\textrm{T}}=100$~\GeV, and if the HLT identifies either one $b$-tagged jet with $E_{\textrm{T}}>225$~\GeV\ or two $b$-tagged
jets with different thresholds of $E_{\textrm{T}}=150$~\GeV\ and $E_{\textrm{T}}=50$~\GeV.
For the single (double) $b$-tagged jet trigger, the operating points correspond to a $b$-quark identification efficiency of 79\% (72\%) in 2015 and 60\% (60\%) in 2016, as measured with a reference $t\bar{t}$ sample.
An inefficiency in the online vertex reconstruction affected a fraction of the data collected during 2016; events from these running periods
were not included in the analysis.
The efficiency of the combination of the two HLT $b$-jet triggers
is shown in Figure~\ref{fig:trigEff} for events passing the final selection described in Section~\ref{sec:selection} and ranges from 80\% for $m_\phi=450$~\GeV\ to 95\% for $m_\phi > 700$~\GeV.  The efficiency of the double $b$-tagged jet trigger falls with increasing
$m_\phi$ since the per-jet $b$-tagging efficiency is dropping with increasing $E_{\textrm{T}}$.  The single $b$-tagged jet trigger efficiency increases with increasing $m_\phi$ because a larger fraction of the second leading jets satisfy
$E_{\textrm{T}}>225$~\GeV.
 
\begin{figure}[ht!]
\begin{center}
\includegraphics[width=0.75\textwidth]{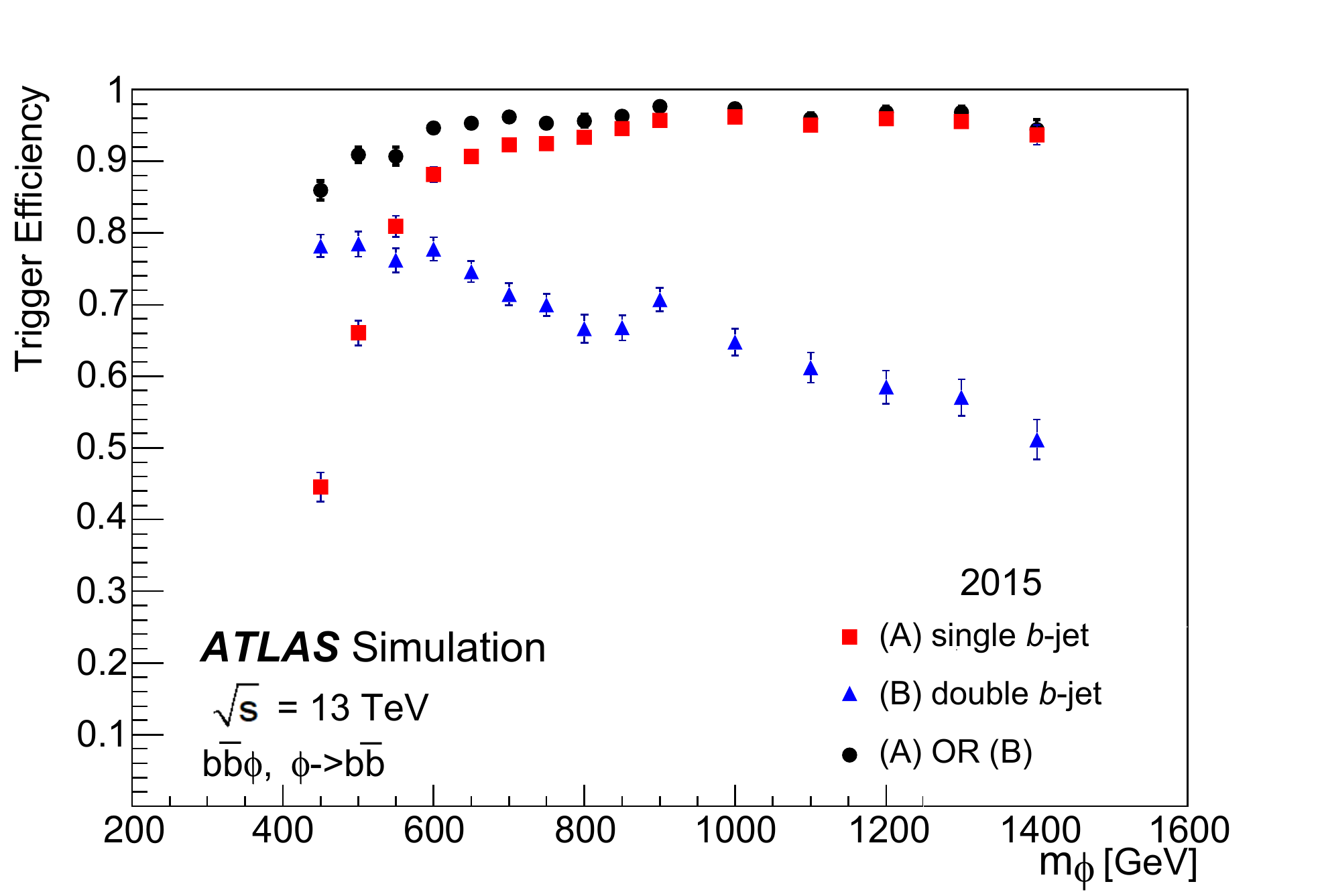}
\includegraphics[width=0.75\textwidth]{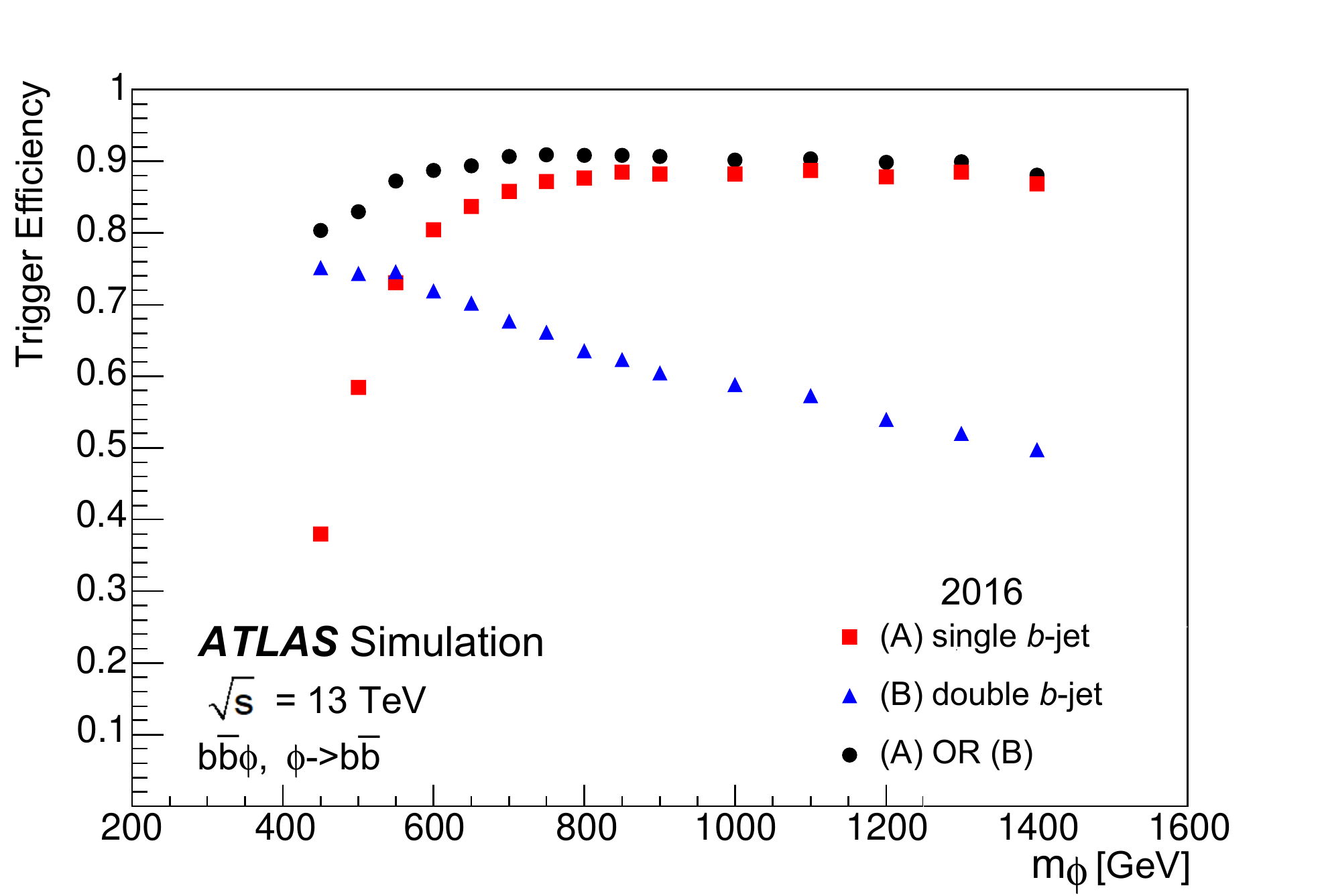}
\caption{Efficiency of the single $b$-tagged jet and double $b$-tagged jet triggers and their logical OR for signal events fulfilling the final selection
of Section~\ref{sec:selection} as a function of the neutral Higgs boson mass for datasets collected in 2015 (top) and 2016 (bottom).
The operating points for the single (double) $b$-tagged jet triggers correspond to $b$-quark identification efficiencies of 79\% (72\%) in 2015 and 60\% (60\%) in 2016.}\label{fig:trigEff}
 
\end{center}
\end{figure}

\subsection{Signal model and background simulations}

Signal events for the subprocesses $bg\to b\phi +0,1$~jet, $gg\to b\bar{b}\phi$, and $q\bar{q}\to b\bar{b}\phi$ with $\phi\to b\bar{b}$  were generated at leading order (LO) for seventeen
$m_\phi$ values from 450 to 1400~\GeV\ using
the \textsc{Sherpa} 2.2.0~\cite{Gleisberg:2008ta} Monte Carlo (MC) program in the 5FS with the  NNPDF30NNLO~\cite{NNPDF} set of parton distribution functions (PDF).
In order to determine the total width at each value of $m_\phi$, a specific MSSM scenario tailored for large values of the branching ratio ($\mathcal{B}$) for $\phi\rightarrow b\bar{b}$ was used in which
$\tan \beta=20$, the higgsino mass parameter $\mu = -800$~\GeV, the generic soft-SUSY-breaking mass parameter
$M_\textrm{SUSY}=1000$~\GeV,
the trilinear Higgs--top-squark coupling $A_t=2000$~\GeV, the $SU(2)$ gaugino mass parameter $M_2=800$~\GeV, and the $SU(3)$ gaugino mass parameter $M_3=1600$~\GeV.
These parameters suppress
$\phi$ boson decays into top quark pairs, top-squark pairs, and electroweak gauginos,
while decays into pairs of $b$-quarks are enhanced through MSSM radiative corrections~\cite{Carena:2005ek}.
The FeynHiggs program~\cite{Heinemeyer:1998yj} was used to calculate
the branching ratios and the cross sections shown in Table~\ref{tab:mssmmodel}, where the
branching ratio $\mathcal{B}(\phi\rightarrow b\bar{b}) > 85\%$ for all $m_{\phi}$ values up to 1400~\GeV.
Given the large
values for $\mathcal{B}(\phi\rightarrow b\bar{b})$ in Table~\ref{tab:mssmmodel}, the total widths derived from this set of MSSM parameters also represent the typical total widths in the flipped scenario of the
2HDM in the alignment limit for the same $m_\phi$ and $\tan\beta$.
The values of the total width in
Table~\ref{tab:mssmmodel} are much smaller than the 10\%--15\% experimental $b\bar{b}$ mass resolution.   Although several decay modes are present in
this MSSM scenario, only the decay mode $\phi\to b\bar{b}$ is simulated in the
generated signal samples.  Since they are ignored in the analysis, additional $\phi$ decay modes that happen  to leak into the $\phi\to b\bar{b}$ acceptance will only  make any limits on $\phi\to b\bar{b}$
more conservative.
 
\begin{table}
\caption{Mass, total width, and branching ratios of the $\phi$ boson of the MSSM scenario used for signal event generation where $\tan \beta=20$, $\mu = -800$~\GeV, $M_\textrm{SUSY}=1000$~\GeV,
$A_t=2000$~\GeV, $M_2=800$~\GeV, and $M_3=1600$~\GeV. The $pp\rightarrow b\bar{b}\phi$ cross section at $\sqrt{s}=13$~\TeV\ is also shown.
Full simulation samples of 300,000 events were produced for each of the mass points.
The FeynHiggs program~\cite{Heinemeyer:1998yj} was used to calculate
the branching ratios and the cross sections for $pp\rightarrow b\bar{b}\phi$ at $\sqrt{s}=13$~\TeV. \label{tab:mssmmodel}}
\centering
\begin{tabular}{rrcccr}\toprule
\upstrut $m_{\phi}$ & $\Gamma_{\phi}(\textrm{total})$ & $\mathcal{B}(\phi\rightarrow b\bar{b})$ & $\mathcal{B}(\phi\rightarrow \tau^+\tau^-)$ & $\mathcal{B}(\phi\rightarrow t\bar{t})$ &
$\sigma(b\bar{b}\phi)$ \downstrut \\
$\lbrack$\GeV$\rbrack$ & $\lbrack$\GeV$\rbrack$ & & & & $\lbrack$fb$\rbrack$  \\\midrule
450 &  4.7 & 0.91 & 0.07 & 0.02 &  2792 \\
500 &  5.1 & 0.91 & 0.07 & 0.02 &  1694 \\
550 &  5.5 & 0.91 & 0.07 & 0.02 &  1066 \\
600 &  5.9 & 0.91 & 0.07 & 0.02 &  693 \\
650 &  6.3 & 0.91 & 0.07 & 0.02 &  463 \\
700 &  6.7 & 0.91 & 0.07 & 0.02 &  317 \\
750 &  7.1 & 0.90 & 0.07 & 0.02 &  222 \\
800 &  7.5 & 0.90 & 0.08 & 0.02 &  158 \\
850 &  7.9 & 0.90 & 0.08 & 0.02 &  115 \\
900 &  8.3 & 0.90 & 0.08 & 0.02 &   85 \\
950 &  8.7 & 0.90 & 0.08 & 0.02 &   66 \\
1000 &  9.1 & 0.90 & 0.08 & 0.02 &   48 \\
1100 &  9.8 & 0.90 & 0.08 & 0.02 &   29 \\
1200 & 10.7 & 0.89 & 0.08 & 0.02 &   18 \\
1300 & 11.6 & 0.87 & 0.08 & 0.02 &   12 \\
1350 & 12.1 & 0.87 & 0.08 & 0.02 &   10 \\
1400 & 12.5 & 0.86 & 0.08 & 0.02 &    8 \\\bottomrule
\end{tabular}
\end{table}

The ATLAS detector response to the generated signal events was modeled using the ATLAS full simulation software~\cite{SOFT-2010-01} based on \textsc{Geant4}~\cite{Agostinelli:2002hh}.
The impact of multiple $pp$ collisions in the same or nearby bunch crossings (pileup)
was simulated by overlaying minimum-bias events on each generated
event. The minimum-bias events were generated with \PYTHIA 8.186~\cite{Sjostrand:2007gs}, using the A2 set of tuned parameters~\cite{ATL-PHYS-PUB-2012-003} and the MSTW2008LO PDF sets~\cite{Martin:2009iq}.
Finally, events were processed using the same reconstruction software as in data.

The background estimate is data-driven, as described in Section~\ref{sec:fit}.  Background MC
samples (referred to later as `multijet MC samples') served as a guide
in developing the background model and
consist of a 	\textsc{Sherpa} 2.1.1 simulation of
multiple $b$-jets, a next-to-leading-order (NLO) \textsc{Powheg}~\cite{POWHEG0,POWHEG01,Frixione:2007nw} simulation of $t\bar t$ production interfaced to \textsc{Pythia} 6.428~\cite{Sjostrand:2006za},
and an LO \textsc{Madgraph}~\cite{Maltoni:2002qb} simulation of the subprocess $jj\to Z\ jj$,
$Z\to b\bar{b}$ interfaced to \textsc{Pythia}~8.205, where $j$ represents a gluon or a $u,d,s,c$ quark/antiquark.
The full ATLAS detector simulation software was used for $t\bar t$ production.
A fast ATLAS detector simulation in which the calorimeter response is
parametrized~\cite{SOFT-2010-01,ATL-PHYS-PUB-2010-013} was used for the multiple $b$-jets and
$jj\to Z\ jj$ samples.
 
The dominant background comes from the production of multiple $b$-jets.
In the 	\textsc{Sherpa} 2.1.1 simulation of multiple $b$-jets all $2\rightarrow 2,3,4$ hard subprocesses with at least one $b$-quark in the final state were generated at LO.
The $c$- and $b$-quarks masses are set to their running Yukawa values to properly simulate gluon splitting into heavy quarks.

\section{Object reconstruction and event selection}
\label{sec:selection}
 
Primary vertex candidates~\cite{PERF-2015-01} are reconstructed using tracks in the inner detector, and the vertex
with the highest sum of the squared transverse momenta of all associated tracks is selected as the hard-scatter primary vertex.
Jets are reconstructed using the \antikt
algorithm~\cite{Cacciari:2008gp} with radius parameter $R=0.4$ from topological clusters of energy in the calorimeter calibrated at the electromagnetic scale~\cite{PERF-2014-07}.
Jets are then calibrated using correction factors derived from simulation and data~\cite{PERF-2016-04}.
In order to suppress  jets
arising from pileup,
jets with transverse momentum $p_{\textrm{T}}<60$~\GeV\ and $|\eta|<2.4$ are removed if they fail to satisfy a requirement imposed by the multivariate jet vertex tagger (JVT) algorithm~\cite{PERF-2014-03},
where the JVT working point provides
a 92\% selection efficiency for hard-scatter jets.
In addition,
events with jets consistent with noise in the calorimeter or noncollision backgrounds are vetoed~\cite{ATLAS-CONF-2015-029}.

Jets containing a $b$-hadron are identified offline using the MV2c20 multivariate classifier~\cite{PERF-2012-04,ATLAS:2016-012},
which combines information from several algorithms. These algorithms are based on impact parameters
of tracks, reconstructed secondary vertices, and a multi-vertex fitter which reconstructs the $b\to c$ hadron decay chain.
A working point with an average $b$-tagging efficiency of 70\%, as determined using simulated $t\bar{t}$ events, is chosen.
The corresponding misidentification rates for $c$-jets and jets originating from light ($u,d,s$) quarks or gluons is 8.2\% and 0.3\%,
respectively.  Jets  tagged as $b$-jets receive an additional energy correction to account for the presence of reconstructed muons in the jet~\cite{HIGG-2016-29}.

Event preselection begins by requiring that the event pass the trigger selection and that there be at least three jets with $p_{\textrm{T}}>20$~\GeV\ and $|\eta|<2.4$.
The leading and second-leading jets (ordered in $p_{\textrm{T}}$) are then required to have $p_{\textrm{T1}}>160$~\GeV\ and
$p_{\textrm{T2}}>60$~\GeV, respectively. The two leading jets must also be $b$-tagged.
Events are considered to be in the signal region, and classified as \textit{bbb}, if there exists at least one additional $b$-tagged jet.
Events with only two
$b$-tagged jets are considered to be in the control region, and are classified as \textit{bbanti}.
For \textit{bbb} events the ``third jet'' is defined to be the third-leading $b$-tagged jet in $p_{\textrm{T}}$, while for \textit{bbanti} events
the third jet is the third-leading jet in $p_{\textrm{T}}$.
The final preselection requirement is that the minimum
$\Delta R$ between the third jet and the two leading jets
must be greater than 0.8.
This requirement
reduces the background from gluon splitting into $b\bar{b}$ in parton showers and
subprocesses such as $bg\to bg^*\to bb\bar{b}$.
 
Events are further classified according to the number of jets.   The 3-jet, 4-jet and 5-jet regions are defined,
where the last one contains events with five or more jets. A larger number of jets often means that significant
final-state radiation (FSR) is present in the $\phi$ boson decay,
making
it more difficult to accurately reconstruct $m_\phi$
from the two highest-$p_{\textrm{T}}$ jets.  For example, the root-mean-square values of the reconstructed signal MC mass distributions for
the two highest-$p_{\textrm{T}}$ jets for the 3-jet, 4-jet and 5-jet regions are 56~GeV, 65~GeV, and 74~GeV,
respectively at  $m_\phi$=600~GeV, and 163~GeV, 196~GeV, and  210~GeV, respectively at  $m_\phi$=1200~GeV.
Studies performed with and without jet multiplicity categorization demonstrate improvement in the expected limits
from this categorization.

Signal sensitivity is enhanced with a transformation of the kinematic variables
$p_{\textrm{T1}}$, $p_{\textrm{T2}}$, and the invariant mass of the two leading $b$-tagged jets, $m_{bb}$.
Two-dimensional distributions of $p_{\textrm{T1}}$ versus $m_{bb}$ and of
$p_{\textrm{T2}}$ versus   $m_{bb}$ for events with the \textit{bbb} classification
are displayed in Figure~\ref{fig:ptvsmbb}.   As $m_\phi$
increases the two high-$p_{\textrm{T}}$ jets from the $\phi$ boson decay produce additional FSR, but the jet radius parameter remains fixed at $R=0.4$.
As a consequence, the reconstructed mass distribution based on the two highest-$p_{\textrm{T}}$ jets smears out and it becomes more difficult to distinguish
signal from background. However, since FSR occurs stochastically, the $\phi$ boson decays with little or no FSR in a subset of the signal events, and these have
reconstructed masses close to the true $m_\phi$ (bottom row of Figure~\ref{fig:ptvsmbb}).  If these events can be
isolated from the others, they offer a chance to improve the sensitivity via improved mass resolution
and signal-to-background ratio.

To isolate events with small FSR and good $m_\phi$ resolution, a principal component analysis (PCA)~\cite{PCAbook} is performed on the three-dimensional distribution of the
variables $m_{bb}$, $p_{\textrm{T1}}$,  and $p_{\textrm{T2}}$ using events drawn from the signal MC sample with the \textit{bbb} classification following preselection.
Separate PCAs
are performed for each of the seventeen simulated
values of $m_\phi$ and  for each of the three $n$-jet regions.
Upon diagonalization of the covariance matrix for $m_{bb}$, $p_{\textrm{T1}}$,  and $p_{\textrm{T2}}$, the first, second, and third principal components
define the variables $m_{bb}'$,  $p_{\textrm{T1}}'$, and $p_{\textrm{T2}}'$, respectively.  The point $(m_{bb}',p_{\textrm{T1}}',p_{\textrm{T2}}')=(0,0,0)$ corresponds to the vector
of mean values for $m_{bb}$, $p_{\textrm{T1}}$,  and $p_{\textrm{T2}}$.
Two-dimensional distributions of $p_{\textrm{T1}}'$ versus $m_{bb}'$ and of $p_{\textrm{T2}}'$ versus $m_{bb}'$
are shown in Figure~\ref{fig:ptpvsmbbp}.
 
\begin{figure}[htbp]
\centering
\begin{tabular}{c}
\includegraphics[width=0.43\linewidth]{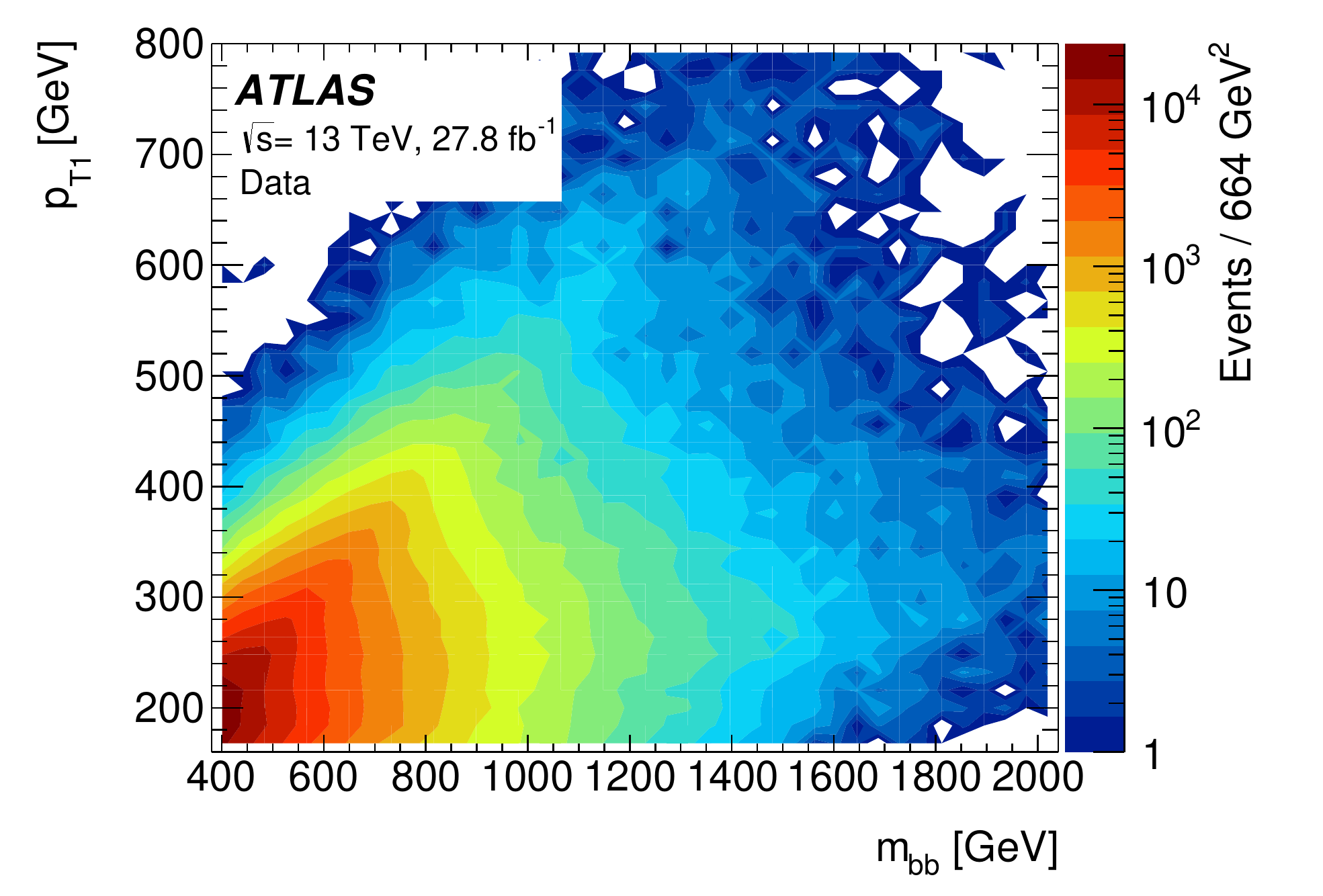}
\includegraphics[width=0.43\linewidth]{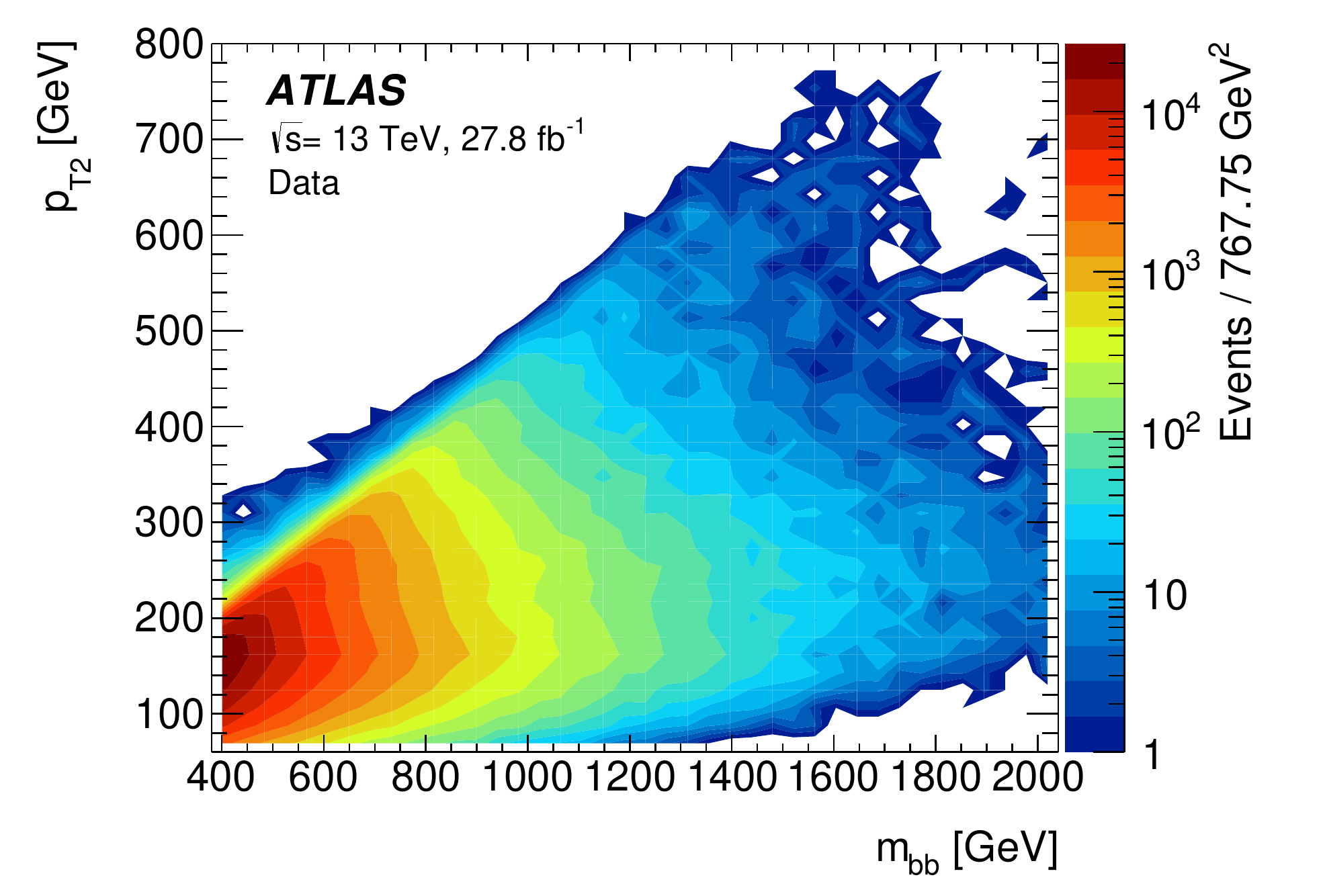}
\\
\includegraphics[width=0.43\linewidth]{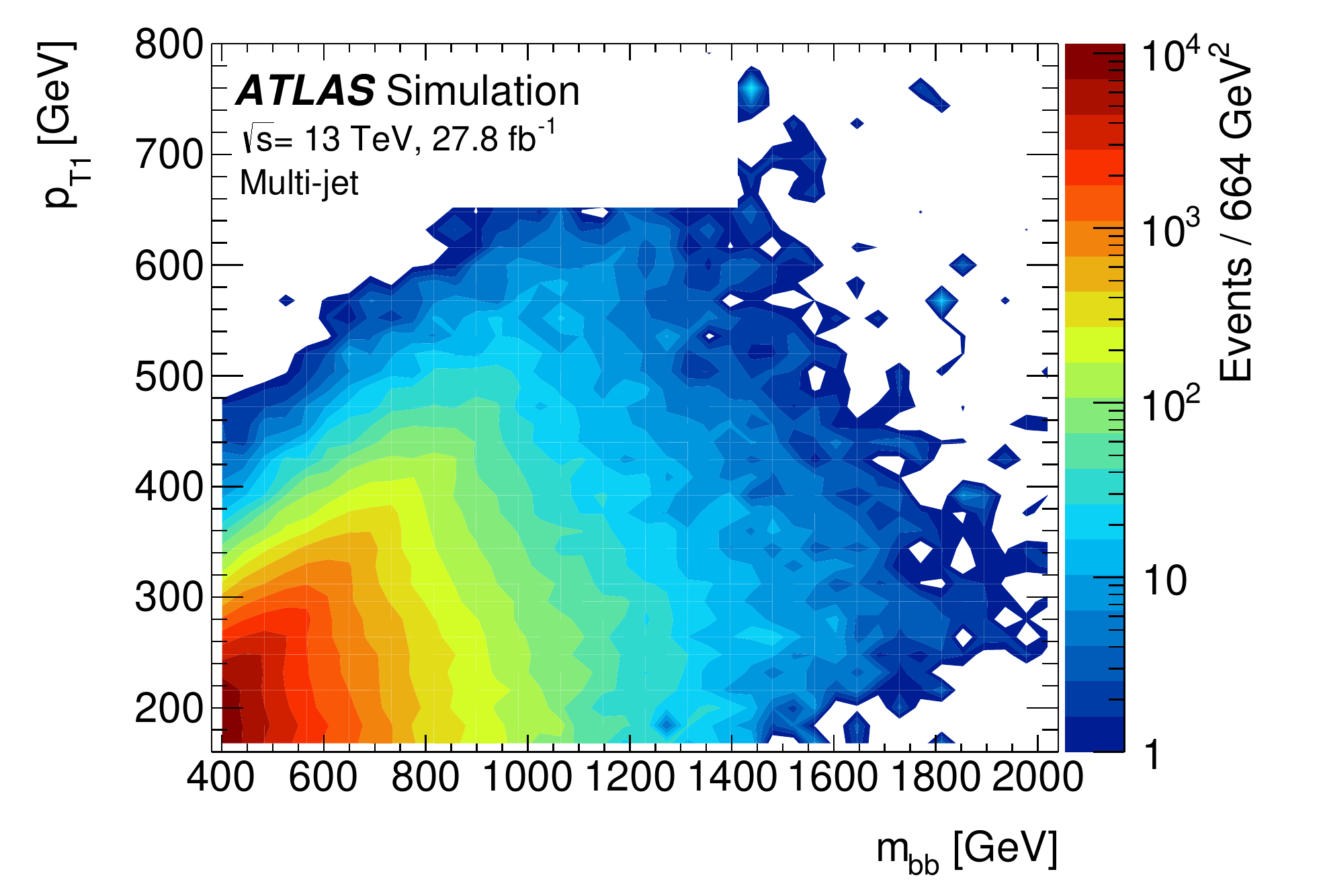}
\includegraphics[width=0.43\linewidth]{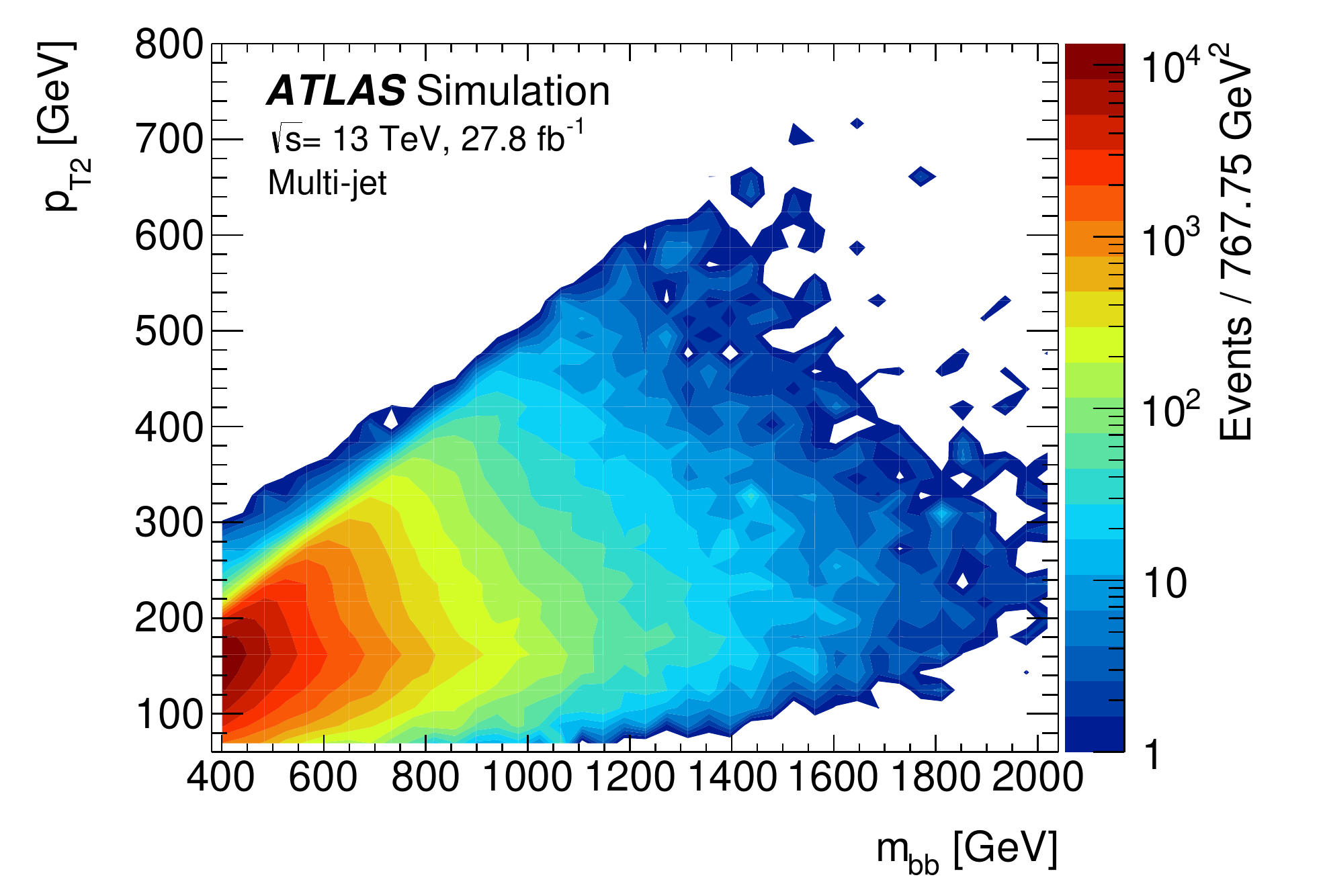}
\\
\includegraphics[width=0.43\linewidth]{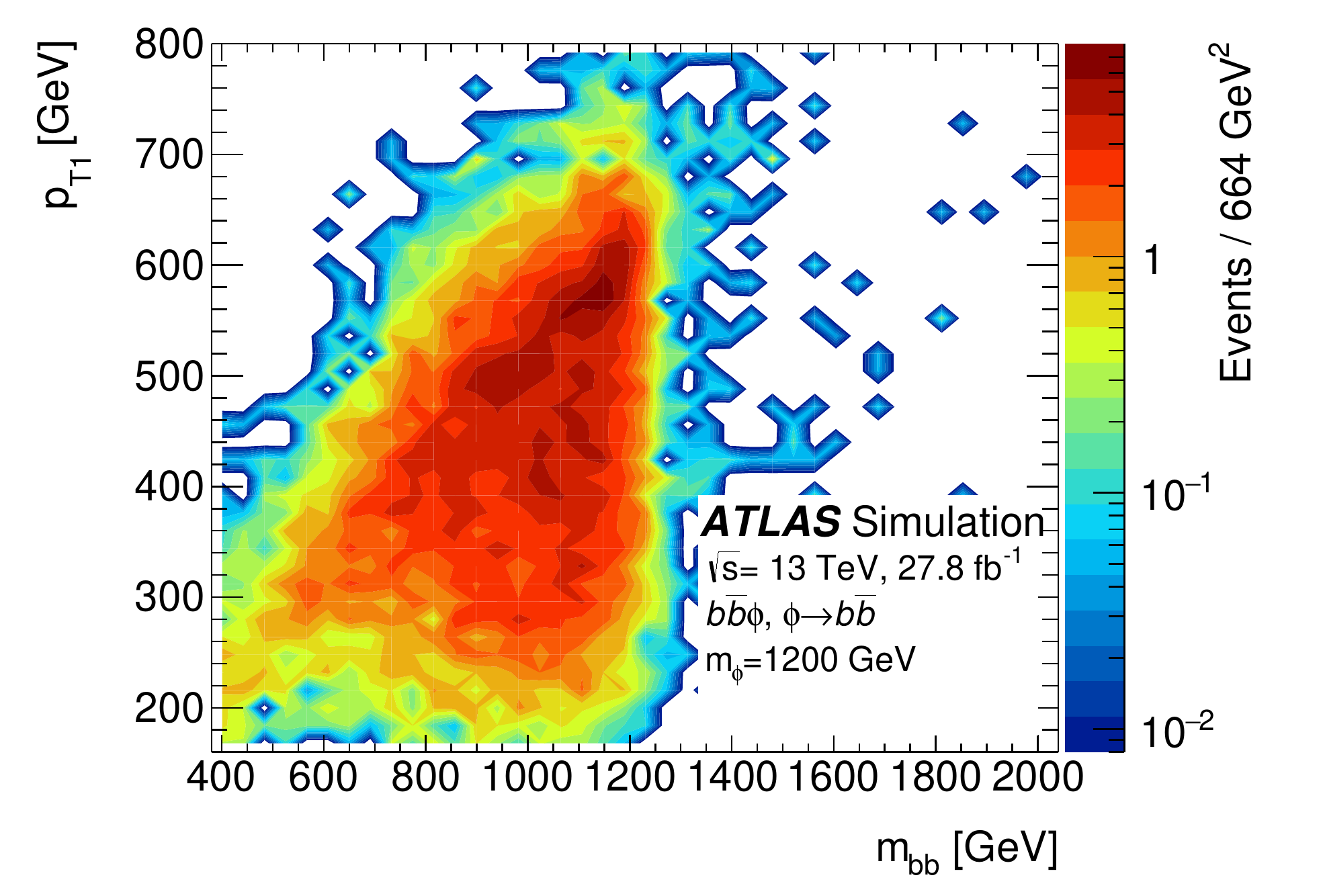}
\includegraphics[width=0.43\linewidth]{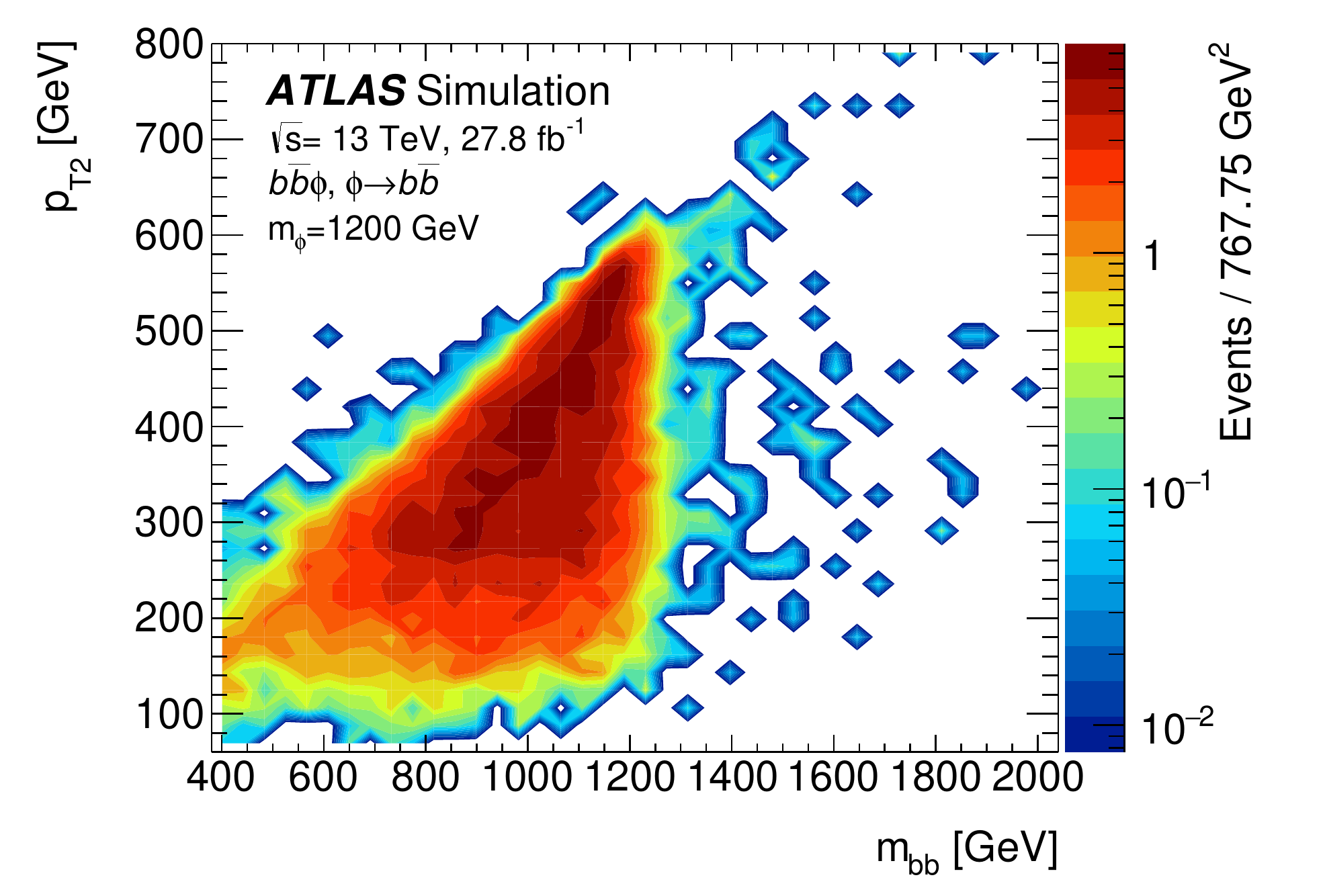}
\\
\end{tabular}
\caption{Two-dimensional distributions of $p_{\textrm{T1}}$ versus $m_{bb}$ (left column) and of $p_{\textrm{T2}}$ versus $m_{bb}$ (right column) for events with the \textit{bbb} classification following preselection, summed
over all three $n$-jet regions.  Plots are shown for data (top row),
the multijet MC sample (middle row), and the $m_\phi=1200$~\GeV\ signal MC sample (bottom row). The  multijet plots are normalized to an integrated luminosity of 27.8~fb$^{-1}$ based on the cross sections provided by the event generators.
The signal plots are normalized to
$\sigma(pp\rightarrow b\bar{b}\phi)\times \mathcal{B}(\phi\rightarrow b\bar{b})=1$~pb. \label{fig:ptvsmbb}}
\end{figure}

\begin{figure}[htbp]
\centering
\begin{tabular}{c}
\includegraphics[width=0.43\linewidth]{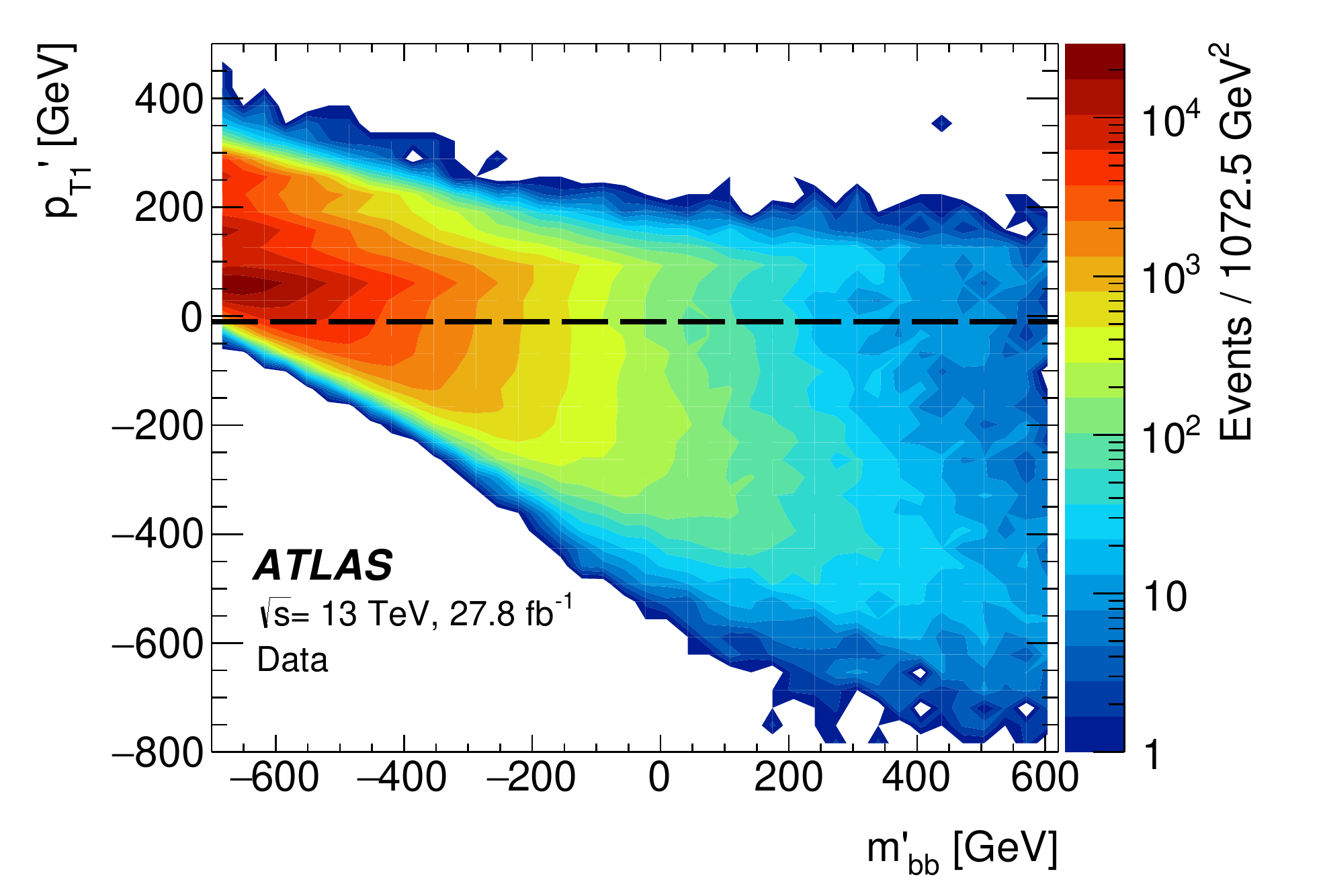}
\includegraphics[width=0.43\linewidth]{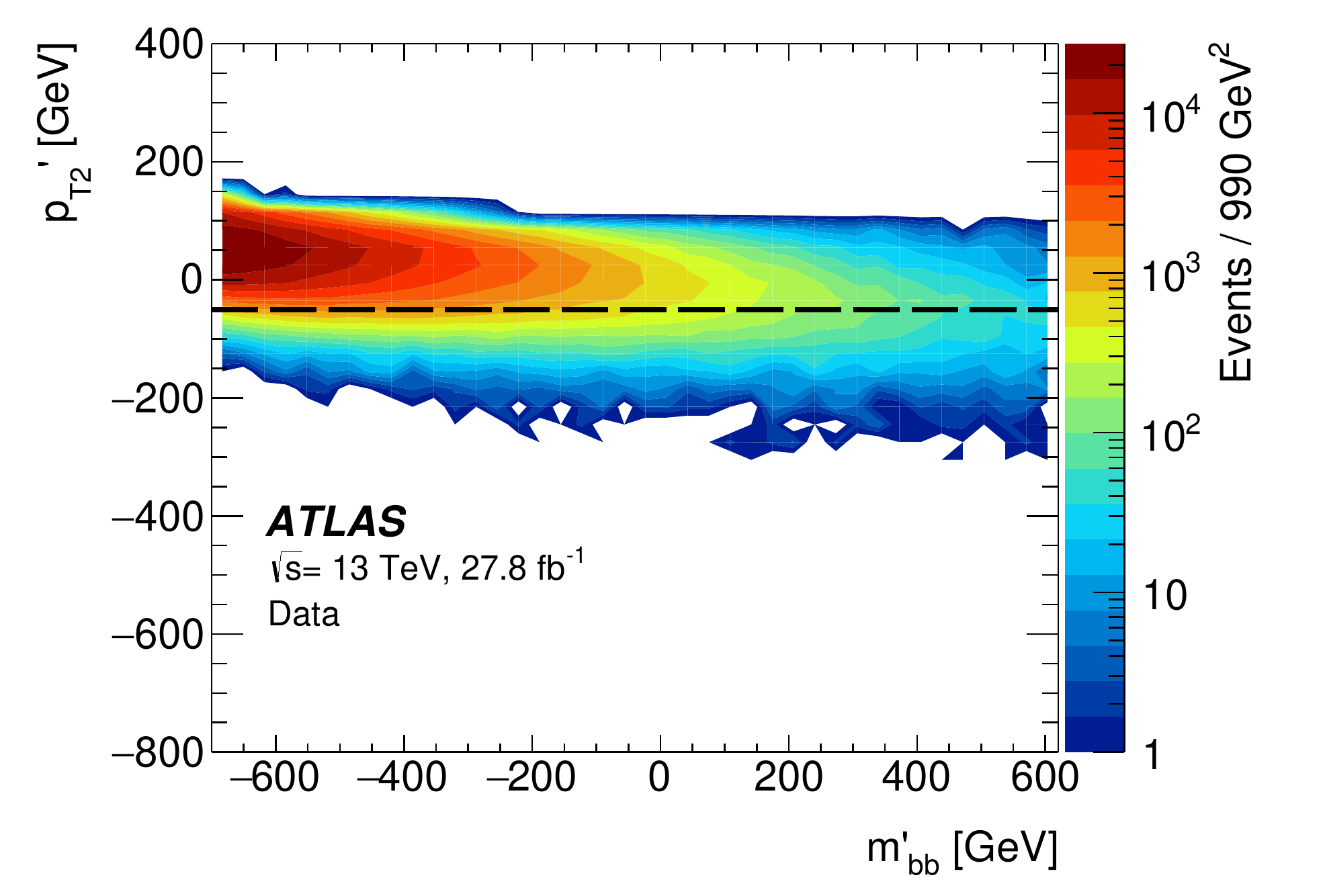}
\\
\includegraphics[width=0.43\linewidth]{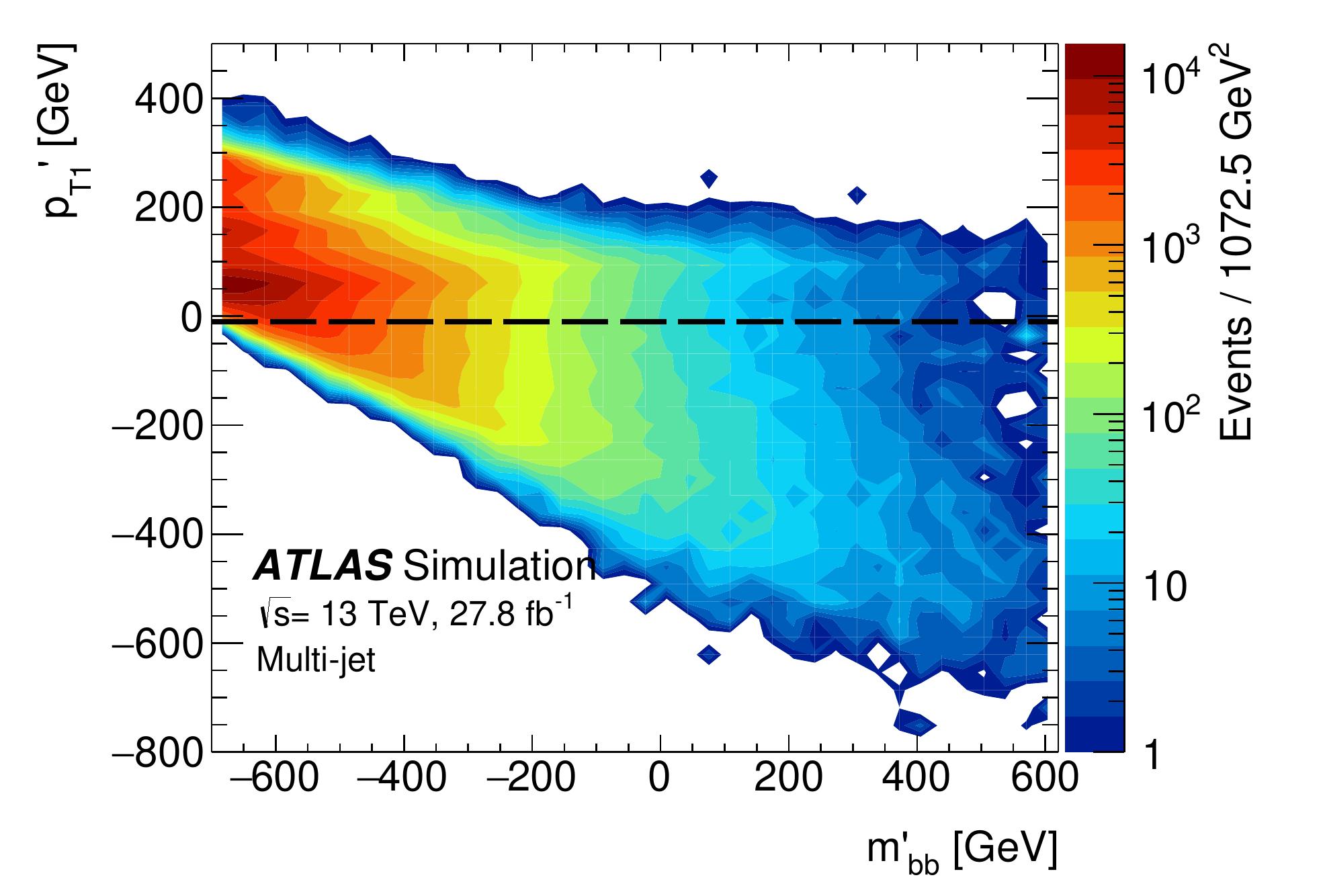}
\includegraphics[width=0.43\linewidth]{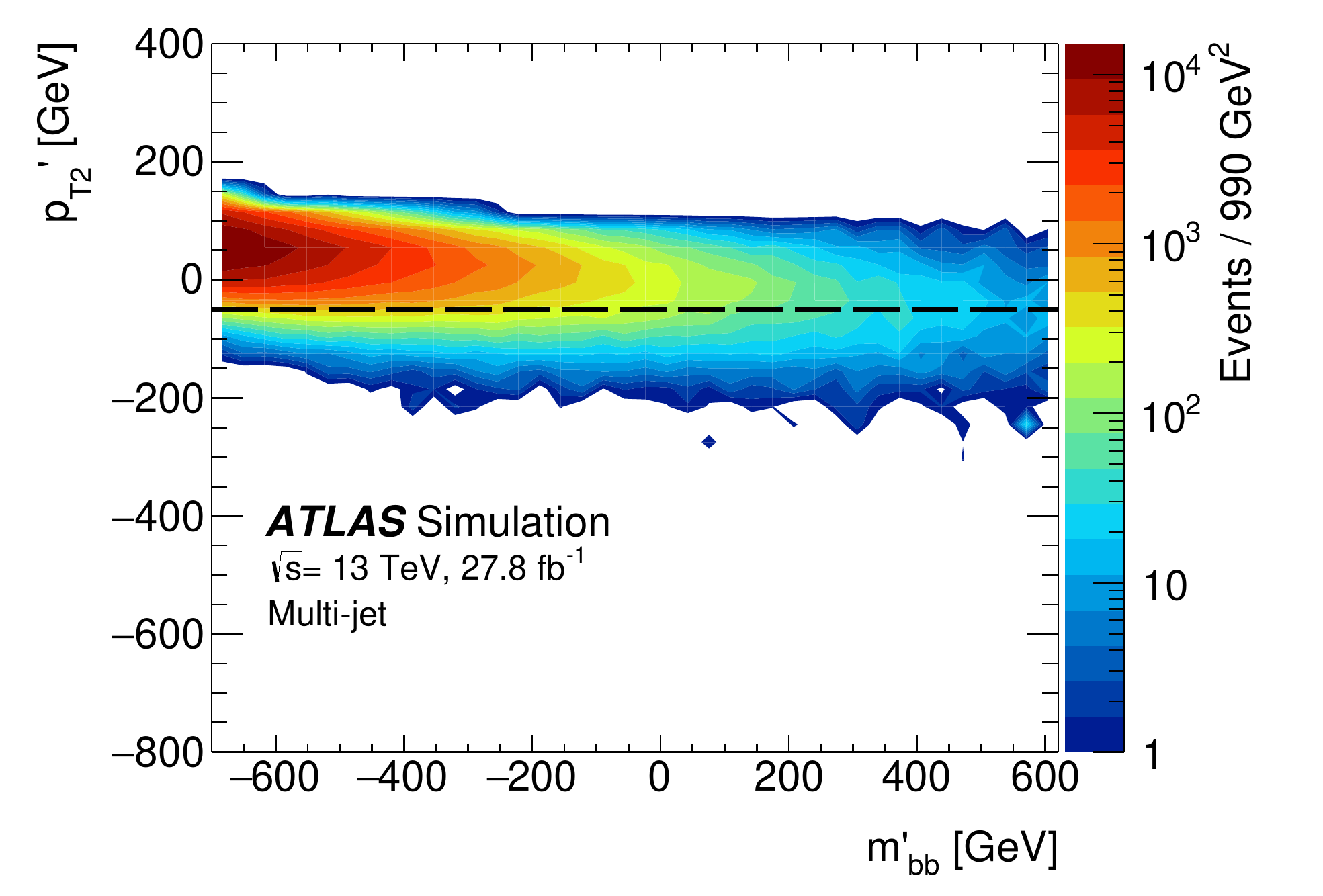}
\\
\includegraphics[width=0.43\linewidth]{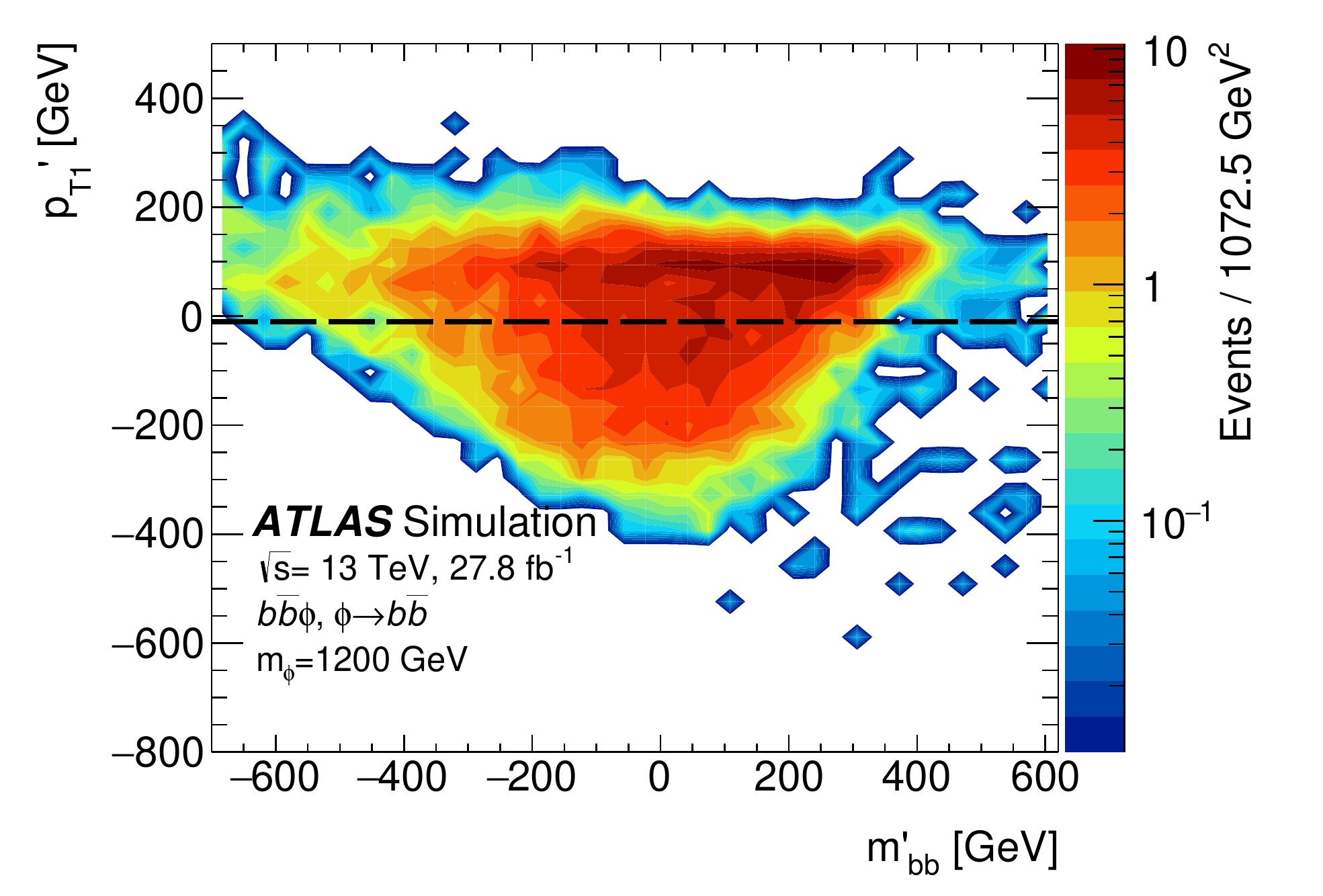}
\includegraphics[width=0.43\linewidth]{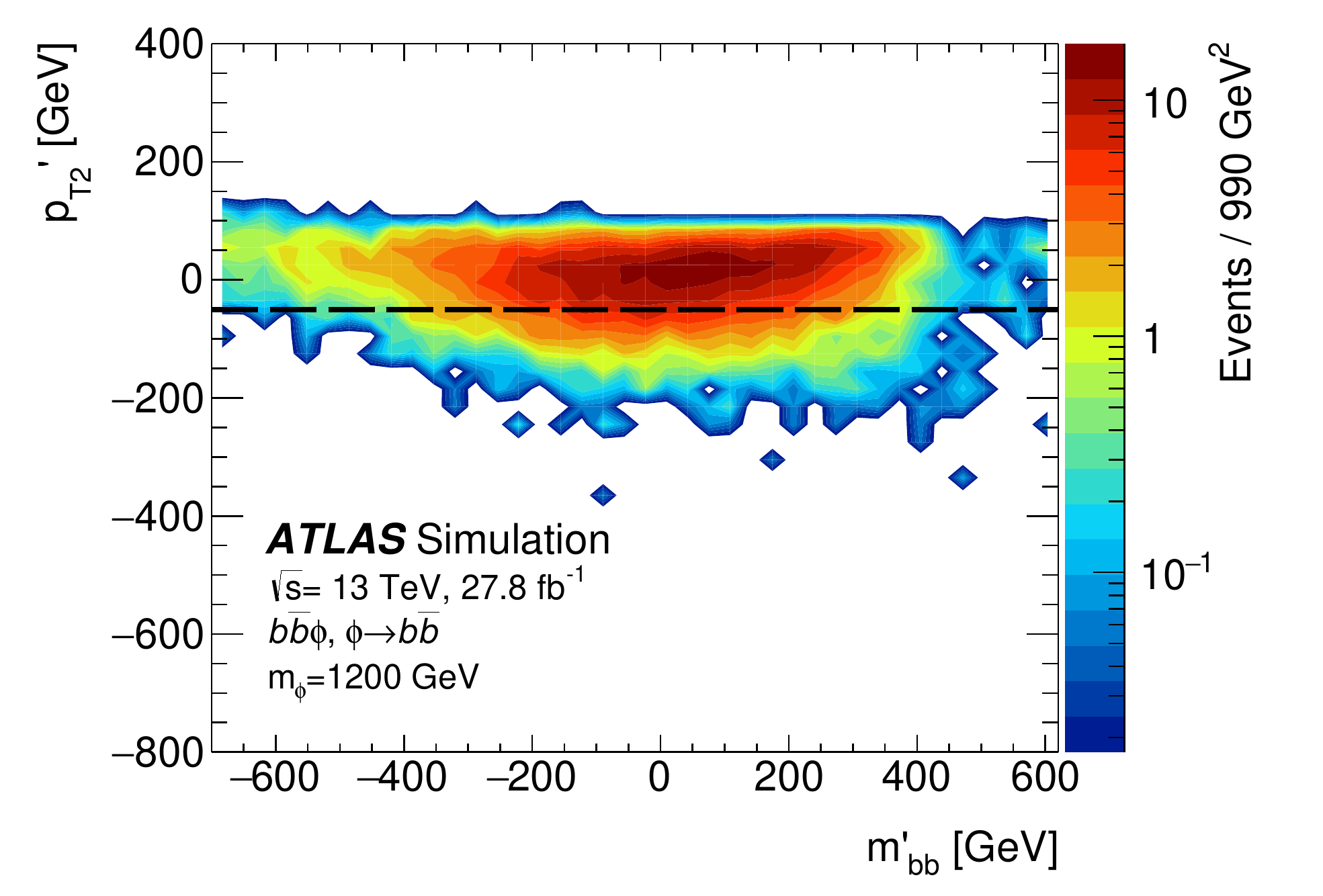}
\\
\end{tabular}
\caption{Two-dimensional distributions of $p_{\textrm{T1}}'$ versus $m_{bb}'$ (left column) and of $p_{\textrm{T2}}'$ versus $m_{bb}'$ (right column) for events with the \textit{bbb} classification following preselection, summed
over all three $n$-jet regions, using the PCA for $m_\phi=1200$~\GeV.   Plots are shown for data (top row),
the multijet MC sample (middle row), and the $m_\phi=1200$~\GeV\ signal MC sample (bottom row).
The  multijet plots are normalized to an integrated luminosity of 27.8~fb$^{-1}$ based on the cross sections provided by the event generators.
The  signal plots are normalized to
$\sigma(pp\rightarrow b\bar{b}\phi)\times \mathcal{B}(\phi\rightarrow b\bar{b})=1$~pb.
The minimum values of  $p_{\textrm{T1}}'$ and $p_{\textrm{T2}}'$ in the final event selection are indicated by horizontal lines.  \label{fig:ptpvsmbbp}}
\end{figure}

The variables $m_{bb}$, $p_{\textrm{T1}}$,  and $p_{\textrm{T2}}$ that are used to calculate the PCA variables are reconstructed signal MC quantities, and are subject to the
experimental systematic uncertainties, such as jet energy scale uncertainty, discussed in Sec.~\ref{sec:systematics}.  These uncertainties can lead to jet-multiplicity
migration and other effects which alter the event populations in the  3-jet, 4-jet and 5-jet regions and change the PCA variables.   Full error propagation is performed to account for
these effects, from variation in $m_{bb}$, $p_{\textrm{T1}}$,  and $p_{\textrm{T2}}$
to jet-multiplicity migration, PCA variable calculation, and, finally, experimental sensitivity to the Higgs boson cross section times branching ratio.

Examples of the relative contributions of $m_{bb}$, $p_{\textrm{T1}}$, and $p_{\textrm{T2}}$ to the
rotated variable $m_{bb}'$ are shown in Table~\ref{tab:eigenvector_elements}
for the 3-jet region.
The largest component of
$m_{bb}'$ comes from $m_{bb}$, regardless of the mass point.  However,
at larger values of $m_\phi$ - where there is greater FSR - the $p_{\textrm{T1}}$ and $p_{\textrm{T2}}$
components become more important.

The final event selection requirements
are $p_{\textrm{T1}}' >-10$~\GeV\ and
$p_{\textrm{T2}}'>-50$~\GeV, independent of $n$-jet and $m_\phi$.
These requirements reduce the background while retaining a large fraction of the signal events in regions of high signal density in $(m_{bb}', p_{\textrm{T1}}', p_{\textrm{T2}}')$ space, as shown in Figure~\ref{fig:ptpvsmbbp}.
Using $m_{bb}'$ instead of $m_{bb}$ as the final discriminant
leads to increased sensitivity, which becomes more pronounced with increasing values of $m_\phi$.
 
\begin{table}
\caption{The squares of the elements $c_{m_{bb}}$, $c_{p_{\textrm{T1}}}$, and $c_{p_{\textrm{T2}}}$ of the first principal component eigenvectors for the PCAs for
$m_\phi$ = 600, 900, and 1200~\GeV\ and each $n$-jet region.
The eigenvectors are normalized to unity.
The principal component is given by $m_{bb}'=c_{m_{bb}}(m_{bb}-\langle m_{bb}\rangle) + c_{p_{\textrm{T1}}}(p_{\textrm{T1}}-\langle p_{\textrm{T1}}\rangle)+c_{p_{\textrm{T2}}}(p_{\textrm{T2}}-\langle p_{\textrm{T2}}\rangle)$,
where $\langle m_{bb}\rangle$, $\langle p_{\textrm{T1}}\rangle$, and $\langle p_{\textrm{T2}}\rangle$ are the mean values for $m_{bb}$, $p_{\textrm{T1}}$,  and $p_{\textrm{T2}}$, respectively.
The transformation into the eigenbasis can provide some physical intuition for the relative contributions of $m_{bb}$, $p_{\textrm{T1}}$, and $p_{\textrm{T2}}$. \label{tab:eigenvector_elements}}
\centering
\begin{tabular}{ r    c c c | c c c | c c c } \toprule
&  \multicolumn{3}{ c| }{3-jet} &  \multicolumn{3}{ c|}{4-jet} &  \multicolumn{3}{ c }{5-jet} \\[3pt]
$m_\phi$ $\lbrack$\GeV$\rbrack$  & $c_{m_{bb}}^2$ & $c_{p_{\textrm{T1}}}^2$ & $c_{p_{\textrm{T2}}}^2$      & $c_{m_{bb}}^2$ & $c_{p_{\textrm{T1}}}^2$ & $c_{p_{\textrm{T2}}}^2$      & $c_{m_{bb}}^2$ & $c_{p_{\textrm{T1}}}^2$ & $c_{p_{\textrm{T2}}}^2$    \\ \midrule
600 &         0.74 & 0.12 & 0.14 &           0.76 & 0.12 & 0.13 &           0.82 & 0.10 & 0.08 \\
900 &         0.55 & 0.20 & 0.26 &           0.64 & 0.16 & 0.20 &           0.74 & 0.14 & 0.13  \\
1200 &         0.45 & 0.25 & 0.30 &           0.59 & 0.20 & 0.21 &           0.72 & 0.15 & 0.14 \\
\bottomrule
\end{tabular}
\end{table}

\section{Statistical analysis}
\label{sec:fit}
 
The presence of a signal is tested with a binned maximum-likelihood fit to the data using
$m'_{bb}$ as the final discriminating variable.
For each of the considered mass points, a fit is performed simultaneously over six categories corresponding to all combinations of the three jet-multiplicity regions (3-jet, 4-jet, and 5-jet) and of the two $b$-tag classifications, \textit{bbb} and \textit{bbanti}.
The shapes and normalizations
for the different  categories consist of a sum of signal and background contributions.
The shapes and normalizations of the signal distributions are obtained from signal MC samples, with the exception of a global normalization factor representing the primary variable of interest,
the heavy Higgs bosons production cross section times branching ratio $\sigma(pp\rightarrow b\bar{b}\phi)\times \mathcal{B}(\phi\rightarrow b\bar{b})$.
The shapes and normalizations of the background distributions are determined by data.  The background shapes are free to take any form satisfying the constraint that the \textit{bbb} and \textit{bbanti} shapes for a specific jet-multiplicity region be identical modulo a second-order polynomial correction factor.
The six background normalization factors float freely in the fit.

Based on the multijet MC sample, the backgrounds for both the \textit{bbb} and \textit{bbanti} regions are dominated by the subprocesses $gg\rightarrow gb\bar{b}$ and $gg\rightarrow ggb\bar{b}$ (such events enter the
\textit{bbb} region via gluon splitting into $b\bar{b}$ in the parton showering of one of the final-state gluons).  However, the subprocesses
$gb\rightarrow bb\bar{b}$ and $gg\rightarrow b\bar{b}b\bar{b}$ uniquely provide a small but non-negligible contribution to the \textit{bbb} background, and the polynomial correction factor accounts for this
and any other differences between the \textit{bbb} and \textit{bbanti} regions.
The $m_{bb}'$ distributions for both the $bbb$ and $bbanti$ classifications are plotted in Figure~\ref{fig:mbbpRatiosQCDAlt2} for the multijet MC sample along with their ratio.
The $bbb$ and $bbanti$ shapes for the 3-jet and 4-jet regions in Figure~\ref{fig:mbbpRatiosQCDAlt2} are nearly identical, while the $bbb/bbanti$ ratio for the
5-jet region appears to have an approximately linear dependence on $m'_{bb}$.
Application of the $\chi^2$ probability test and the $F$-test~\cite{Roe:2001aa}  to the simulated multijet $m'_{bb}$ distributions over all values of
$m_\phi$
demonstrates that a first-order polynomial is sufficient to describe the ratio of the simulated multijet $bbb$ and $bbanti$ background shapes for signal masses $m_\phi < 1200$~\GeV, while a second-order polynomial is needed for $m_\phi \geq 1200$~\GeV.

\begin{figure}[htbp]
\centering
\begin{tabular}{c}
\includegraphics[width=0.43\linewidth]{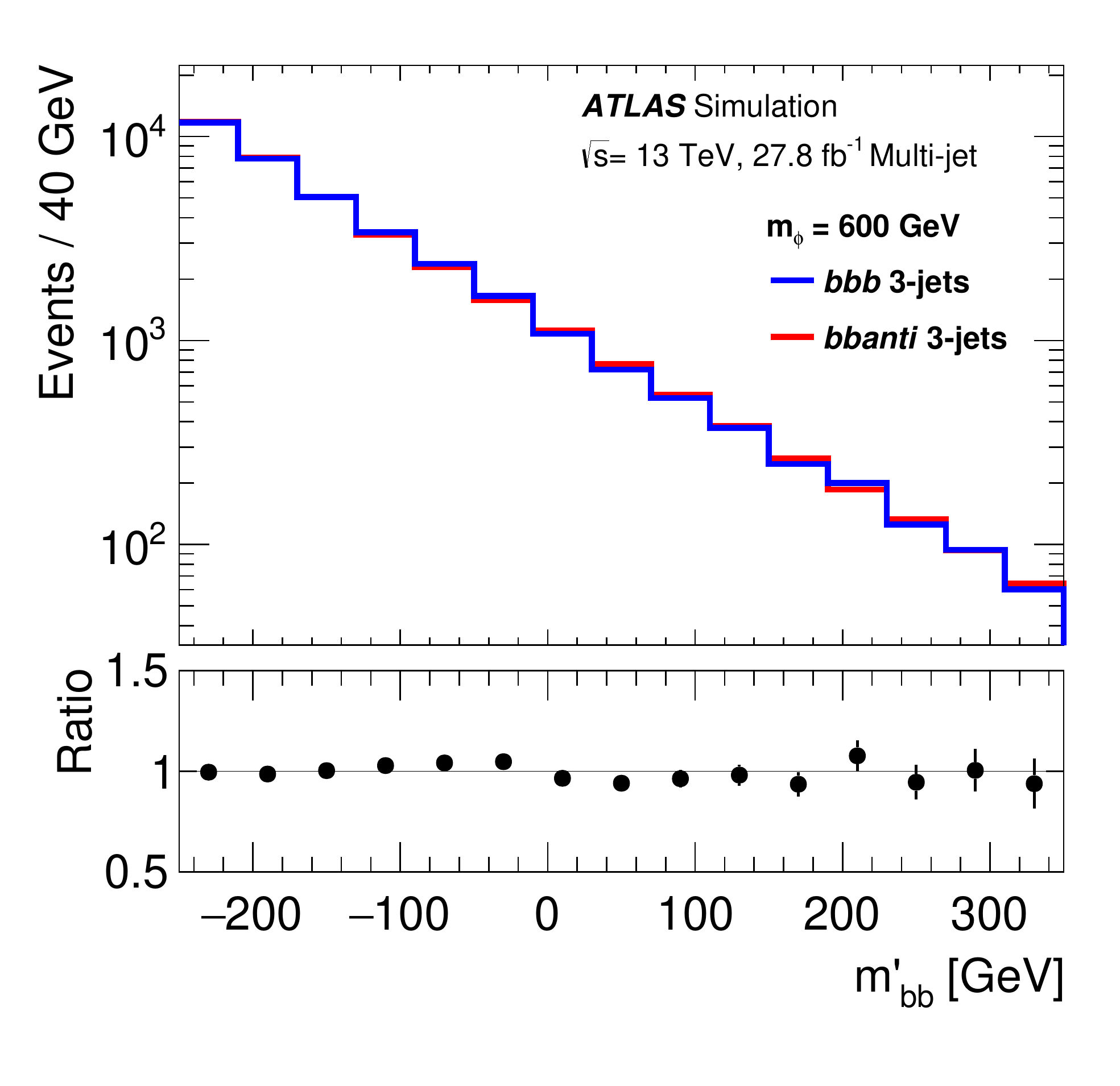}
\includegraphics[width=0.43\linewidth]{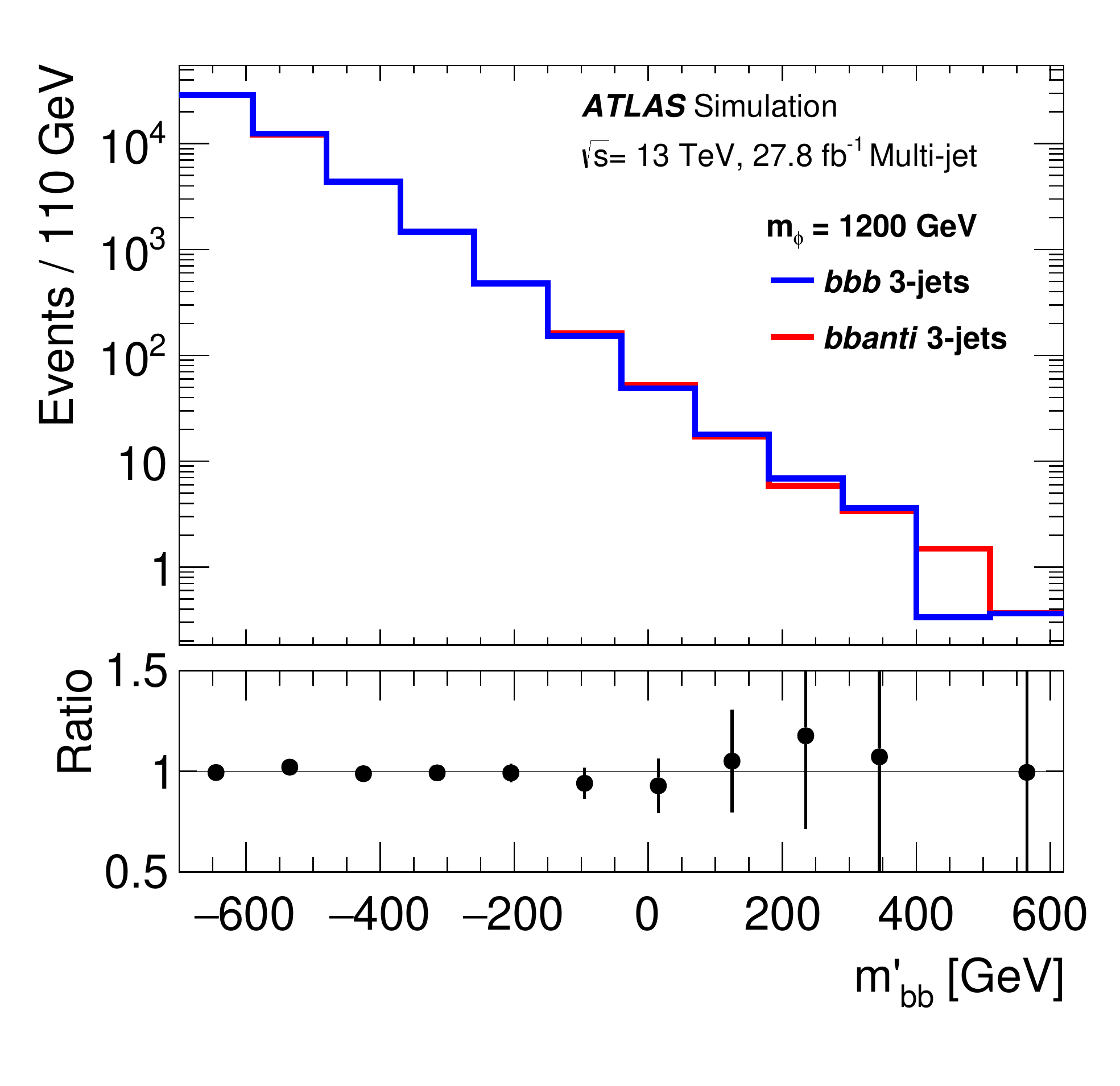}
\\
\includegraphics[width=0.43\linewidth]{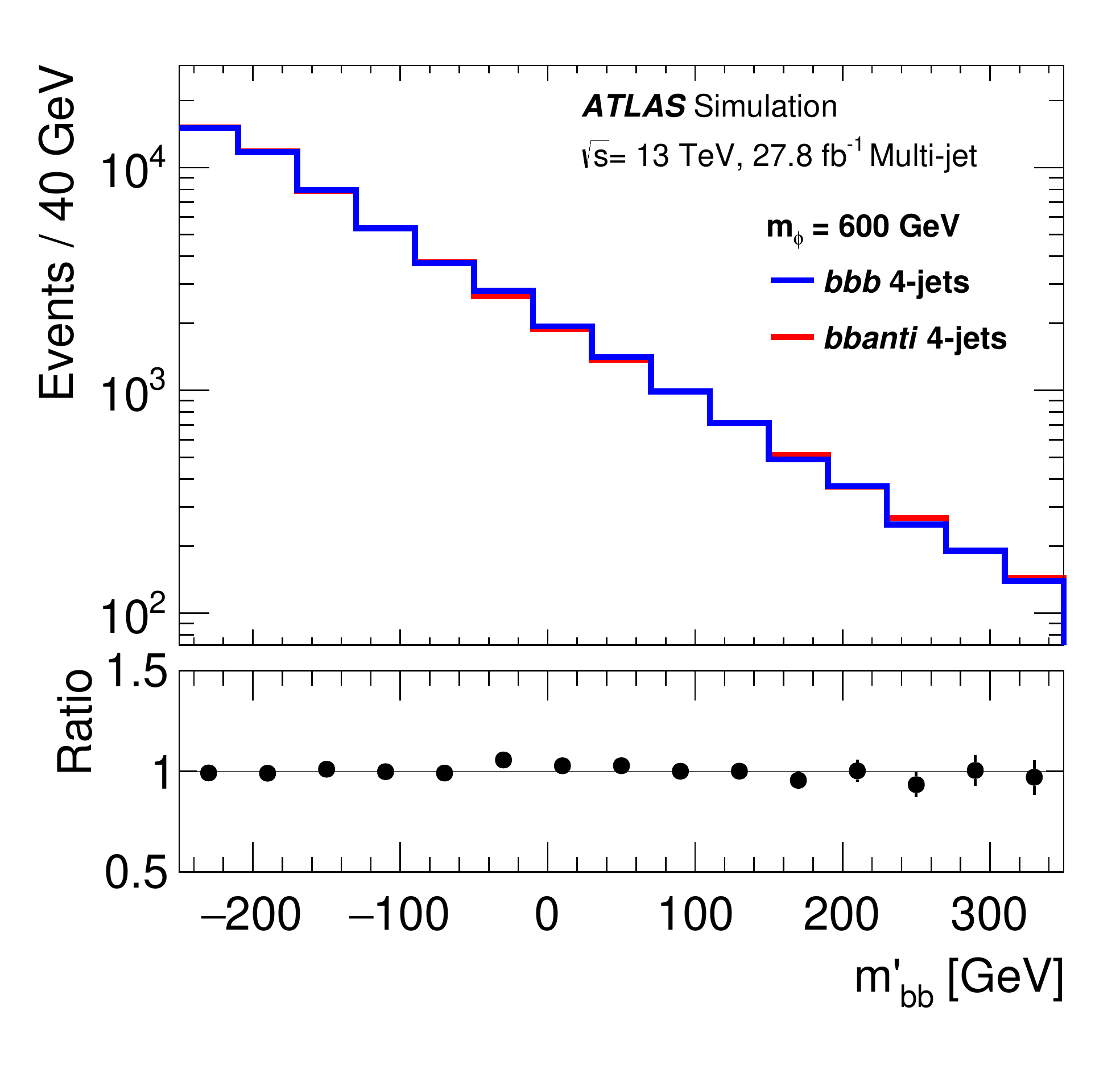}
\includegraphics[width=0.43\linewidth]{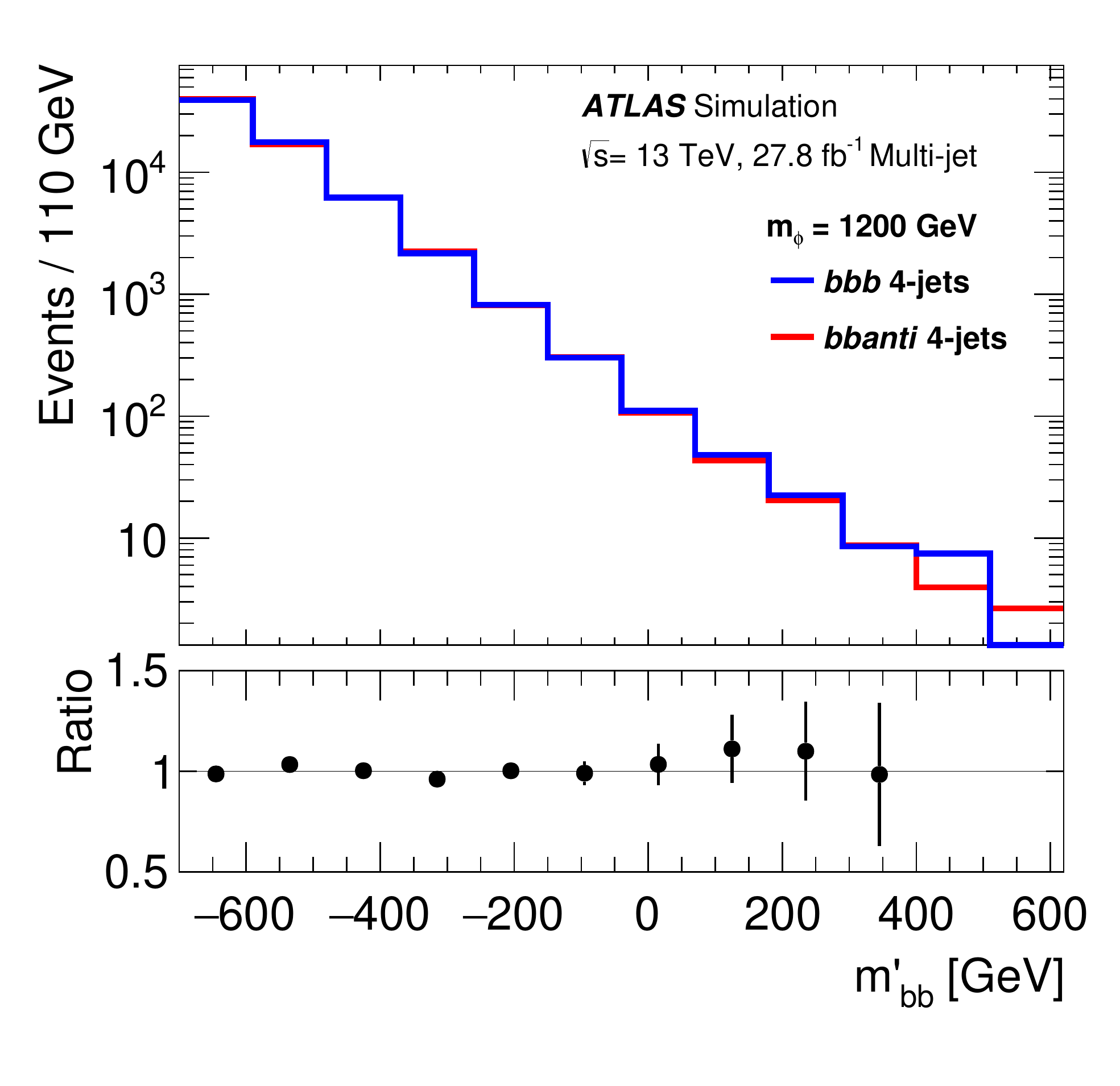}
\\
\includegraphics[width=0.43\linewidth]{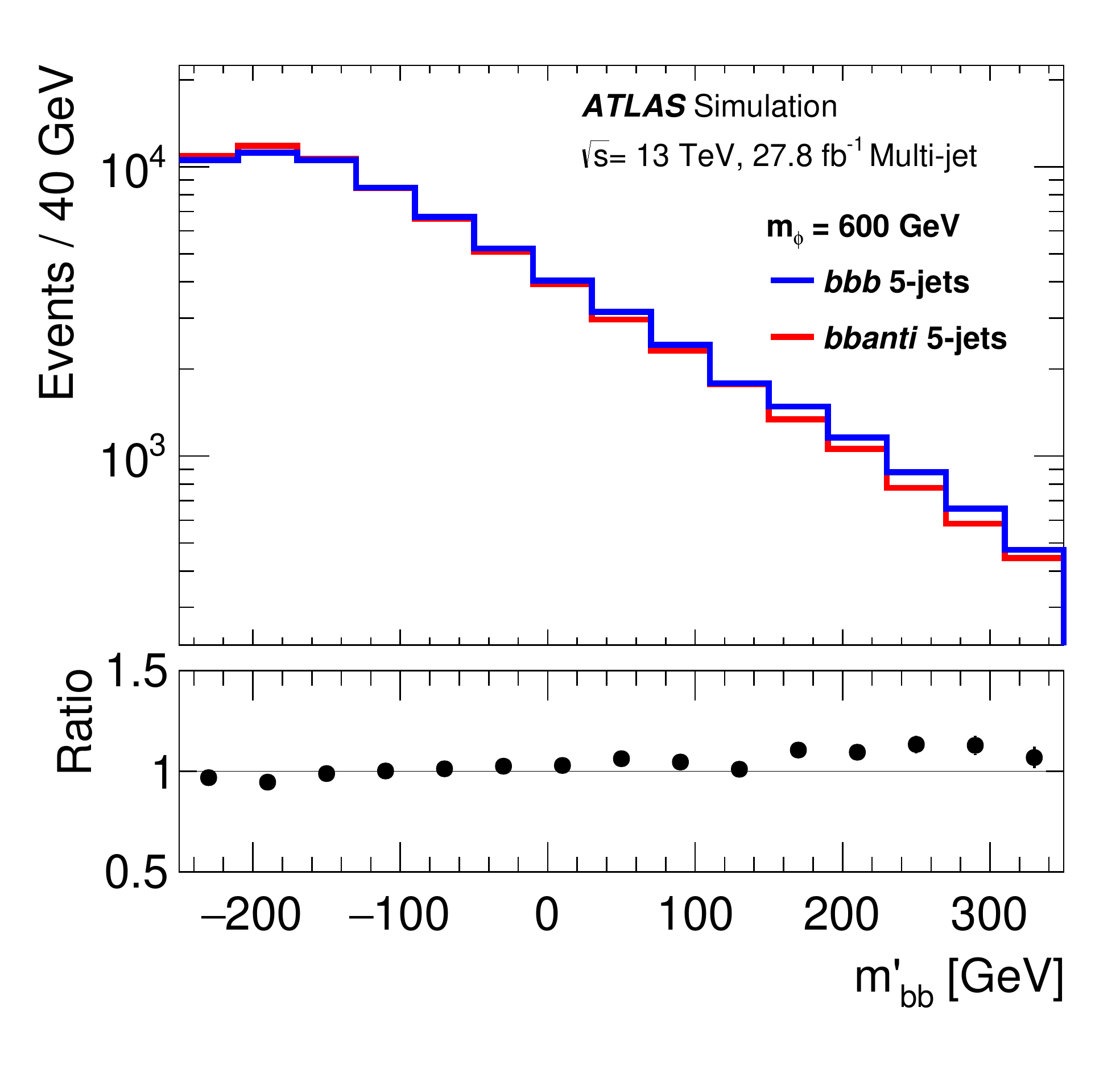}
\includegraphics[width=0.43\linewidth]{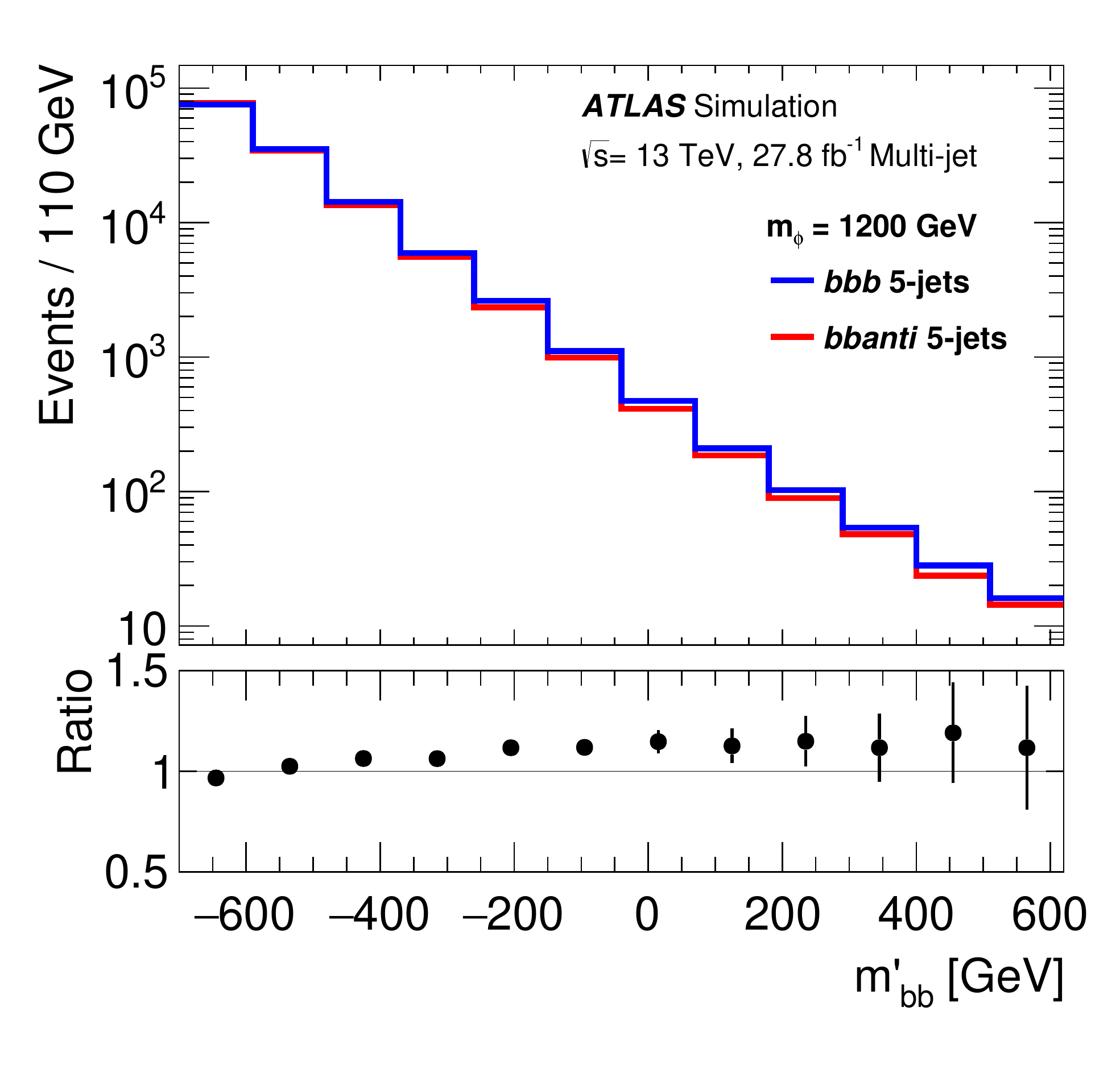}
\\
\end{tabular}
\caption{Distributions of $m_{bb}'$ in simulated multijet events with the \textit{bbb} and \textit{bbanti}
classifications following all selection requirements for the 3-jet (top row), 4-jet (middle row), and 5-jet (bottom row) regions.
The definition of the PCA variable $m_{bb}'$ depends on the mass hypothesis $m_\phi$
and distributions are shown for $m_\phi=600$~\GeV\ (left column) and $m_\phi=1200$~\GeV\ (right column).
For each case, the \textit{bbanti} distribution is normalized to an integrated luminosity of 27.8~fb$^{-1}$ based on the cross section provided by the event generator, and the \textit{bbb}
distribution is normalized to the same area as the \textit{bbanti} distribution. The ratios of \textit{bbb} to \textit{bbanti}
are also shown. \label{fig:mbbpRatiosQCDAlt2}}
\end{figure}

\begin{figure}[htbp]
\centering
\begin{tabular}{c}
\includegraphics[width=0.43\linewidth]{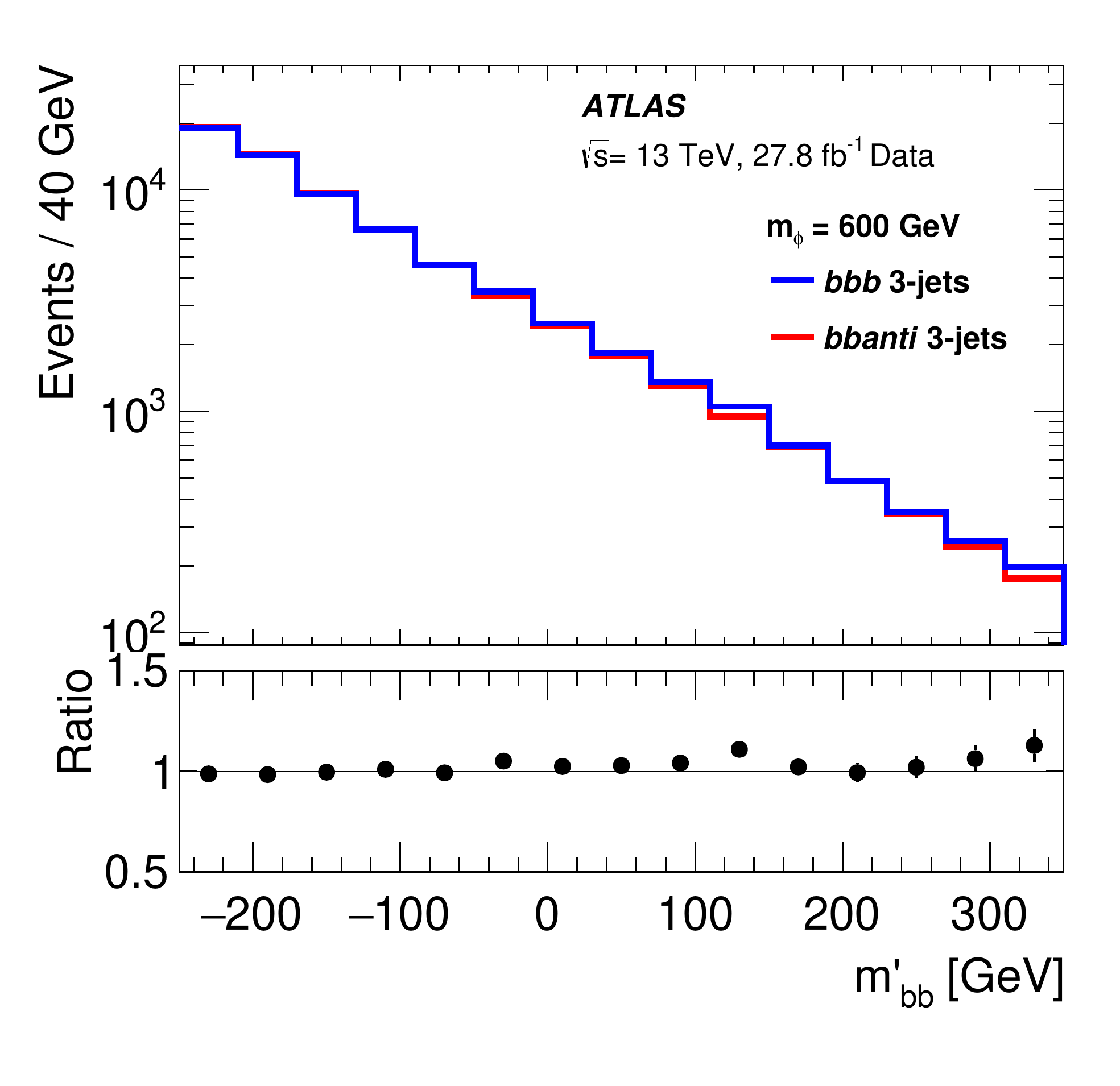}
\includegraphics[width=0.43\linewidth]{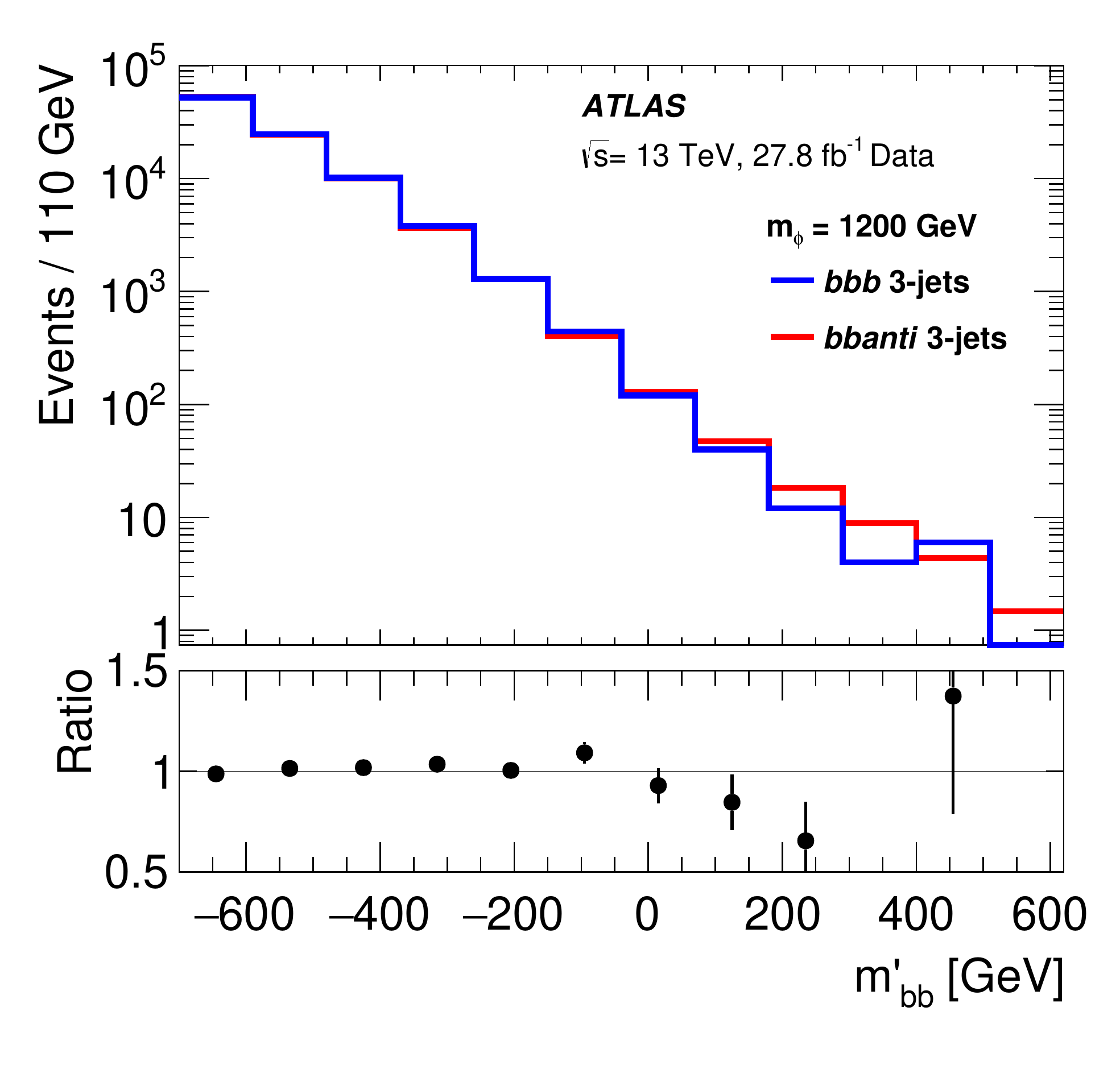}
\\
\includegraphics[width=0.43\linewidth]{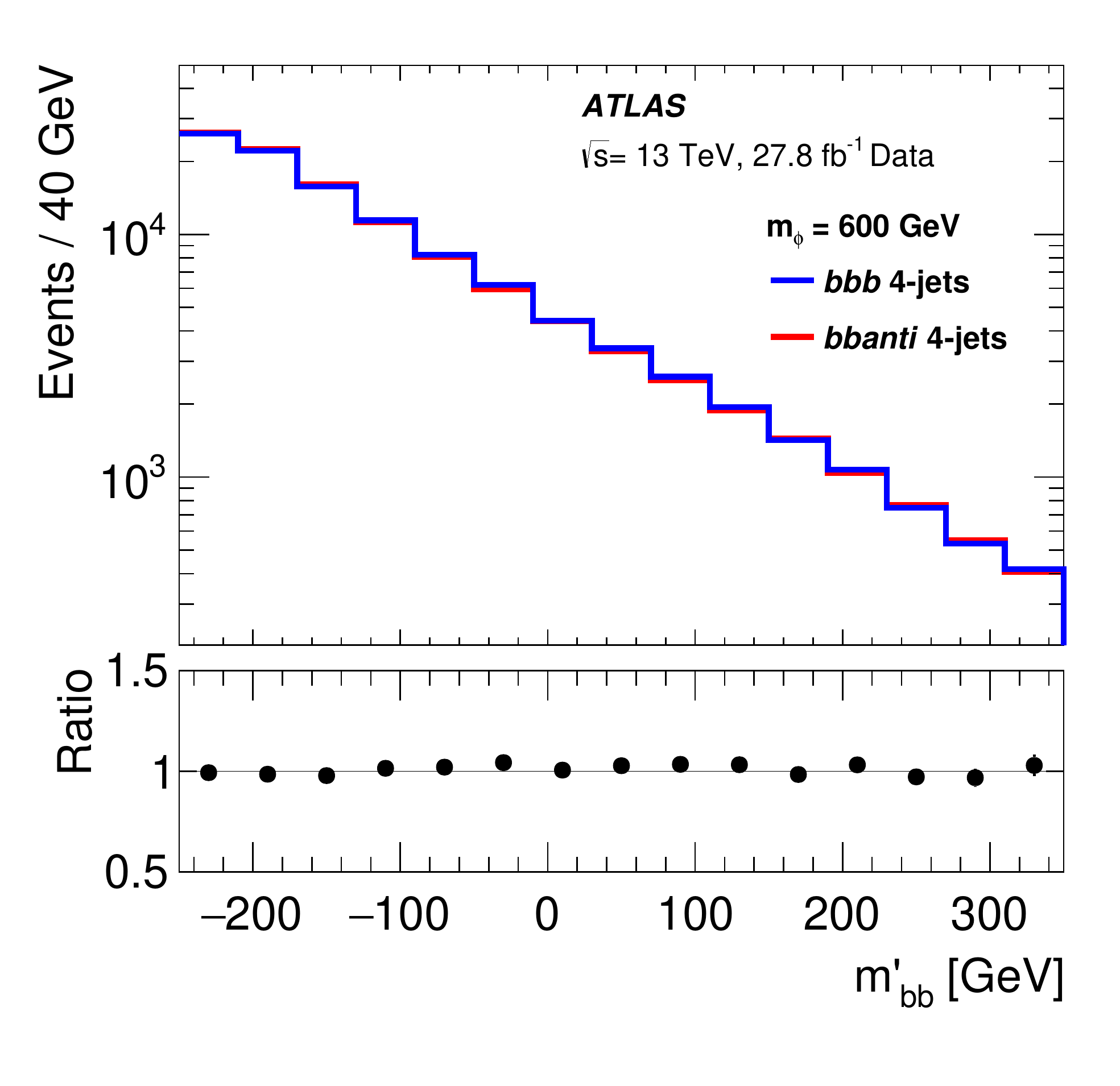}
\includegraphics[width=0.43\linewidth]{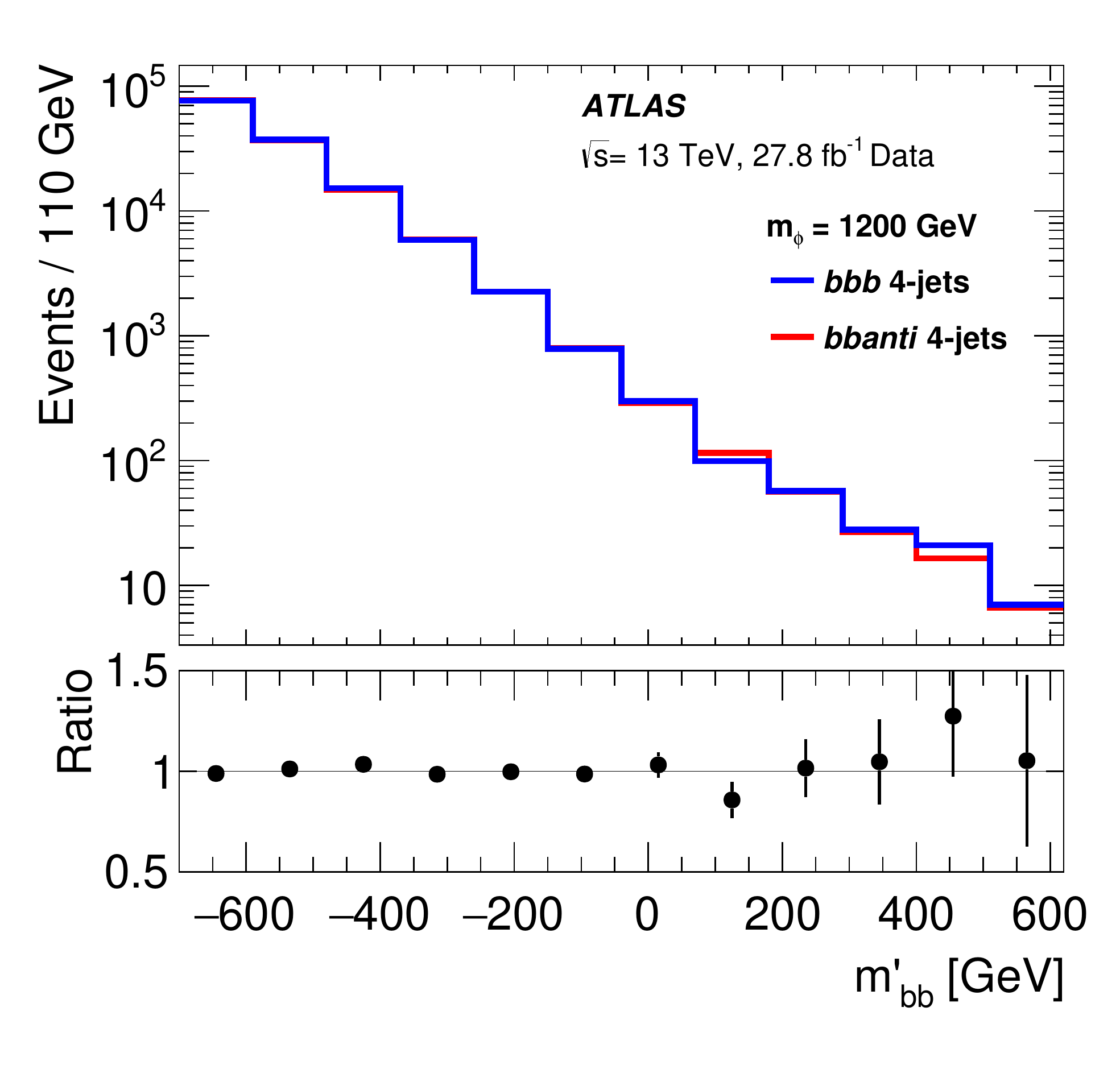}
\\
\includegraphics[width=0.43\linewidth]{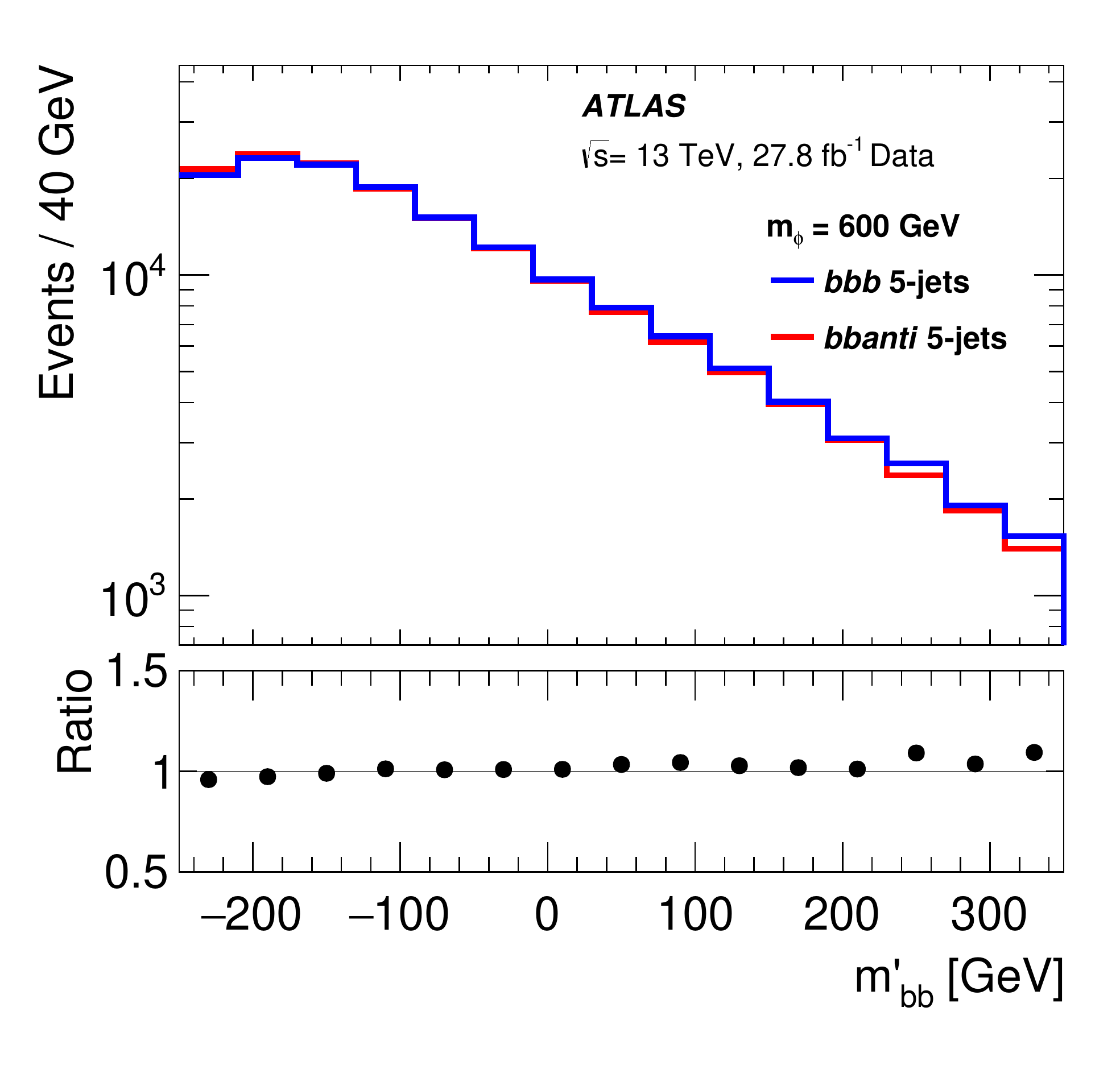}
\includegraphics[width=0.43\linewidth]{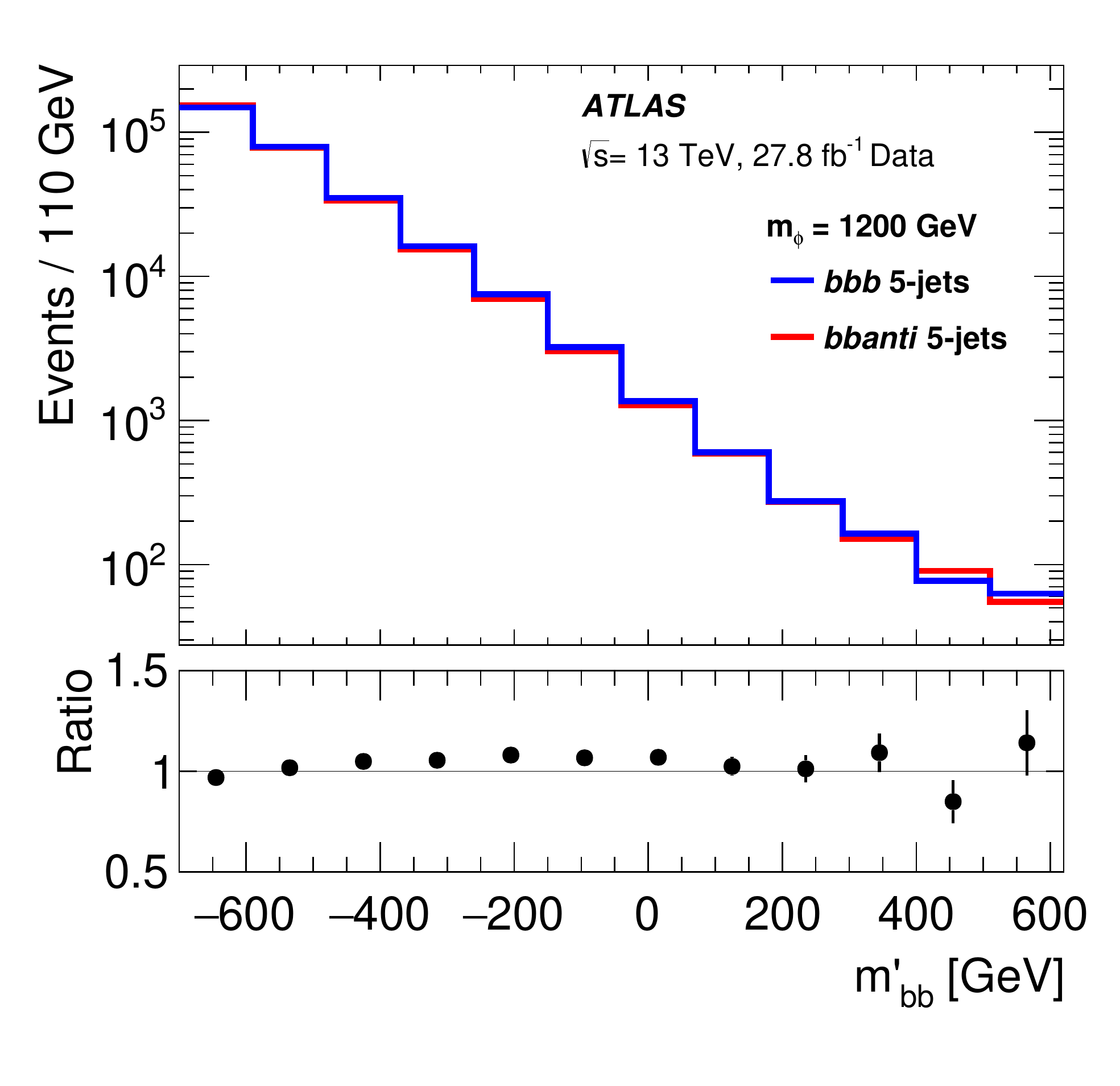}
\\
\end{tabular}
\caption{Distributions of $m_{bb}'$ in data with the \textit{bbb} and \textit{bbanti}
classifications following all selection requirements for the 3-jet (top row), 4-jet (middle row), and 5-jet (bottom row) regions.
The definition of the PCA variable $m_{bb}'$ depends on the mass hypothesis $m_\phi$
and distributions are shown for $m_\phi=600$~\GeV\ (left column) and $m_\phi=1200$~\GeV\ (right column).
For each case, the \textit{bbb}
distribution is normalized to the \textit{bbanti} distribution. The ratios of \textit{bbb} to \textit{bbanti}
are also shown. \label{fig:mbbpRatiosDataAlt2}}
\end{figure}

The data $bbb$ and $bbanti$ shapes, as well as their ratio, after applying the selection on $p_{\textrm{T1}}'$ and $p_{\textrm{T2}}'$ are qualitatively similar to the multijet MC distributions, as shown in Figure~\ref{fig:mbbpRatiosDataAlt2}.
Applying  $\chi^2$ probability and $F$-test criteria to the data $m'_{bb}$ distributions over all values of
$m_\phi$, it is found that a first-order polynomial is sufficient for the 3-jet region with masses $m_\phi < 1200$~\GeV\ and for the 4-jet and 5-jet regions with
masses  $m_\phi  < 800$~\GeV.  For all other jet-category/mass combinations, a second-order polynomial is needed. Potential signal contamination does not affect the results of the tests.

The binned maximum-likelihood fit is performed using the RooFit~\cite{RooFit} framework and the \textsc{HistFactory}~\cite{HistFactory} software tool.
A product of Poisson probability terms over the bins of the $m'_{bb}$ distributions involving the numbers of
data events $n_{i,j}$ and the sum of expected signal and background yields  $\nu_{i,j}$ in each category $i$ and mass bin $j$ forms the binned likelihood function.
It accounts for the effects of floating background normalization and systematic uncertainties and is
 
\begin{equation*}
P(\boldsymbol{n},\boldsymbol a | \mu, \boldsymbol{N}, \boldsymbol\gamma, \boldsymbol\alpha) = \prod_{i\, \in\, \text{categories}}\ \prod_{j\, \in\, \text{bins}}\mathrm{Pois}(n_{i,j}|\nu_{i,j}) \prod_{p\, \in\, \text{sys. nuis. param.}}f_p(a_p|\alpha_p,\sigma_p),
\label{eqn:poiss1}
\end{equation*}
\begin{eqnarray*}
\nu_{i,j} &=& N_{i}\cdot \gamma_{i,j} +\mu \cdot S_{i,j}\cdot \beta_{i,j}(\alpha_p) \\
\gamma_{i,j} &=&
\begin{cases}
B_{k,j}\cdot (Q_k\cdot P_j+A_k\cdot L_j+1)\ \ , & t=\textit{bbb} \\
B_{k,j}\ \ , &  t=\textit{bbanti}.
\end{cases}
\label{eqn:poiss2}
\end{eqnarray*}
The index $t$ runs over the two flavor categories \textit{bbb} and \textit{bbanti},  $k$ runs over the three jet-multiplicity regions,
$i=(t,k)$ runs over the six categories,
and $p$ runs over the systematic error nuisance parameters. The boldfaced symbols represent vectors of parameters whereas the symbols of the same type in lightface represent individual parameters (usually containing indices).
The template histograms, $S_{i,j}$, are taken directly from signal simulation for the given mass point
and are normalized to the event yields expected for a one picobarn signal.
Thus, the signal strength parameter $\mu$, which is common to all categories, is $\sigma(pp\rightarrow b\bar{b}\phi)\times \mathcal{B}(\phi\rightarrow b\bar{b})$ in picobarns.

Within the  \textsc{HistFactory} framework,  the second-order polynomial correction is implemented with the histograms $L_j$ and $P_j$, which are binned in $m'_{bb}$ and have bin contents given by the bin center value and bin center
value squared, respectively.
The normalization parameters for these histograms, $A_k$ and  $Q_k$ , correspond to the linear  and parabolic parameters, respectively, for the jet-multiplicity region $k$.
The signal strength parameter $\mu$, the six background normalization parameters $N_i$,
the background shape parameters $B_{k,j}$, the linear parameters $A_k$, and the parabolic parameters $Q_k$, are freely floating parameters in the fit,
with the exception that, for a fixed jet-multiplicity region $k$, the sum over bins $j$ of $B_{k,j}$ is constrained to unity.

The fit model contains nuisance parameters accounting for the statistical uncertainty of the MC signal samples and for
systematic variations of the shapes and normalizations of the histogram templates used in the fit,
as described in Section~\ref{sec:systematics}. The variable  $\beta_{i,j}$ represents the systematic variation in the bin content as a function of the nuisance parameters $\alpha_p$.
The nuisance parameters $\alpha_{p}$
are constrained within the allowed systematic variations by the $f_{p}(a_{p}|\alpha_{p},\sigma_{p})$ terms, where $a_p$ are auxiliary measurements and
$\sigma_p$ denotes the uncertainty in $\alpha_p$. Individual sources of uncertainties are considered uncorrelated.
 
The statistical uncertainty related to the size of MC signal samples is estimated with a
variation of the Barlow--Beeston method~\cite{barlowbeeston}. In this analysis, each bin in each category is given a single
Poisson-constrained nuisance parameter associated with the signal MC prediction for the number of events entering the bin and the total statistical uncertainty in that bin.

\section{Systematic uncertainties}
\label{sec:systematics}
This analysis relies on the prediction of the shapes and normalizations of the discriminating variable $m'_{bb}$ for the searched signal.
The signal uncertainties are divided into two categories: experimental and those related to the theoretical modeling of the signal.
The background model is validated through statistical analysis of the data and
tests utilizing the multijet MC sample.
 
\subsection{Systematic uncertainties of the background model}

The background model is validated through the $\chi^2$ probability and $F$-test analyses described in Sec.~\ref{sec:fit}.
As a cross-check, the fit procedure is applied to a small multijet MC sample with an equivalent integrated luminosity of 6.8 fb$^{-1}$ for eight $m_{\phi}$ hypotheses.
Events in MC samples are weighted to reflect data-based corrections for pileup, flavor tagging and trigger efficiency. In order to use the data fit procedure
without modification, a special unweighted version of the MC sample is produced using acceptance-rejection sampling and the total weight of each event.
The results are summarized in Table~\ref{tab:qcdfit}.
The eight separate fits should be uncorrelated given the 10\%--15\% mass resolution for a heavy Higgs boson. With the assumption of no correlation,
the $\chi^2$ per degree of freedom is 1.09 for the eight measurements, indicating that no statistically significant spurious signal is found by the analysis.

\begin{table}[ht!]
\caption{Best-fit values for $\sigma(pp\rightarrow b\bar{b}\phi)\times \mathcal{B}(\phi\rightarrow b\bar{b})$ in the multijet MC sample with a total uncertainty $\Delta (\sigma \times \mathcal{B})$ for each mass point. The multijet MC sample has an equivalent
integrated luminosity of 6.8~fb$^{-1}$.     \label{tab:qcdfit}  }
\centering
\begin{tabular}{rcc}
\toprule
Mass $\lbrack$\GeV$\rbrack$ & $\sigma \times \mathcal{B}$ $\lbrack$pb$\rbrack$ & $\Delta (\sigma \times \mathcal{B})$ $\lbrack$pb$\rbrack$  \\
\midrule
450  & --0.99 & 2.42 \\
550  & 0.77 & 1.44 \\
650  & --0.75 & 0.80 \\
750  & 0.42 & 0.62 \\
850  & --0.31 & 0.50 \\
1000  & --0.24 & 0.49 \\
1200  &  0.71 & 0.35 \\
1400  & --0.29 & 0.20 \\
 
\bottomrule
\end{tabular}
 
\end{table}

\subsection{Experimental systematic uncertainties of the signal}
 
The dominant experimental systematic uncertainties are related to the calibration of the $b$-tagging efficiencies in simulation relative to those measured in data for $p_{\textrm{T}}<300$~\GeV.
They are extrapolated to $p_{\textrm{T}}>300$~\GeV\ using MC simulation taking into account uncertainties in the jet modeling and detector response affecting $b$-tagging performance.
This calibration is performed separately for \textit{b-}jets, \textit{c-}jets, and light-flavor jets and as a function of jet $p_{\textrm{T}}$ and $|\eta|$~\cite{PERF-2012-04}.
Uncertainties in the cross sections for processes used in the $b$-tagging calibration, modeling of the jet kinematics and flavor composition in the simulated signal samples, detector simulation, and event reconstruction are included~\cite{myPERF-2016-05,ATLAS-CONF-2018-001,ATLAS-CONF-2018-006}.
These uncertainties are decomposed into uncorrelated components resulting in three components for \textit{b-}jets and \textit{c-}jets, and five components for light-flavor jets.
 
Simulation-to-data efficiency differences are also corrected for the trigger, specifically for $b$-tagged jets\cite{mcclymontThesis2018}. Since the background estimation is data-driven, this scaling only affects
signal.
Scale factors obtained by comparing data and simulated dilepton $t\bar{t}$ events,
which are enriched in $b$-jets, are used to correct simulation-to-data efficiency differences in the $b$-jet trigger for
$p_{\textrm{T}}<240$~\GeV.  For $p_{\textrm{T}}>240$~\GeV, due to the limited size
of the $t\bar{t}$ data sample, extrapolation based on simulation is used, as described in detail in  Ref.~\cite{mcclymontThesis2018}.
The systematic uncertainties in these scale factors include mismodeling of the fraction of \textit{b-}jets in simulation, mismodeling of the \textit{b-}jet trigger efficiency for non-\textit{b-}jets,
simulation statistical uncertainty, data statistical uncertainty for $p_{\textrm{T}} < 240$~\GeV, uncertainty in the simulation-based extrapolation to $p_{\textrm{T}} > 240$~\GeV, and uncertainties in the dependence
of the $b$-jet trigger efficiency on jet $p_{\textrm{T}}$ and $\eta$.
The $b$-jet trigger was calibrated relative to a set of offline $b$-tagging operating points to correctly take into account correlations between the $b$-jet trigger and offline $b$-tagging.
 
Systematic uncertainties in the jet energy scale and jet energy resolution are based on measurements with data~\cite{PERF-2016-04, PERF-2011-04}. All sources of the jet energy scale uncertainty are decomposed into 21 uncorrelated components that are treated as independent.
 
The uncertainty in the combined 2015+2016 integrated luminosity is 2.1\%. It is derived from the calibration of the luminosity scale using $x$-$y$ beam-separation scans, following a methodology similar to that detailed in Ref.~\cite{DAPR-2013-01}, and using the LUCID-2 detector for the baseline luminosity measurements \cite{LUCID2}.

\subsection{Theoretical systematic uncertainties of the signal}
The uncertainty related to the choice of generator for the signal hard process and showering model
is estimated by comparing the nominal sample with the one obtained
by reweighting the nominal sample to the NLO generator \textsc{Madgraph5}\_aMC@NLO~\cite{Alwall:2014hca,Wiesemann:2014ioa} with a 4 flavor scheme (4FS) PDF interfaced to the \textsc{Pythia 8.205} parton showering model.
The particle-level Higgs boson mass, Higgs boson $p_{\textrm{T}}$,  and the $p_{\textrm{T}}$ values of the two leading \textit{b-}tagged jets are used for sequential reweighting.
The uncertainty from the PDF set used in the nominal sample is computed using the standard-deviation method described in Ref.~\cite{NNPDF}.  Variations in PDF parameters can modify the $\phi$ boson detection efficiency through changes to the relative contribution of the four
Higgs production Feynman diagrams
of Figure~\ref{fig:fd}, as well as the distributions of kinematic variables, such as $p_{\textrm{T}}$ and $\eta$ for the third $b$-jet.   The PDF uncertainty is likely overestimated, but remains small compared to the statistical uncertainty.

\def\ra{\rightarrow}
\def\bb{b\overline{b}}
\def\tanb{{\tan\beta}}
\section{Results}
\label{sec:results}

The search for a single heavy neutral Higgs boson $\phi$ produced in association with $b$-quarks and decaying into a $b\bar{b}$ pair shows no significant excess above the SM background for any of the analyzed mass points. The postfit \textit{bbb} category distributions of the rotated $bb$ invariant mass $m'_{bb}$ are shown in Figure~\ref{fig:postfit} together with the $m'_{bb}$ distribution extracted from the \textit{bbanti} region (prefit background).
\begin{figure}[ht!]
\centering
 
\includegraphics[width=0.45\textwidth]{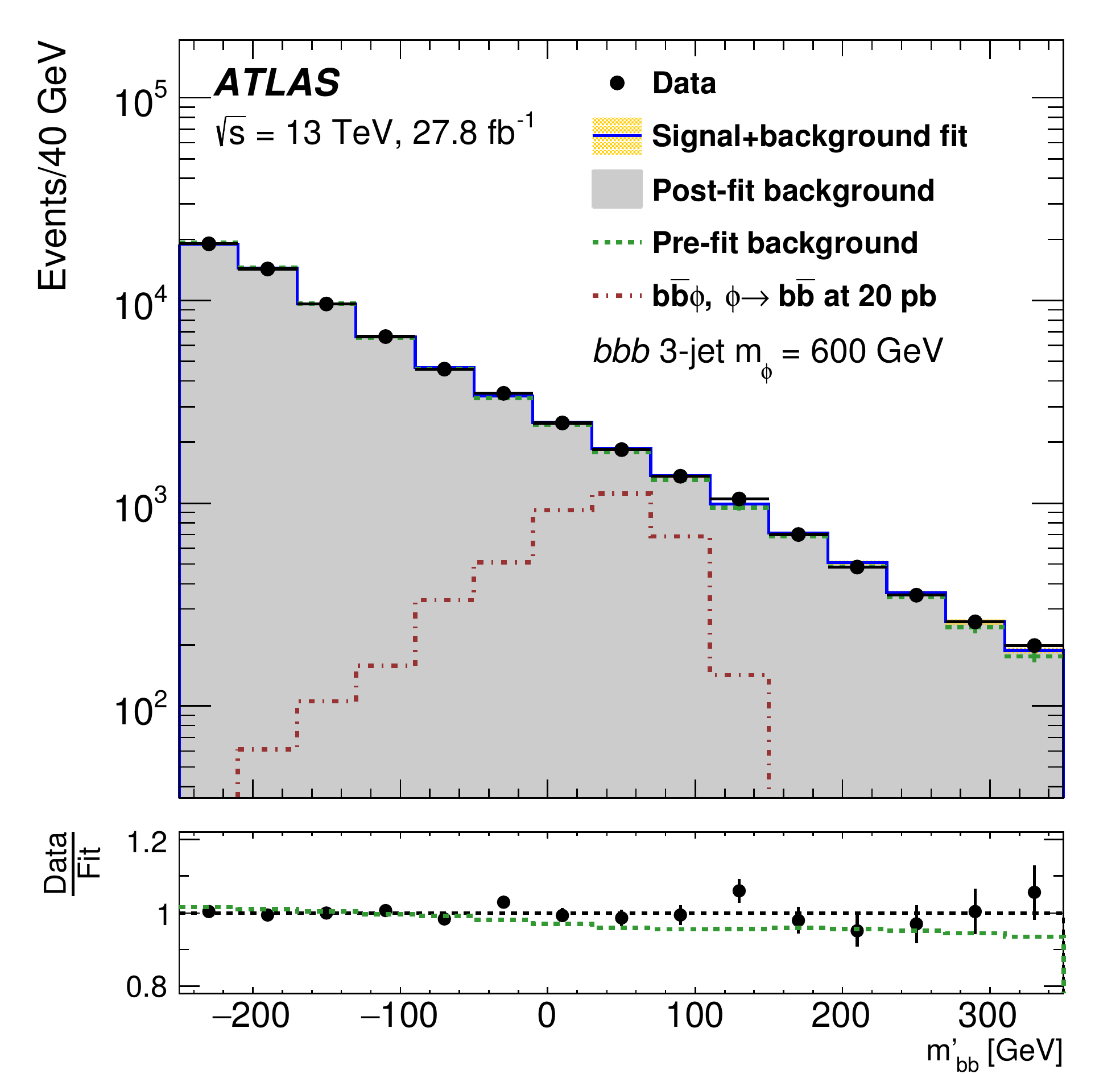}
\includegraphics[width=0.45\textwidth]{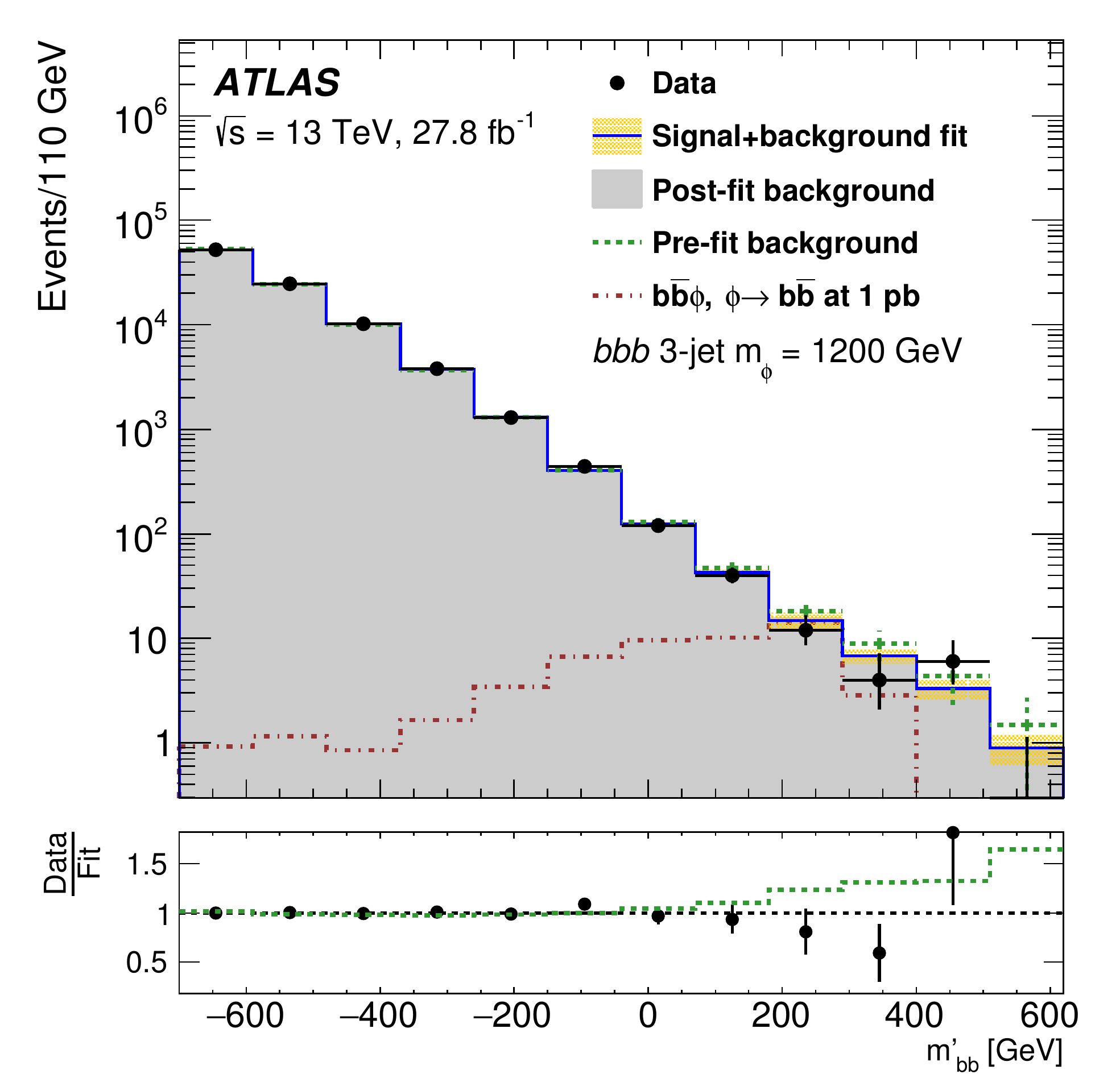}\\
\includegraphics[width=0.45\textwidth]{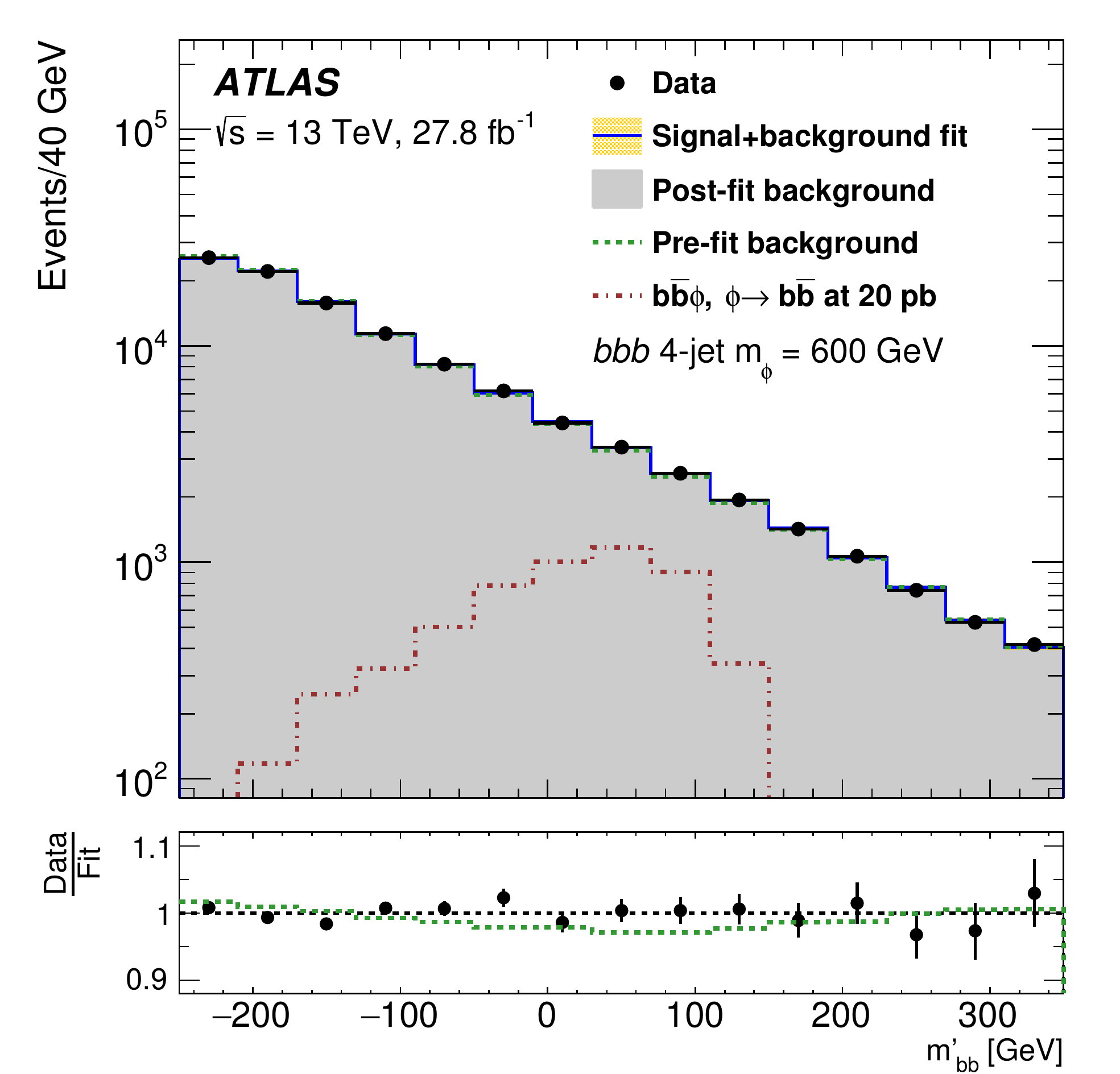}
\includegraphics[width=0.45\textwidth]{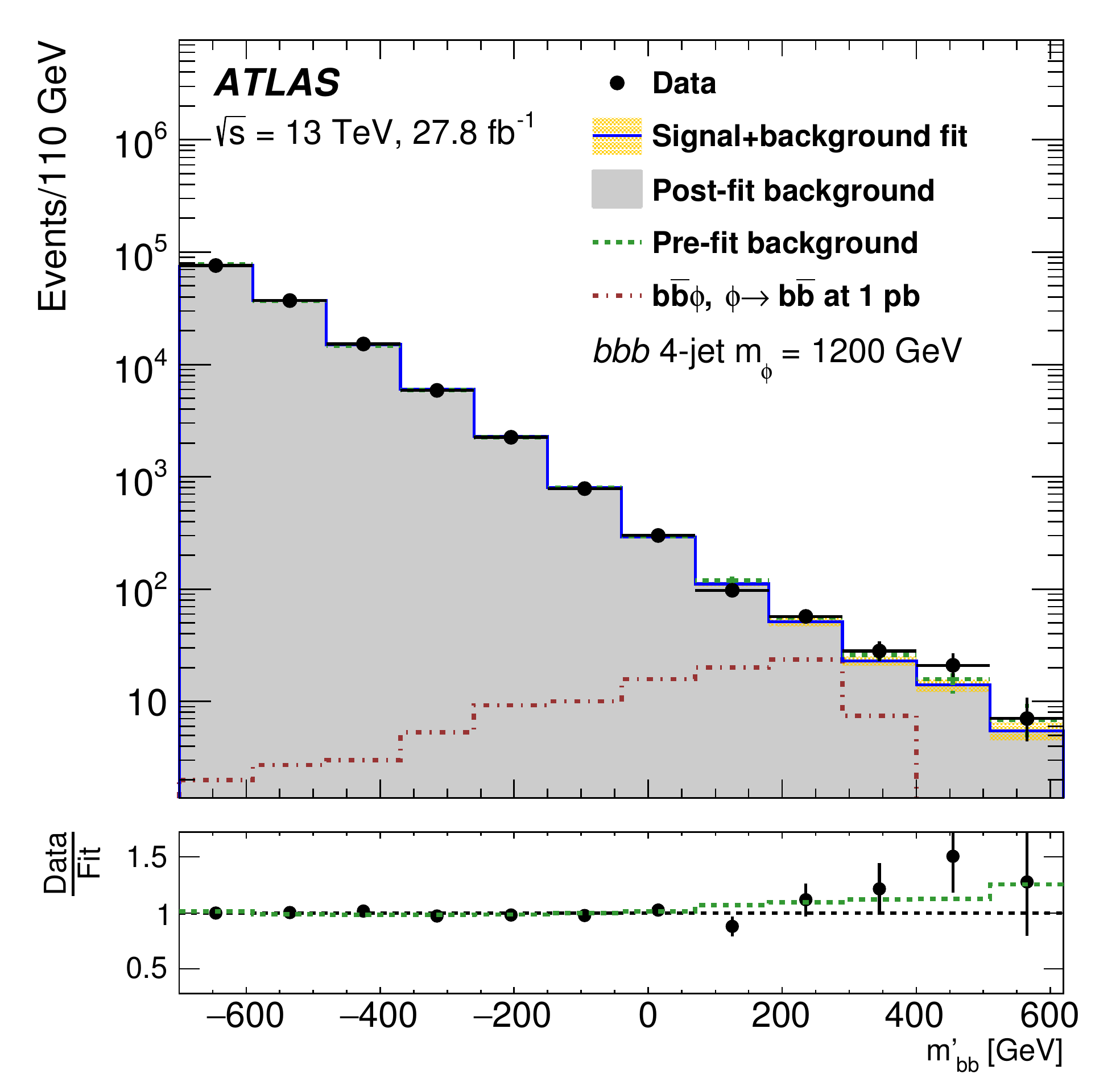}\\
\includegraphics[width=0.45\textwidth]{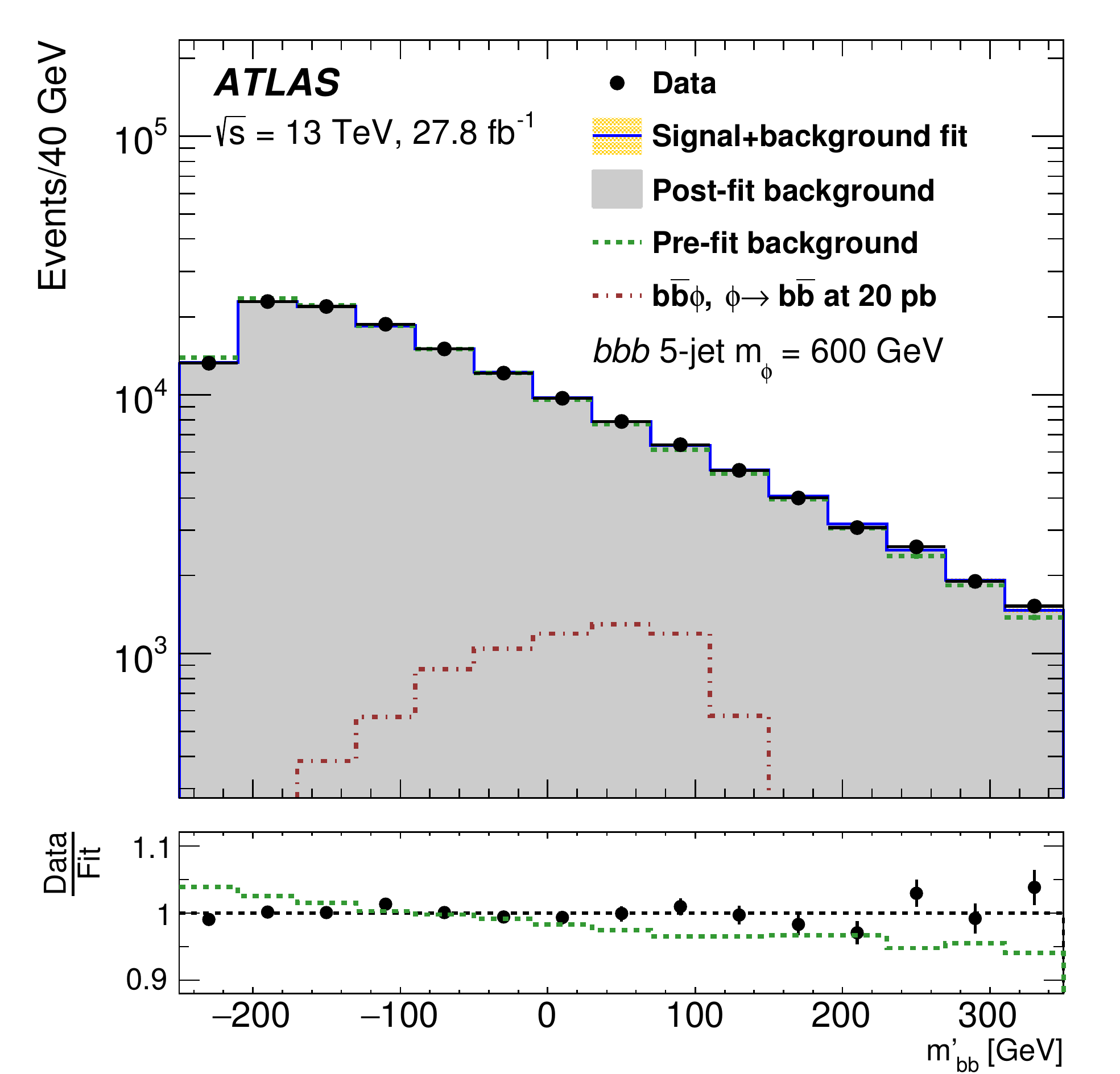}
\includegraphics[width=0.45\textwidth]{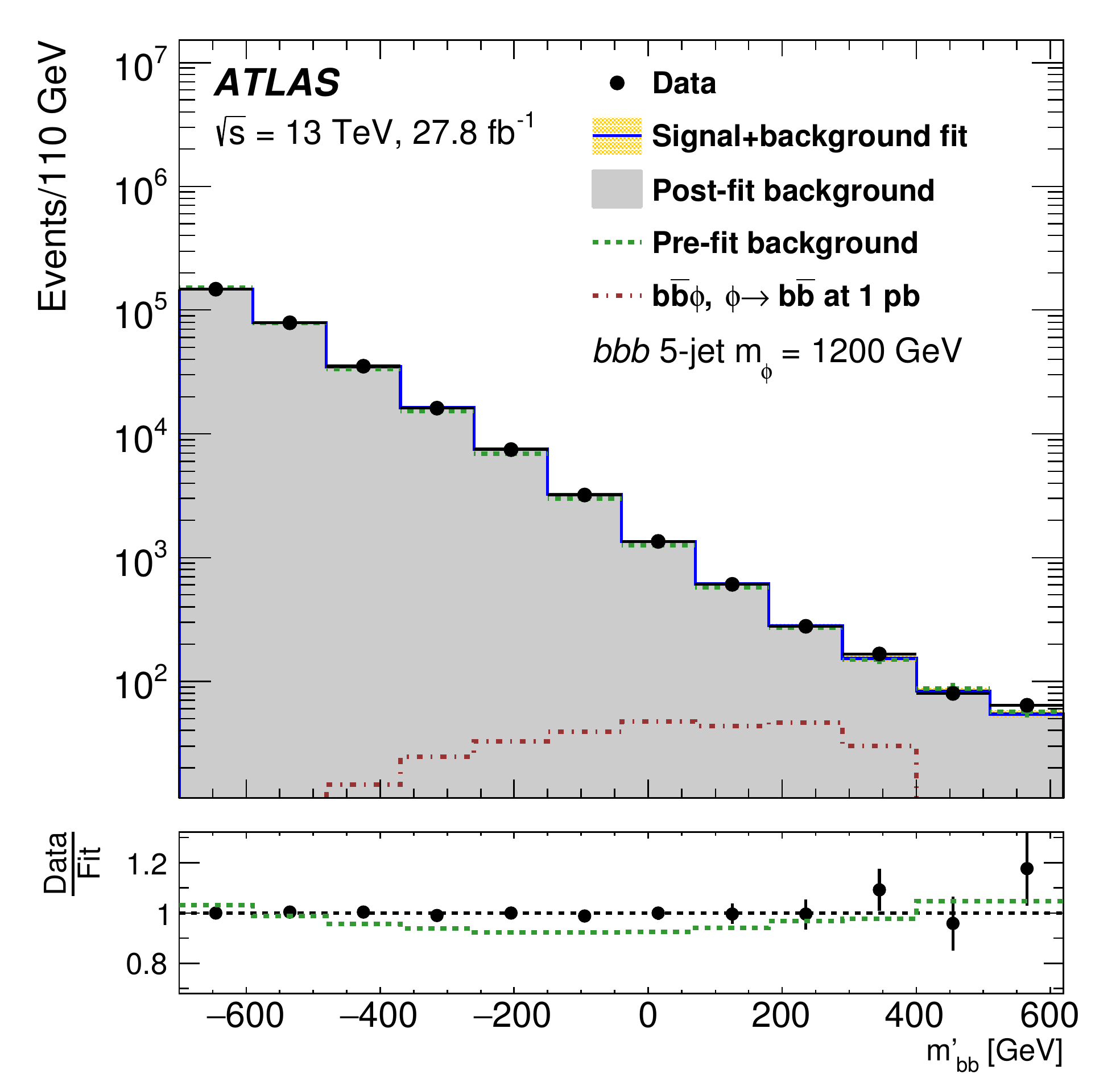}\\
\caption{Postfit \textit{bbb} category distributions of $m'_{bb}$ for the 600~\GeV\ (left) and 1200~\GeV\ (right) mass points in the 3-jet (top),
4-jet (middle), and 5-jet (bottom) categories. The prefit background shape is also shown in the top panels, and its ratio to the postfit shape is shown in the bottom panels (green dashed line). The signal shape (red dashed line) is overlaid for illustration.}\label{fig:postfit}
\end{figure}

\subsection{Cross section limits}

Since no significant excess over the background expectation is observed,
upper limits on the production of a single heavy neutral Higgs boson $\phi$ decaying into a $b\bar{b}$ pair are set. Figure~\ref{fig:limit} presents the observed and expected limits for $\sigma(pp\rightarrow b\bar{b}\phi)\times\mathcal{B}(\phi\rightarrow b\bar{b})$
at the $95\%$ confidence level (C.L.). The limits at specific mass points are calculated with the $\text{C.L.}_{\text{s}}$ method~\cite{CLS}.
The expected and observed limits between mass points are given by straight lines connecting the mass point limits.
Linear interpolations of the PCA parameters for adjacent mass points are used to calculate the observed $95\%$ C.L. limits for masses between the mass points. The straight lines in Figure~\ref{fig:limit} are consistent with these observed limits to within
the accuracy of the method, which grows from 0.00 to 0.56 times the $+1\sigma$ expected limit band as the mass is varied from  the mass point itself to a position midway between adjacent mass points.

\begin{figure}[ht!]
\begin{center}
\includegraphics[width=0.95\textwidth]{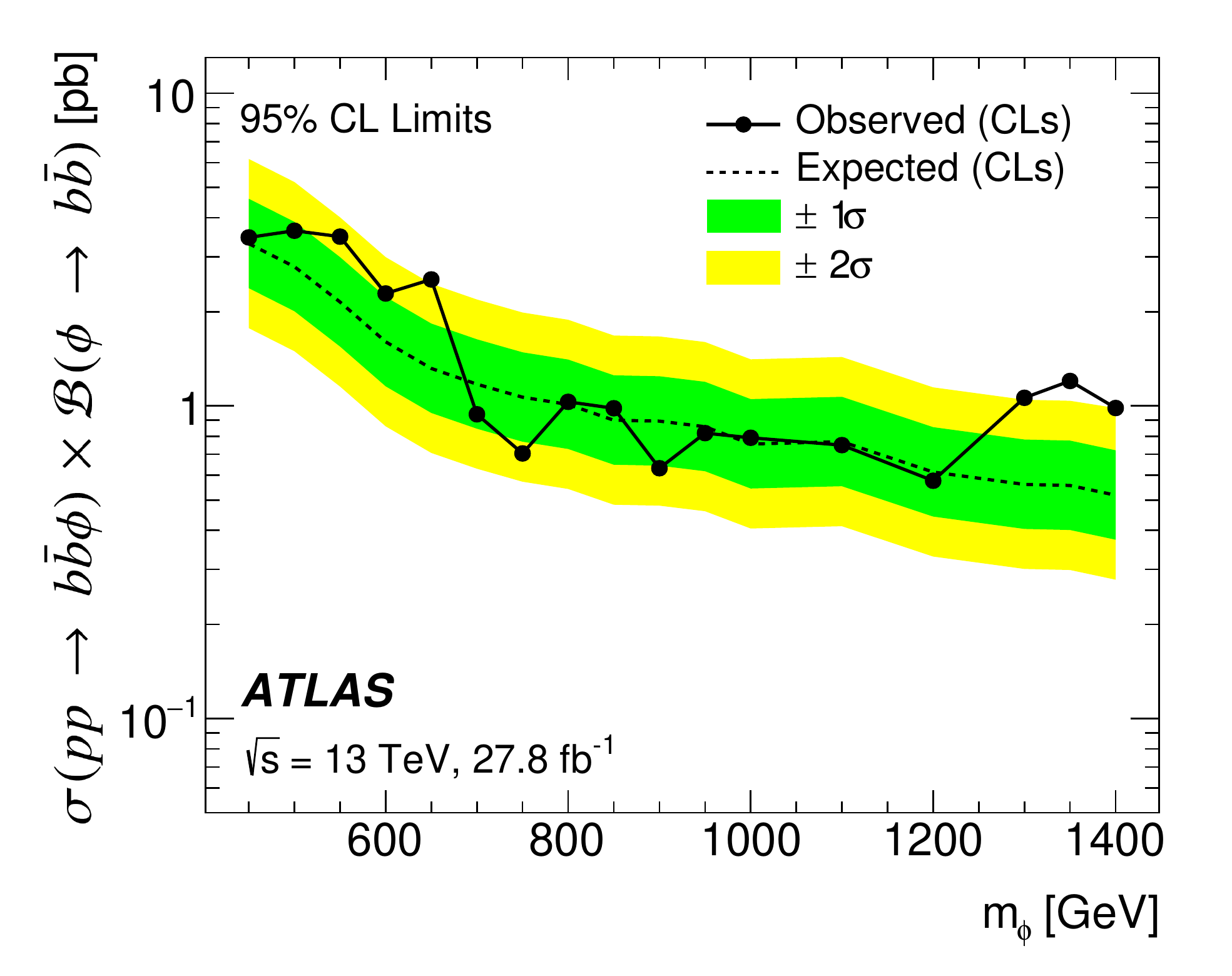}
\caption{Observed and expected upper limits on $\sigma(pp\rightarrow b\bar{b}\phi)\times\mathcal{B}(\phi\rightarrow b\bar{b})$ at $95\%$ C.L. as a function of the Higgs boson mass in 27.8~fb$^{-1}$ of $pp$ collision data at $\sqrt{s}=13$~\TeV. }\label{fig:limit}
\end{center}
\end{figure}

The leading sources of uncertainty affecting the best-fit value of $\sigma\times \mathcal{B}$
for two of the mass points, 600 and 1200~\GeV, are given in Table~\ref{tab:fullsys},
together with their relative importance. The impact of the given source of uncertainty is obtained by first fixing all the nuisance parameters related to other systematic uncertainties to their best-fit values and then allowing only the nuisance parameters ascribed to the considered source of uncertainty to float in the fit.
The uncertainty is dominated by the statistical error, which improves significantly between
$m_\phi=600$~and~1200~\GeV\ due to the sharp drop in the background level.
The systematic
uncertainties with the largest impact on the sensitivity are related
to the flavor-tagging calibration of the offline $b$-tagging algorithm and $b$-jet trigger and to jet reconstruction.
 
\begin{table}[ht!]
\caption{Grouped contributions to the systematic uncertainty of the best-fit value of $\sigma\times \mathcal{B}$. The best-fit values for $m_{\phi} = 600$ and 1200~\GeV\ are 0.76 and $-0.1$~pb, respectively. \label{tab:fullsys}}
\centering
\begin{tabular}{lll}
\toprule
& $m_\phi=600$ \GeV & $m_\phi=1200$ \GeV  \\
Source of uncertainty & $\Delta(\sigma\times \mathcal{B})$ [pb] & $\Delta(\sigma\times \mathcal{B})$ [pb] \\
\midrule
Total  & 0.80 & 0.29\\
Statistical  & 0.77 & 0.26\\
Systematic  & 0.20 & 0.11\\
\midrule
\multicolumn{3}{l}{Experimental uncertainties}\\
\midrule
Jet-related  & 0.05 & 0.05\\
Flavor-tagging  & 0.12 & 0.05\\
Trigger  & 0.04 & 0.05\\
Luminosity  & 0.02 & 0.01\\
\midrule
\multicolumn{3}{l}{Theoretical and modeling uncertainties}\\
\midrule
Generator  & 0.03 & 0.03\\
PDF  & 0.08 & 0.04\\
MC statistical  & 0.09 & 0.04\\
\bottomrule
\end{tabular}
\end{table}

\subsection{Model interpretations}

The two 2HDM scenarios with enhanced $pp\ra b\bar{b}\phi$ production and $\phi\ra\bb$ decay at large $\tanb$ are Type II and Type Y (flipped).
The most commonly analyzed scenario is Type~II since the Higgs sector of the MSSM is a Type~II 2HDM. The results of this search are interpreted
in the context of the MSSM for the hMSSM scenario~\cite{Djouadi:2013uqa} and for the $m_h^{\mathrm{mod}+}$ and $m_h^{\mathrm{mod}-}$ scenarios~\cite{Carena:2013ytb}.
The Higgs boson production cross-sections and branching
ratios are  calculated using the procedures outlined in the LHC Higgs Cross-section Working Group report~\cite{deFlorian:2016spz}.
The cross-sections for Higgs boson production through $b\bar{b}$ fusion~\cite{Harlander:2011aa}
are determined by matching the 5FS~\cite{Harlander:2003ai,Harlander:2012pb} and 4FS~\cite{Dittmaier:2003ej, Dawson:2003kb} cross-section calculations.  For the hMSSM
scenario, the Higgs boson masses and branching ratios are calculated using \textsc{HDecay}~\cite{Djouadi:1997yw, Djouadi:2018xqq}.
For the $m_h^{\mathrm{mod}+}$ and $m_h^{\mathrm{mod}-}$ scenarios, the Higgs boson masses and couplings are calculated with \textsc{FeynHiggs}~\cite{Heinemeyer:1998yj, Heinemeyer:1998np, Degrassi:2002fi, Frank:2006yh, Hahn:2013ria},
and the branching ratios are calculated by combining the most precise results from \textsc{FeynHiggs}, \textsc{HDecay}, and PROPHECY4f~\cite{Bredenstein:2006rh,Bredenstein:2006ha}.

The 95\% C.L. exclusion limits on $\tanb$ as a function of $m_A$ are shown
in Figure~\ref{Fig:MSSM-tanb-mA} for the hMSSM benchmark. The hMSSM scenario is well
defined and broadly representative of the remaining parameter space with SUSY partners too heavy for direct
detection at the LHC. As an indication of the sensitivity variation for different MSSM scenarios, Figure~\ref{Fig:MSSM-tanb-mA} also displays the expected sensitivities for the
$m_h^{\mathrm{mod}+}$ and $m_h^{\mathrm{mod}-}$ scenarios, in which top squark
mixing parameters are chosen to allow a wide range of $\tanb$ values while maintaining compatibility with  $m_h=125$~\GeV.
The hMSSM limits obtained by this search are comparable
to the limits obtained in the charged Higgs boson search in the $H^+\ra \tau\nu$ decay channel by ATLAS and CMS Collaborations~\cite{Aaboud:2018gjj,Sirunyan:2019hkq} but not as stringent as the hMSSM limits obtained in the
ATLAS and CMS searches for heavy neutral Higgs bosons decaying via $\phi\ra \tau^+\tau^-$~\cite{Aaboud:2017sjh,Sirunyan:2018zut}.

\begin{figure}[h]
\begin{center}
\includegraphics[width=0.9\textwidth]{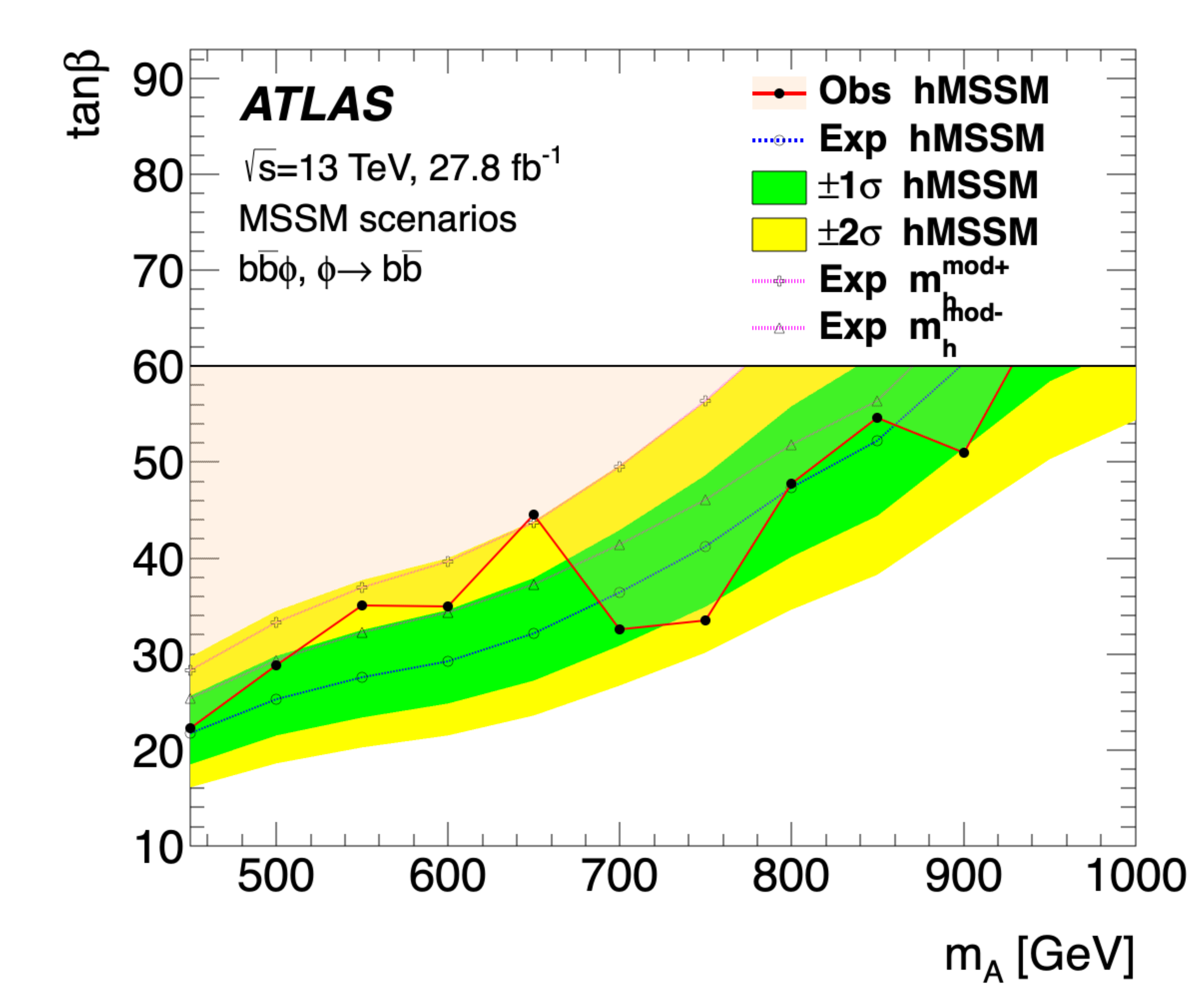}
\caption{\label{Fig:MSSM-tanb-mA}Observed and expected 95\% C.L. exclusion limits for the hMSSM scenario as a
function of $m_A$.  The expected
sensitivities for the $m_h^{\textrm{mod+}}$ and $m_h^{\mathrm{mod-}}$ scenarios are also shown.
The observed  95\% C.L. limits for the $m_h^{\textrm{mod+}}$ and $m_h^{\mathrm{mod-}}$ scenarios follow the same pattern with respect to their expected limits as the hMSSM observed limits.
Limits are not shown for $\tan{\beta}>60$ since the Higgs boson coupling becomes nonperturbative
for very large values of $\tan{\beta}$ in the considered models.}
\end{center}
\end{figure}
 
The 95\% C.L. $\tanb$
exclusion limits for this search assuming the Type Y (flipped) 2HDM scenario are presented, first in
Figure~\ref{Fig:2HDM-tanb-cba} for the specific $\phi$ mass of 450~\GeV\ as a function of $\cos(\beta-\alpha)$
and then in Figure~\ref{Fig:2HDM-tanb-mA} as a function of $m_\phi$ in the alignment limit $\cos(\beta-\alpha)=0$. For these limits, it is assumed that
the flipped 2HDM is \textit{CP-}conserving with $m_h=125$~\GeV\ and $m_{H}=m_{H^\pm}=m_A$.
The model grid points are generated using SusHi~\cite{Harlander:2012pb} and
2HDMC~\cite{Eriksson:2009ws}.  These limits complement the flipped 2HDM limits
obtained from the searches for $A\ra Zh$ in ATLAS~\cite{EXOT-2016-10}, which exclude regions with
$|\cos(\beta-\alpha)| \gtrsim 0.2$ or $\tan\beta \lesssim 4$, and from the search for  $A\ra ZH$ in ATLAS~\cite{Aaboud:2018eoy}, which excludes regions with $m_A-m_H \gtrsim 100$~\GeV.

\begin{figure}
\begin{center}
\includegraphics[width=0.8\textwidth]{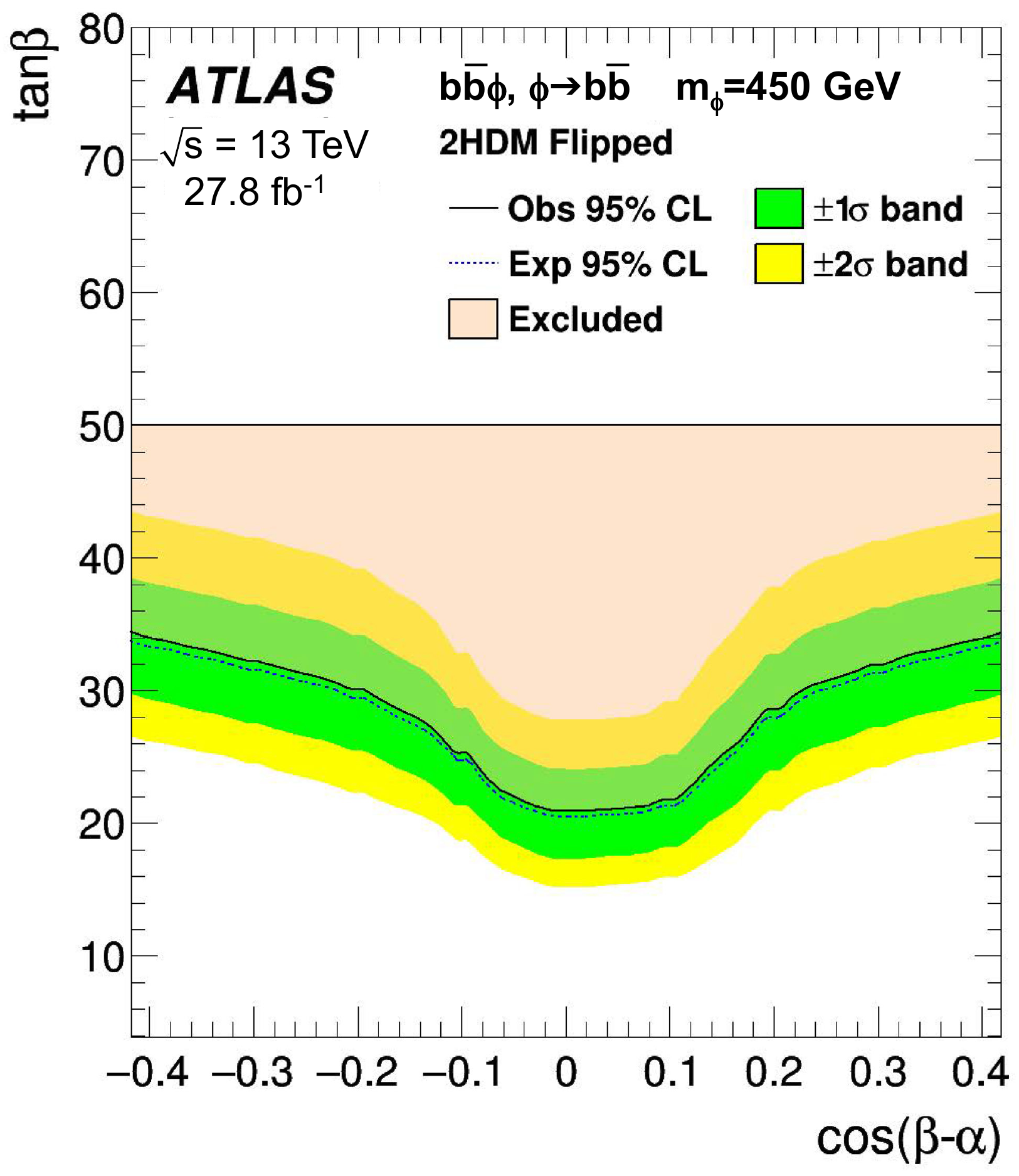}
\caption{\label{Fig:2HDM-tanb-cba}Observed and expected 95\% C.L. exclusion limits for the flipped 2HDM
scenario at $m_\phi$ = 450~\GeV\ as a function of $\cos(\beta-\alpha)$.   Limits are not shown for $\tan{\beta}>50$ since the Higgs boson coupling becomes nonperturbative
for very large values of $\tan{\beta}$ in this model.}
\end{center}
\end{figure}
 
\begin{figure}
\begin{center}
\includegraphics[width=0.8\textwidth]{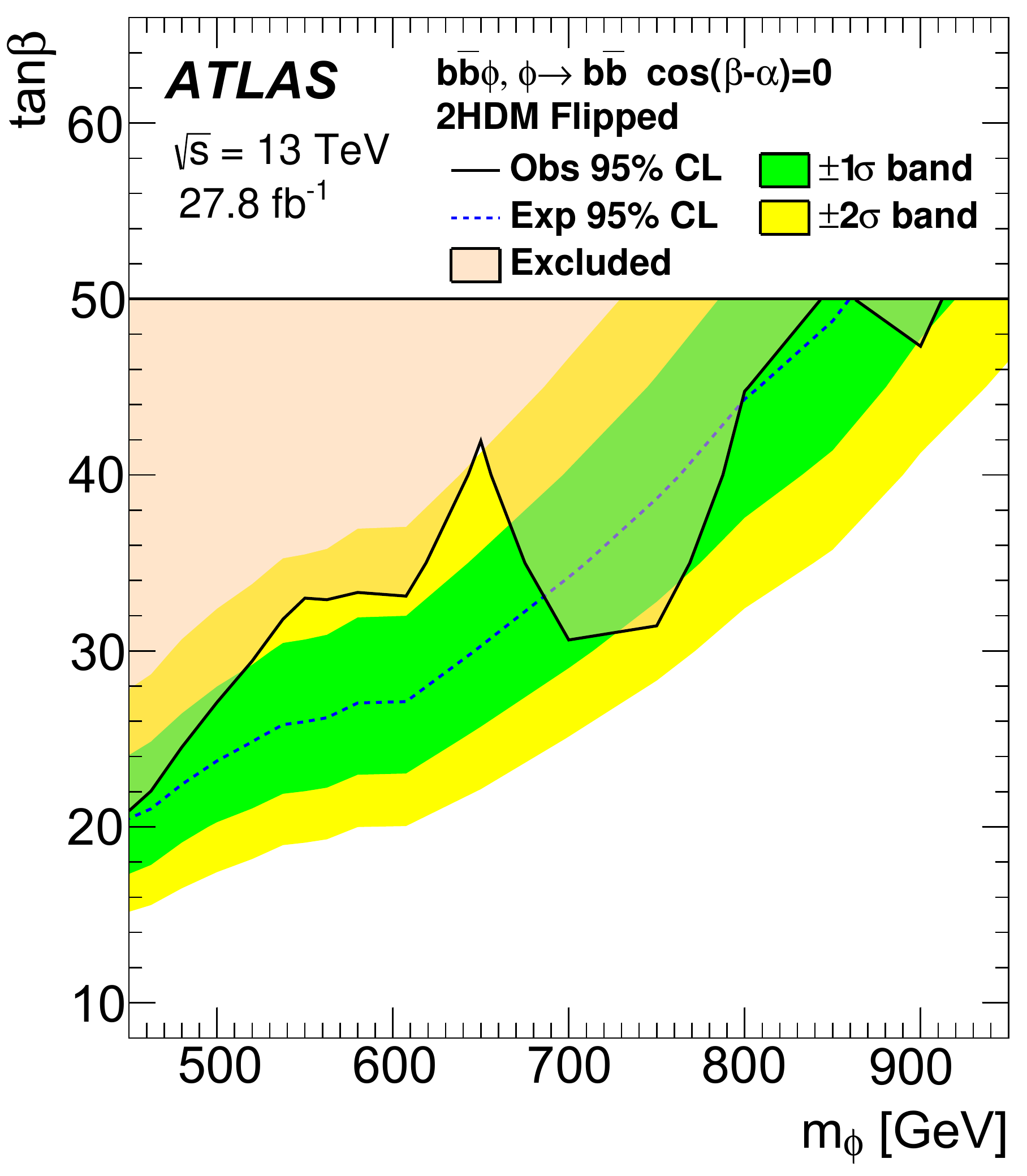}
\caption{\label{Fig:2HDM-tanb-mA}Observed and expected 95\% C.L. exclusion limits for the flipped 2HDM
scenario in the alignment limit as a function of $m_\phi$.  Limits are not shown for $\tan{\beta}>50$ since the Higgs boson coupling becomes nonperturbative
for very large values of $\tan{\beta}$ in this model.}
\end{center}
\end{figure}

\FloatBarrier
\section{Conclusions}
\label{sec:conclusions}
 
A search for heavy neutral Higgs bosons produced in association with at least one $b$-quark and decaying into a pair of $b$-quarks was performed using 27.8~fb$^{-1}$ of 13~TeV $pp$ collision data recorded by the ATLAS detector at the LHC in 2015 and 2016. The data are compatible with SM expectations, yielding no significant excess of events in the mass range 450--1400~\GeV.
Upper limits on the cross section times branching ratio were derived as a function of the mass of the heavy Higgs boson. The 95\% C.L. upper limits
are in the range 0.6--4.0~pb.
Compared to heavy neutral Higgs boson searches utilizing $\phi\to \tau^+\tau^-$ or $A\to Zh$ decays, these limits
expand the excluded Type Y (flipped) 2HDM parameter space into regions with $|\cos(\beta-\alpha)| \approx 0$ and $\tan\beta \gtrsim 20$.
 
\appendix

--------------------------------------------
\section*{Acknowledgments}
 

We thank CERN for the very successful operation of the LHC, as well as the
support staff from our institutions without whom ATLAS could not be
operated efficiently.
 
We acknowledge the support of ANPCyT, Argentina; YerPhI, Armenia; ARC, Australia; BMWFW and FWF, Austria; ANAS, Azerbaijan; SSTC, Belarus; CNPq and FAPESP, Brazil; NSERC, NRC and CFI, Canada; CERN; CONICYT, Chile; CAS, MOST and NSFC, China; COLCIENCIAS, Colombia; MSMT CR, MPO CR and VSC CR, Czech Republic; DNRF and DNSRC, Denmark; IN2P3-CNRS and CEA-DRF/IRFU, France; SRNSFG, Georgia; BMBF, HGF and MPG, Germany; GSRT, Greece; RGC and Hong Kong SAR, China; ISF and Benoziyo Center, Israel; INFN, Italy; MEXT and JSPS, Japan; CNRST, Morocco; NWO, Netherlands; RCN, Norway; MNiSW and NCN, Poland; FCT, Portugal; MNE/IFA, Romania; MES of Russia and NRC KI, Russia Federation; JINR; MESTD, Serbia; MSSR, Slovakia; ARRS and MIZ\v{S}, Slovenia; DST/NRF, South Africa; MICINN, Spain; SRC and Wallenberg Foundation, Sweden; SERI, SNSF and Cantons of Bern and Geneva, Switzerland; MOST, Taiwan; TAEK, Turkey; STFC, United Kingdom; DOE and NSF, United States of America. In addition, individual groups and members have received support from BCKDF, CANARIE, Compute Canada and CRC, Canada; ERC, ERDF, Horizon 2020, Marie Sk{\l}odowska-Curie Actions and COST, European Union; Investissements d'Avenir Labex, Investissements d'Avenir Idex and ANR, France; DFG and AvH Foundation, Germany; Herakleitos, Thales and Aristeia programmes co-financed by EU-ESF and the Greek NSRF, Greece; BSF-NSF and GIF, Israel; La Caixa Banking Foundation, CERCA Programme Generalitat de Catalunya and PROMETEO and GenT Programmes Generalitat Valenciana, Spain; G\"{o}ran Gustafssons Stiftelse, Sweden; The Royal Society and Leverhulme Trust, United Kingdom.
 
The crucial computing support from all WLCG partners is acknowledged gratefully, in particular from CERN, the ATLAS Tier-1 facilities at TRIUMF (Canada), NDGF (Denmark, Norway, Sweden), CC-IN2P3 (France), KIT/GridKA (Germany), INFN-CNAF (Italy), NL-T1 (Netherlands), PIC (Spain), ASGC (Taiwan), RAL (UK) and BNL (USA), the Tier-2 facilities worldwide and large non-WLCG resource providers. Major contributors of computing resources are listed in Ref.~\cite{ATL-SOFT-PUB-2020-001}.
 

\clearpage

\printbibliography

@Article{myPERF-2016-05,
    author         = "{ATLAS Collaboration}",
    title          = "{Measurements of $b$-jet tagging efficiency with the ATLAS detector using $t\bar{t}$ events at $\sqrt{s} = 13\;\mbox{TeV}$}",
    journal        = "JHEP",
    volume         = "08",
    year           = "2018",
    pages          = "089",
    doi            = "10.1007/JHEP08(2018)089",
    reportNumber   = "CERN-EP-2018-047",
    eprint         = "1805.01845",
    archivePrefix  = "arXiv",
    primaryClass   = "hep-ex",
}

@Article{CLS,
author="Cowan, Glen
and Cranmer, Kyle
and Gross, Eilam
and Vitells, Ofer",
title="Asymptotic formulae for likelihood-based tests of new physics",
journal="Eur. Phys. J. C",
year="2011",
month="Feb",
day="09",
volume="71",
number="2",
pages="1554",
%issn="1434-6052",
doi="10.1140/epjc/s10052-011-1554-0",
eprint         = "1007.1727",
archivePrefix  = "arXiv",
primaryClass   = "hep-ex",
related="CLSe",
relatedstring="Erratum:",
%url="https://doi.org/10.1140/epjc/s10052-011-1554-0"
}

@article{NNPDF,
      author         = "Ball, Richard D. and others",
      title          = "{Parton distributions for the LHC run II}",
      collaboration  = "NNPDF",
      journal        = "JHEP",
      volume         = "04",
      year           = "2015",
      pages          = "040",
      doi            = "10.1007/JHEP04(2015)040",
      eprint         = "1410.8849",
      archivePrefix  = "arXiv",
      primaryClass   = "hep-ph",
      reportNumber   = "EDINBURGH-2014-15, IFUM-1034-FT, CERN-PH-TH-2013-253,
                        OUTP-14-11P, CAVENDISH-HEP-14-11",
      SLACcitation   = "%%CITATION = ARXIV:1410.8849;%%"
}

@article{barlowbeeston,
    title = "Fitting using finite Monte Carlo samples",
    journal        = "Comput. Phys. Commun.",
    volume = "77",
    number = "2",
    pages = "219",
    year = "1993",
    doi = "10.1016/0010-4655(93)90005-W",
    author = "Roger Barlow and Christine Beeston",
}

@Booklet{HistFactory,
      author         = "Cranmer, Kyle and Lewis, George and Moneta, Lorenzo and
                        Shibata, Akira and Verkerke, Wouter",
      title          = "{HistFactory: A tool for creating statistical models for
                        use with RooFit and RooStats}",
      collaboration  = "ROOT",
      year           = "2012",
      howpublished   = "{CERN-OPEN-2012-016}",
      Url           = "https://cds.cern.ch/record/1456844/",
      SLACcitation   = "%%CITATION = CERN-OPEN-2012-016;%%"
}

@Booklet{RooFit,
      author         = "Verkerke, Wouter and Kirkby, David P.",
      title          = "{The RooFit toolkit for data modeling}",
      booktitle      = "{Statistical Problems in Particle Physics, Astrophysics
                        and Cosmology (PHYSTAT 05): Proceedings, Oxford, UK,
                        September 12-15, 2005}",
      year           = "2003",
      eprint         = "physics/0306116",
      archivePrefix  = "arXiv",
      reportNumber   = "CHEP-2003-MOLT007",
      SLACcitation   = "%%CITATION = PHYSICS/0306116;%%"
}

@article{Englert:1964et,
      author         = "Englert, F. and Brout, R.",
      title          = "{Broken Symmetry and the Mass of Gauge Vector Mesons}",
      journal        = "Phys. Rev. Lett.",
      volume         = "13",
      year           = "1964",
      pages          = "321-323",
      doi            = "10.1103/PhysRevLett.13.321",
      SLACcitation   = "%%CITATION = PRLTA,13,321;%%"
}

@article{Higgs:1964ia,
      author         = "Higgs, Peter W.",
      title          = "{Broken symmetries, massless particles and gauge fields}",
      journal        = "Phys. Lett.",
      volume         = "12",
      year           = "1964",
      pages          = "132-133",
      doi            = "10.1016/0031-9163(64)91136-9",
      SLACcitation   = "%%CITATION = PHLTA,12,132;%%"
}

@article{Higgs:1964pj,
      author         = "Higgs, Peter W.",
      title          = "{Broken Symmetries and the Masses of Gauge Bosons}",
      journal        = "Phys. Rev. Lett.",
      volume         = "13",
      year           = "1964",
      pages          = "508-509",
      doi            = "10.1103/PhysRevLett.13.508",
      SLACcitation   = "%%CITATION = PRLTA,13,508;%%"
}

@article{Heinemeyer:1998yj,
      author         = "Heinemeyer, S. and Hollik, W. and Weiglein, G.",
      title          = "{FeynHiggs: A program for the calculation of the masses
                        of the neutral CP-even Higgs bosons in the MSSM}",
      journal        = "Comput. Phys. Commun.",
      volume         = "124",
      year           = "2000",
      pages          = "76-89",
      doi            = "10.1016/S0010-4655(99)00364-1",
      eprint         = "hep-ph/9812320",
      archivePrefix  = "arXiv",
%      primaryClass   = "hep-ph",
      reportNumber   = "KA-TP-16-1998, DESY-98-193, CERN-TH-98-389",
      SLACcitation   = "%%CITATION = HEP-PH/9812320;%%"
}

@article{Gunion:1989we,
      author         = "Gunion, John F. and Haber, Howard E. and Kane, Gordon L.
                        and Dawson, Sally",
      title          = "{The Higgs Hunter's Guide}",
      journal        = "Front. Phys.",
      volume         = "80",
      year           = "2000",
%      pages          = "1-404",
%      pages          = "404pp.",
      reportNumber   = "SCIPP-89/13, UCD-89-4, BNL-41644",
      url           = "https://cds.cern.ch/record/425736",
      SLACcitation   = "%%CITATION = FRPHA,80,1;%%"
}

@article{Carena:2005ek,
      author         = "Carena, Marcela and Heinemeyer, S. and Wagner, C. E. M.
                        and Weiglein, G.",
      title          = "{MSSM Higgs boson searches at the Tevatron and the LHC:
                        Impact of different benchmark scenarios}",
      journal        = "Eur. Phys. J. C",
      volume         = "45",
      year           = "2006",
      pages          = "797-814",
      doi            = "10.1140/epjc/s2005-02470-y",
      eprint         = "hep-ph/0511023",
      archivePrefix  = "arXiv",
%      primaryClass   = "hep-ph",
      reportNumber   = "ANL-HEP-PR-05-109, CERN-PH-TH-2005-158, DCPT-05-128,
                        EFI-05-19, FERMILAB-PUB-05-370-T, IPPP-05-64",
      SLACcitation   = "%%CITATION = HEP-PH/0511023;%%"
}

@article{Gleisberg:2008ta,
      author         = "Gleisberg, T. and Hoeche, Stefan. and Krauss, F. and
                        Schonherr, M. and Schumann, S. and Siegert, F. and Winter,
                        J.",
      title          = "{Event generation with SHERPA 1.1}",
      journal        = "JHEP",
      volume         = "02",
      year           = "2009",
      pages          = "007",
      doi            = "10.1088/1126-6708/2009/02/007",
      eprint         = "0811.4622",
      archivePrefix  = "arXiv",
      primaryClass   = "hep-ph",
      reportNumber   = "FERMILAB-PUB-08-477-T, SLAC-PUB-13420, ZU-TH-17-08,
                        DCPT-08-138, IPPP-08-69, EDINBURGH-2008-30, MCNET-08-14",
      SLACcitation   = "%%CITATION = ARXIV:0811.4622;%%"
}

@article{Frixione:2007nw,
      author         = "Frixione, Stefano and Nason, Paolo and Ridolfi, Giovanni",
      title          = "{A positive-weight next-to-leading-order Monte Carlo for
                        heavy flavour hadroproduction}",
      journal        = "JHEP",
      volume         = "09",
      year           = "2007",
      pages          = "126",
      doi            = "10.1088/1126-6708/2007/09/126",
      eprint         = "0707.3088",
      archivePrefix  = "arXiv",
      primaryClass   = "hep-ph",
      reportNumber   = "BICOCCA-FT-07-12, GEF-TH-19-2007",
      SLACcitation   = "%%CITATION = ARXIV:0707.3088;%%"
}

@article{Sjostrand:2006za,
      author         = "Sj{\"o}strand, Torbjorn and Mrenna, Stephen and Skands, Peter
                        Z.",
      title          = "{PYTHIA 6.4 physics and manual}",
      journal        = "JHEP",
      volume         = "05",
      year           = "2006",
      pages          = "026",
      doi            = "10.1088/1126-6708/2006/05/026",
      eprint         = "hep-ph/0603175",
      archivePrefix  = "arXiv",
%      primaryClass   = "hep-ph",
      reportNumber   = "FERMILAB-PUB-06-052-CD-T, LU-TP-06-13",
      SLACcitation   = "%%CITATION = HEP-PH/0603175;%%"
}

@article{LUCID2,
      author         = "Avoni, G. and others",
      title          = "{The new LUCID-2 detector for luminosity measurement and
                        monitoring in ATLAS}",
      journal        = "JINST",
      volume         = "13",
      year           = "2018",
      number         = "07",
      pages          = "P07017",
      doi            = "10.1088/1748-0221/13/07/P07017",
      SLACcitation   = "%%CITATION = JINST,13,P07017;%%"
}

@article{Sjostrand:2007gs,
      author         = "Sj{\"o}strand, Torbjorn and Mrenna, Stephen and Skands, Peter
                        Z.",
      title          = "{A brief introduction to PYTHIA 8.1}",
      journal        = "Comput. Phys. Commun.",
      volume         = "178",
      year           = "2008",
      pages          = "852-867",
      doi            = "10.1016/j.cpc.2008.01.036",
      eprint         = "0710.3820",
      archivePrefix  = "arXiv",
      primaryClass   = "hep-ph",
      reportNumber   = "CERN-LCGAPP-2007-04, LU-TP-07-28,
                        FERMILAB-PUB-07-512-CD-T",
      SLACcitation   = "%%CITATION = ARXIV:0710.3820;%%"
}

@article{Maltoni:2002qb,
      author         = "Maltoni, Fabio and Stelzer, Tim",
      title          = "{MadEvent: automatic event generation with MadGraph}",
      journal        = "JHEP",
      volume         = "02",
      year           = "2003",
      pages          = "027",
      doi            = "10.1088/1126-6708/2003/02/027",
      eprint         = "hep-ph/0208156",
      archivePrefix  = "arXiv",
%      primaryClass   = "hep-ph",
      SLACcitation   = "%%CITATION = HEP-PH/0208156;%%"
}

@article{Agostinelli:2002hh,
      author         = "Agostinelli, S. and others",
      title          = "{GEANT4--A simulation toolkit}",
      collaboration  = "GEANT4",
      journal        = "Nucl. Instrum. Meth. A",
      volume         = "506",
      year           = "2003",
      pages          = "250-303",
      doi            = "10.1016/S0168-9002(03)01368-8",
      reportNumber   = "SLAC-PUB-9350, FERMILAB-PUB-03-339",
      SLACcitation   = "%%CITATION = NUIMA,A506,250;%%"
}

@article{Martin:2009iq,
      author         = "Martin, A. D. and Stirling, W. J. and Thorne, R. S. and
                        Watt, G.",
      title          = "{Parton distributions for the LHC}",
      journal        = "Eur. Phys. J. C",
      volume         = "63",
      year           = "2009",
      pages          = "189-285",
      doi            = "10.1140/epjc/s10052-009-1072-5",
      eprint         = "0901.0002",
      archivePrefix  = "arXiv",
      primaryClass   = "hep-ph",
      reportNumber   = "IPPP-08-95, DCPT-08-190, CAVENDISH-HEP-08-16",
      SLACcitation   = "%%CITATION = ARXIV:0901.0002;%%"
}

@article{Cacciari:2008gp,
      author         = "Cacciari, Matteo and Salam, Gavin P. and Soyez, Gregory",
      title          = "{The anti-$k_t$ jet clustering algorithm}",
      journal        = "JHEP",
      volume         = "04",
      year           = "2008",
      pages          = "063",
      doi            = "10.1088/1126-6708/2008/04/063",
      eprint         = "0802.1189",
      archivePrefix  = "arXiv",
      primaryClass   = "hep-ph",
      reportNumber   = "LPTHE-07-03",
      SLACcitation   = "%%CITATION = ARXIV:0802.1189;%%"
}

@Booklet{ATLAS:2016-012,
      author         = "{ATLAS Collaboration}",
      title          = "{Optimisation of the ATLAS b-tagging performance for the 2016 LHC Run}",
      year           = "2016",
      howpublished   = "{ATLAS-PHYS-PUB-2016-012}",
      url            = "https://cds.cern.ch/record/2160731",
}

@Article{POWHEG0,
      author         = "Nason, Paolo",
      title          = "{A New method for combining NLO QCD with shower Monte
                        Carlo algorithms}",
      journal        = "JHEP",
      volume         = "11",
      pages          = "040",
      doi            = "10.1088/1126-6708/2004/11/040",
      year           = "2004",
      eprint         = "hep-ph/0409146",
      archivePrefix  = "arXiv",
}

@Article{POWHEG01,
      author         = "Frixione, Stefano and Nason, Paolo and Oleari, Carlo",
      title          = "{Matching NLO QCD computations with Parton Shower
                        simulations: the POWHEG method}",
      journal        = "JHEP",
      volume         = "11",
      pages          = "070",
      doi            = "10.1088/1126-6708/2007/11/070",
      year           = "2007",
      eprint         = "0709.2092",
      archivePrefix  = "arXiv",
      primaryClass   = "hep-ph",
}

@ARTICLE{Pythia,
author       = "T. Sj{\"o}strand, S. Mrenna, and P. Z. Skands",
title        = "PYTHIA 6.4 physics and manual",
year         = "2006",
url          = "http://arxiv.org/abs/hep-ph/0603175"
}

@article{Djouadi:2013uqa,
      author         = "Djouadi, A. and Maiani, L. and Moreau, G. and Polosa, A.
                        and Quevillon, J. and Riquer, V.",
      title          = "{The post-Higgs MSSM scenario: habemus MSSM?}",
      journal        = "Eur. Phys. J. C",
      volume         = "73",
      year           = "2013",
      pages          = "2650",
      doi            = "10.1140/epjc/s10052-013-2650-0",
      eprint         = "1307.5205",
      archivePrefix  = "arXiv",
      primaryClass   = "hep-ph",
      reportNumber   = "LPT-ORSAY-13-56",
      SLACcitation   = "%%CITATION = ARXIV:1307.5205;%%"
}

@article{Carena:2013ytb,
      author         = "Carena, M. and Heinemeyer, S. and Stål, O. and Wagner,
                        C. E. M. and Weiglein, G.",
      title          = "{MSSM Higgs boson searches at the LHC: benchmark
                        scenarios after the discovery of a Higgs-like particle}",
      journal        = "Eur. Phys. J. C",
      volume         = "73",
      year           = "2013",
      number         = "9",
      pages          = "2552",
      doi            = "10.1140/epjc/s10052-013-2552-1",
      eprint         = "1302.7033",
      archivePrefix  = "arXiv",
      primaryClass   = "hep-ph",
      reportNumber   = "ANL-HEP-PR-13-12, EFI-13-2, DESY-13-024,
                        FERMILAB-PUB-13-041-T",
      SLACcitation   = "%%CITATION = ARXIV:1302.7033;%%"
}

@Booklet{Capeans:1291633,
     author       = "{ATLAS Collaboration}",
     title        = "ATLAS Insertable B-Layer Technical Design Report",
     howpublished = "ATLAS-TDR-19",
     year         = "2010",
     url          = "https://cds.cern.ch/record/1291633",
     addendum     = "{\textit{ATLAS Insertable B-Layer Technical Design Report Addendum}}, ATLAS-TDR-19-ADD-1, 2012, {\scriptsize{URL}}: \url{https://cds.cern.ch/record/1451888}",
}

@article{Abbott:2018ikt,
      author         = "B. Abbott and others",
      title          = "{Production and integration of the ATLAS Insertable B-Layer}",
      journal        = "JINST",
      volume         = "13",
      year           = "2018",
      number         = "05",
      pages          = "T05008",
      doi            = "10.1088/1748-0221/13/05/T05008",
      eprint         = "1803.00844",
      archivePrefix  = "arXiv",
      primaryClass   = "physics.ins-det",
      SLACcitation   = "%%CITATION = ARXIV:1803.00844;%%"
}

@article{Guralnik:1964eu,
      author         = "Guralnik, G. S. and Hagen, C. R. and Kibble, T. W. B.",
      title          = "{Global Conservation Laws and Massless Particles}",
      journal        = "Phys. Rev. Lett.",
      volume         = "13",
      year           = "1964",
      pages          = "585-587",
      doi            = "10.1103/PhysRevLett.13.585",
      SLACcitation   = "%%CITATION = PRLTA,13,585;%%"
}

@article{Alwall:2014hca,
      author         = "Alwall, J. and Frederix, R. and Frixione, S. and Hirschi,
                        V. and Maltoni, F. and Mattelaer, O. and Shao, H. -S. and
                        Stelzer, T. and Torrielli, P. and Zaro, M.",
      title          = "{The automated computation of tree-level and
                        next-to-leading order differential cross sections, and
                        their matching to parton shower simulations}",
      journal        = "JHEP",
      volume         = "07",
      year           = "2014",
      pages          = "079",
      doi            = "10.1007/JHEP07(2014)079",
      eprint         = "1405.0301",
      archivePrefix  = "arXiv",
      primaryClass   = "hep-ph",
      reportNumber   = "CERN-PH-TH-2014-064, CP3-14-18, LPN14-066, MCNET-14-09,
                        ZU-TH-14-14",
      SLACcitation   = "%%CITATION = ARXIV:1405.0301;%%"
}

@article{Wiesemann:2014ioa,
      author         = "Wiesemann, M. and Frederix, R. and Frixione, S. and
                        Hirschi, V. and Maltoni, F. and Torrielli, P.",
      title          = "{Higgs production in association with bottom quarks}",
      journal        = "JHEP",
      volume         = "02",
      year           = "2015",
      pages          = "132",
      doi            = "10.1007/JHEP02(2015)132",
      eprint         = "1409.5301",
      archivePrefix  = "arXiv",
      primaryClass   = "hep-ph",
      reportNumber   = "CERN-PH-TH-2014-182, CP3-14-64, LPN14-114, MCNET-14-20,
                        ZU-TH-33-14",
      SLACcitation   = "%%CITATION = ARXIV:1409.5301;%%"
}

@article{Harlander:2011aa,
      author         = "Harlander, Robert and Kr{\"a}mer, Michael and Schumacher,
                        Markus",
      title          = "{Bottom-quark associated Higgs-boson production:
                        reconciling the four- and five-flavour scheme approach}",
      year           = "2011",
      eprint         = "1112.3478",
      archivePrefix  = "arXiv",
      primaryClass   = "hep-ph",
      reportNumber   = "CERN-PH-TH-2011-134, FR-PHENO-2011-009, TTK-11-17,
                        WUB-11-04",
      SLACcitation   = "%%CITATION = ARXIV:1112.3478;%%"
}

@booklet{Asner:2013psa,
      author         = "Asner, D. M. and others",
      title          = "ILC Higgs White Paper",
      booktitle      = "Proceedings, 2013 Community Summer Study on the Future
                        of U.S. Particle Physics: Snowmass on the Mississippi
                        (CSS2013): Minneapolis, MN, USA, July 29-August 6, 2013",
      %url            = "http://www.slac.stanford.edu/econf/C1307292/docs/submittedArxivFiles/1310.0763.pdf",
      year           = "2013",
      eprint         = "1310.0763",
      archivePrefix  = "arXiv",
      primaryClass   = "hep-ph",
      SLACcitation   = "%%CITATION = ARXIV:1310.0763;%%"
}

@article{Schael:2006cr,
      author         = "{ALEPH, DELPHI, L3 and OPAL Collaborations}",
      title          = "{Search for neutral MSSM Higgs bosons at LEP}",
      collaboration  = "DELPHI, OPAL, ALEPH, LEP Working Group for Higgs Boson
                        Searches, L3",
      journal        = "Eur. Phys. J. C",
      volume         = "47",
      year           = "2006",
      pages          = "547-587",
      doi            = "10.1140/epjc/s2006-02569-7",
      eprint         = "hep-ex/0602042",
      archivePrefix  = "arXiv",
      primaryClass   = "hep-ex",
      reportNumber   = "CERN-PH-EP-2006-001",
      SLACcitation   = "%%CITATION = HEP-EX/0602042;%%"
}

@article{Sirunyan:2018zut,
      author         = "{CMS Collaboration}",
      title          = "{Search for additional neutral MSSM Higgs bosons in the
                        $\tau\tau$ final state in proton-proton collisions at
                        $\sqrt{s}=$ 13 TeV}",
      collaboration  = "CMS",
      journal        = "JHEP",
      volume         = "09",
      year           = "2018",
      pages          = "007",
      doi            = "10.1007/JHEP09(2018)007",
      eprint         = "1803.06553",
      archivePrefix  = "arXiv",
      primaryClass   = "hep-ex",
      reportNumber   = "CMS-HIG-17-020, CERN-EP-2018-026",
      SLACcitation   = "%%CITATION = ARXIV:1803.06553;%%"
}

@article{Aaboud:2017sjh,
      author         = "{ATLAS Collaboration}",
      title          = "{Search for additional heavy neutral Higgs and gauge
                        bosons in the ditau final state produced in 36 fb$^{−1}$
                        of pp collisions at $ \sqrt{s}=13 $ TeV with the ATLAS
                        detector}",
      collaboration  = "ATLAS",
      journal        = "JHEP",
      volume         = "01",
      year           = "2018",
      pages          = "055",
      doi            = "10.1007/JHEP01(2018)055",
      eprint         = "1709.07242",
      archivePrefix  = "arXiv",
      primaryClass   = "hep-ex",
      reportNumber   = "CERN-EP-2017-199",
      SLACcitation   = "%%CITATION = ARXIV:1709.07242;%%"
}

@article{Sirunyan:2018taj,
      author         = "{CMS Collaboration}",
      title          = "{Search for beyond the standard model Higgs bosons
                        decaying into a $\mathrm{b\overline{b}}$ pair in pp
                        collisions at $\sqrt{s} =$ 13 TeV}",
      collaboration  = "CMS",
      journal        = "JHEP",
      volume         = "08",
      year           = "2018",
      pages          = "113",
      doi            = "10.1007/JHEP08(2018)113",
      eprint         = "1805.12191",
      archivePrefix  = "arXiv",
      primaryClass   = "hep-ex",
      reportNumber   = "CMS-HIG-16-018, CERN-EP-2018-124",
      SLACcitation   = "%%CITATION = ARXIV:1805.12191;%%"
}

@article{Roe:2001aa,
      author         = "Roe, B.P.",
      title          = "{Probability and Statistics in Experimental Physics}",
      journal        = "Springer-Verlag",
      volume         = "",
      year           = "2001",
      pages          = "264pp.",
      reportNumber   = "",
      SLACcitation   = "%%CITATION = FRPHA,80,1;%%"
}

@article{Branco:2011iw,
      author         = "Branco, G. C. and Ferreira, P. M. and Lavoura, L. and
                        Rebelo, M. N. and Sher, Marc and Silva, Joao P.",
      title          = "{Theory and phenomenology of two-Higgs-doublet models}",
      journal        = "Phys. Rept.",
      volume         = "516",
      year           = "2012",
      pages          = "1-102",
      doi            = "10.1016/j.physrep.2012.02.002",
      eprint         = "1106.0034",
      archivePrefix  = "arXiv",
      primaryClass   = "hep-ph",
      SLACcitation   = "%%CITATION = ARXIV:1106.0034;%%"
}

@article{deFlorian:2016spz,
      author         = "de Florian, D. and others",
      title          = "{Handbook of LHC Higgs Cross Sections: 4. Deciphering the
                        nature of the Higgs sector}",
      collaboration  = "LHC Higgs Cross Section Working Group",
      journal        = "CERN-2017-002-M",
      year           = "{2017}",
      
      eprint         = "1610.07922",
      archivePrefix  = "arXiv",
      primaryClass   = "hep-ph",
      reportNumber   = "FERMILAB-FN-1025-T, CERN-2017-002-M",
      SLACcitation   = "%%CITATION = ARXIV:1610.07922;%%",
      doi            = "10.23731/CYRM-2017-002"
}

@article{Harlander:2012pb,
      author         = "Harlander, Robert V. and Liebler, Stefan and Mantler,
                        Hendrik",
      title          = "{SusHi: A program for the calculation of Higgs production
                        in gluon fusion and bottom-quark annihilation in the
                        Standard Model and the MSSM}",
      journal        = "Comput. Phys. Commun.",
      volume         = "184",
      year           = "2013",
      pages          = "1605-1617",
      doi            = "10.1016/j.cpc.2013.02.006",
      eprint         = "1212.3249",
      archivePrefix  = "arXiv",
      primaryClass   = "hep-ph",
      reportNumber   = "WUB-12-28, LPN12-134",
      SLACcitation   = "%%CITATION = ARXIV:1212.3249;%%"
}

@article{Eriksson:2009ws,
      author         = "Eriksson, David and Rathsman, Johan and St{\aa}l, Oscar",
      title          = "{2HDMC -- two-Higgs-doublet model calculator}",
      journal        = "Comput. Phys. Commun.",
      volume         = "181",
      year           = "2010",
      pages          = "189-205",
      doi            = "10.1016/j.cpc.2009.09.011",
      eprint         = "0902.0851",
      archivePrefix  = "arXiv",
      primaryClass   = "hep-ph",
      SLACcitation   = "%%CITATION = ARXIV:0902.0851;%%"
}

@article{Aaboud:2018gjj,
      author         = "{ATLAS Collaboration}",
      title          = "{Search for charged Higgs bosons decaying via $H^{\pm}
                        \to \tau^{\pm}\nu_{\tau}$ in the $\tau$+jets and
                        $\tau$+lepton final states with 36 fb$^{-1}$ of $pp$
                        collision data recorded at $\sqrt{s} = 13$ TeV with the
                        ATLAS experiment}",
      collaboration  = "ATLAS",
      journal        = "JHEP",
      volume         = "09",
      year           = "2018",
      pages          = "139",
      doi            = "10.1007/JHEP09(2018)139",
      eprint         = "1807.07915",
      archivePrefix  = "arXiv",
      primaryClass   = "hep-ex",
      reportNumber   = "CERN-EP-2018-148",
      SLACcitation   = "%%CITATION = ARXIV:1807.07915;%%"
}

@article{Aaboud:2018eoy,
      author         = "{ATLAS Collaboration}",
      title          = "{Search for a heavy Higgs boson decaying into a $Z$ boson
                        and another heavy Higgs boson in the $\ell\ell bb$ final
                        state in $pp$ collisions at $\sqrt{s}=13$ TeV with the
                        ATLAS detector}",
      collaboration  = "ATLAS",
      journal        = "Phys. Lett. B",
      volume         = "783",
      year           = "2018",
      pages          = "392-414",
      doi            = "10.1016/j.physletb.2018.07.006",
      eprint         = "1804.01126",
      archivePrefix  = "arXiv",
      primaryClass   = "hep-ex",
      reportNumber   = "CERN-EP-2018-030",
      SLACcitation   = "%%CITATION = ARXIV:1804.01126;%%"
}

@article{Harlander:2003ai,
      author         = "Harlander, Robert V. and Kilgore, William B.",
      title          = "{Higgs boson production in bottom quark fusion at
                        next-to-next-to leading order}",
      journal        = "Phys. Rev. D",
      volume         = "68",
      year           = "2003",
      pages          = "013001",
      doi            = "10.1103/PhysRevD.68.013001",
      eprint         = "hep-ph/0304035",
      archivePrefix  = "arXiv",
%      primaryClass   = "hep-ph",
      reportNumber   = "BNL-HET-03-4, CERN-TH-2003-067",
      SLACcitation   = "%%CITATION = HEP-PH/0304035;%%"
}

@article{Dittmaier:2003ej,
      author         = "Dittmaier, Stefan and Krämer, Michael and Spira,
                        Michael",
      title          = "{Higgs radiation off bottom quarks at the Fermilab Tevatron and
                        the CERN LHC}",
      journal        = "Phys. Rev. D",
      volume         = "70",
      year           = "2004",
      pages          = "074010",
      doi            = "10.1103/PhysRevD.70.074010",
      eprint         = "hep-ph/0309204",
      archivePrefix  = "arXiv",
%      primaryClass   = "hep-ph",
      reportNumber   = "EDINBURGH-2003-13, MPP-2003-49, PSI-PR-03-14",
      SLACcitation   = "%%CITATION = HEP-PH/0309204;%%"
}

@article{Dawson:2003kb,
      author         = "Dawson, S. and Jackson, C. B. and Reina, L. and
                        Wackeroth, D.",
      title          = "{Exclusive Higgs boson production with bottom quarks at
                        hadron colliders}",
      journal        = "Phys. Rev. D",
      volume         = "69",
      year           = "2004",
      pages          = "074027",
      doi            = "10.1103/PhysRevD.69.074027",
      eprint         = "hep-ph/0311067",
      archivePrefix  = "arXiv",
%      primaryClass   = "hep-ph",
      reportNumber   = "BNL-HET-03-17, FSU-HEP-2003-1015, UB-HET-03-05",
      SLACcitation   = "%%CITATION = HEP-PH/0311067;%%"
}

@article{Djouadi:1997yw,
      author         = "Djouadi, A. and Kalinowski, J. and Spira, M.",
      title          = "{HDECAY: A Program for Higgs boson decays in the standard
                        model and its supersymmetric extension}",
      journal        = "Comput. Phys. Commun.",
      volume         = "108",
      year           = "1998",
      pages          = "56-74",
      doi            = "10.1016/S0010-4655(97)00123-9",
      eprint         = "hep-ph/9704448",
      archivePrefix  = "arXiv",
%      primaryClass   = "hep-ph",
      reportNumber   = "DESY-97-079, IFT-96-29, PM-97-04",
      SLACcitation   = "%%CITATION = HEP-PH/9704448;%%"
}

@article{Djouadi:2018xqq,
      author         = "Djouadi, Abdelhak and Kalinowski, Jan and M{\"u}hlleitner,
                        Margarete and Spira, Michael",
      title          = "{HDECAY: Twenty$++$ years after}",
      journal        = "Comput. Phys. Commun.",
      volume         = "238",
      year           = "2019",
      pages          = "214-231",
      doi            = "10.1016/j.cpc.2018.12.010",
      eprint         = "1801.09506",
      archivePrefix  = "arXiv",
      primaryClass   = "hep-ph",
      reportNumber   = "CERN-TH-2017-262, LPT-Orsay-18-04, KA-TP-03-2018,
                        PSI-PR-18-02, LPT-ORSAY-18-04",
      SLACcitation   = "%%CITATION = ARXIV:1801.09506;%%"
}

@article{Heinemeyer:1998np,
      author         = "Heinemeyer, S. and Hollik, W. and Weiglein, G.",
      title          = "{The masses of the neutral CP-even Higgs bosons in the
                        MSSM: Accurate analysis at the two loop level}",
      journal        = "Eur. Phys. J. C",
      volume         = "9",
      year           = "1999",
      pages          = "343-366",
      doi            = "10.1007/s100529900006",
      eprint         = "hep-ph/9812472",
      archivePrefix  = "arXiv",
%      primaryClass   = "hep-ph",
      reportNumber   = "KA-TP-17-1998, DESY-98-194, CERN-TH-98-405",
      SLACcitation   = "%%CITATION = HEP-PH/9812472;%%"
}

@article{Degrassi:2002fi,
      author         = "Degrassi, G. and Heinemeyer, S. and Hollik, W. and
                        Slavich, P. and Weiglein, G.",
      title          = "{Towards high precision predictions for the MSSM Higgs
                        sector}",
      journal        = "Eur. Phys. J. C",
      volume         = "28",
      year           = "2003",
      pages          = "133-143",
      doi            = "10.1140/epjc/s2003-01152-2",
      eprint         = "hep-ph/0212020",
      archivePrefix  = "arXiv",
%      primaryClass   = "hep-ph",
      reportNumber   = "DCPT-02-126, IPPP-02-63, LMU-11-02, MPI-PHT-2002-73,
                        RM3-TH-02-19",
      SLACcitation   = "%%CITATION = HEP-PH/0212020;%%"
}

@article{Frank:2006yh,
      author         = "Frank, M. and Hahn, T. and Heinemeyer, S. and Hollik, W.
                        and Rzehak, H. and Weiglein, G.",
      title          = "{The Higgs boson masses and mixings of the complex MSSM
                        in the Feynman-diagrammatic approach}",
      journal        = "JHEP",
      volume         = "02",
      year           = "2007",
      pages          = "047",
      doi            = "10.1088/1126-6708/2007/02/047",
      eprint         = "hep-ph/0611326",
      archivePrefix  = "arXiv",
%      primaryClass   = "hep-ph",
      reportNumber   = "DCPT-06-160, IPPP-06-80, MPP-2006-158, PSI-PR-06-14",
      SLACcitation   = "%%CITATION = HEP-PH/0611326;%%"
}

@article{Hahn:2013ria,
      author         = "Hahn, T. and Heinemeyer, S. and Hollik, W. and Rzehak, H.
                        and Weiglein, G.",
      title          = "{High-Precision Predictions for the Light CP-Even Higgs
                        Boson Mass of the Minimal Supersymmetric Standard Model}",
      journal        = "Phys. Rev. Lett.",
      volume         = "112",
      year           = "2014",
      number         = "14",
      pages          = "141801",
      doi            = "10.1103/PhysRevLett.112.141801",
      eprint         = "1312.4937",
      archivePrefix  = "arXiv",
      primaryClass   = "hep-ph",
      reportNumber   = "DESY-13-248, FR-PHENO-2013-018, MPP-2013-317",
      SLACcitation   = "%%CITATION = ARXIV:1312.4937;%%"
}

@article{Bredenstein:2006rh,
      author         = "Bredenstein, A. and Denner, Ansgar and Dittmaier, S. and
                        Weber, M. M.",
      title          = "{Precise predictions for the Higgs-boson decay H --->
                        WW/ZZ ---> 4 leptons}",
      journal        = "Phys. Rev. D",
      volume         = "74",
      year           = "2006",
      pages          = "013004",
      doi            = "10.1103/PhysRevD.74.013004",
      eprint         = "hep-ph/0604011",
      archivePrefix  = "arXiv",
%      primaryClass   = "hep-ph",
      reportNumber   = "MPP-2005-24, PSI-PR-06-05",
      SLACcitation   = "%%CITATION = HEP-PH/0604011;%%"
}

@article{Bredenstein:2006ha,
      author         = "Bredenstein, A. and Denner, Ansgar and Dittmaier, S. and
                        Weber, M. M.",
      title          = "{Radiative corrections to the semileptonic and hadronic
                        Higgs-boson decays H ---> W W / Z Z ---> 4 fermions}",
      journal        = "JHEP",
      volume         = "02",
      year           = "2007",
      pages          = "080",
      doi            = "10.1088/1126-6708/2007/02/080",
      eprint         = "hep-ph/0611234",
      archivePrefix  = "arXiv",
      primaryClass   = "hep-ph",
      reportNumber   = "KEK-CP-187, KEK-PREPRINT-2006-47, MPP-2006-138,
                        PSI-PR-06-11, WUB-06-08",
      SLACcitation   = "%%CITATION = HEP-PH/0611234;%%"
}

@article{Sirunyan:2019hkq,
      author         = "{CMS Collaboration}",
      title          = "{Search for charged Higgs bosons in the H$^{\pm} \to
                        \tau^{\pm}\nu_\tau$ decay channel in proton-proton
                        collisions at $\sqrt{s}=$ 13 TeV}",
      collaboration  = "CMS",
      year           = "2019",
      eprint         = "1903.04560",
      archivePrefix  = "arXiv",
      primaryClass   = "hep-ex",
      reportNumber   = "CMS-HIG-18-014, CERN-EP-2019-025",
      SLACcitation   = "%%CITATION = ARXIV:1903.04560;%%"
}

@article{PCAbook,
      author         = "Jolliffe, I.",
      title          = "{Principal Component Analysis}",
      journal        = "in M. Lovric (Ed.), International Encyclopedia of Statistical Science. Springer, Berlin, Heidelberg,  ",
      doi            = {10.1007/978-3-642-04898-2_455},
      year           = "2011",
}

@phdthesis{mcclymontThesis2018,
  location = {{England}},
  title = {Searches for New Physics using Pairs of Jets Containing b-quarks at the ATLAS Detector},
  url = {http://discovery.ucl.ac.uk/10052993/1/e-Thesis.pdf},
  institution = {{University College London}},
  author = {Laurie Iain McClymont},
}

@Article{PERF-2007-01,
    author         = "{ATLAS Collaboration}",
    title          = "{The ATLAS Experiment at the CERN Large Hadron Collider}",
    journal        = "JINST",
    volume         = "3",
    year           = "2008",
    pages          = "S08003",
    doi            = "10.1088/1748-0221/3/08/S08003",
    primaryClass   = "hep-ex",
}

@Article{SOFT-2010-01,
    author         = "{ATLAS Collaboration}",
    title          = "{The ATLAS Simulation Infrastructure}",
    journal        = "Eur. Phys. J. C",
    volume         = "70",
    year           = "2010",
    pages          = "823",
    doi            = "10.1140/epjc/s10052-010-1429-9",
    eprint         = "1005.4568",
    archivePrefix  = "arXiv",
    primaryClass   = "physics.ins-det",
}

@Article{PERF-2011-04,
    author         = "{ATLAS Collaboration}",
    title          = "{Jet energy resolution in proton--proton collisions at \(\sqrt{s} = 7\,\text{TeV}\) recorded in 2010 with the ATLAS detector}",
    journal        = "Eur. Phys. J. C",
    volume         = "73",
    year           = "2013",
    pages          = "2306",
    doi            = "10.1140/epjc/s10052-013-2306-0",
    reportNumber   = "CERN-PH-EP-2012-191",
    eprint         = "1210.6210",
    archivePrefix  = "arXiv",
    primaryClass   = "hep-ex",
}

@Article{HIGG-2012-27,
    author         = "{ATLAS Collaboration}",
    title          = "{Observation of a new particle in the search for the Standard Model Higgs boson with the ATLAS detector at the LHC}",
    journal        = "Phys. Lett. B",
    volume         = "716",
    year           = "2012",
    pages          = "1",
    doi            = "10.1016/j.physletb.2012.08.020",
    reportNumber   = "CERN-PH-EP-2012-218",
    eprint         = "1207.7214",
    archivePrefix  = "arXiv",
    primaryClass   = "hep-ex",
}

@Article{PERF-2012-04,
    author         = "{ATLAS Collaboration}",
    title          = "{Performance of \(b\)-jet identification in the ATLAS experiment}",
    journal        = "JINST",
    volume         = "11",
    year           = "2016",
    pages          = "P04008",
    doi            = "10.1088/1748-0221/11/04/P04008",
    reportNumber   = "CERN-PH-EP-2015-216",
    eprint         = "1512.01094",
    archivePrefix  = "arXiv",
    primaryClass   = "hep-ex",
}

@Article{DAPR-2013-01,
    author         = "{ATLAS Collaboration}",
    title          = "{Luminosity determination in \(pp\) collisions at \(\sqrt{s} = 8\,\text{TeV}\) using the ATLAS detector at the LHC}",
    journal        = "Eur. Phys. J. C",
    volume         = "76",
    year           = "2016",
    pages          = "653",
    doi            = "10.1140/epjc/s10052-016-4466-1",
    eprint         = "1608.03953",
    archivePrefix  = "arXiv",
    primaryClass   = "hep-ex",
}

@Article{PERF-2014-03,
    author         = "{ATLAS Collaboration}",
    title          = "{Performance of pile-up mitigation techniques for jets in \(pp\) collisions at \(\sqrt{s} = 8\,\text{TeV}\) using the ATLAS detector}",
    journal        = "Eur. Phys. J. C",
    volume         = "76",
    year           = "2016",
    pages          = "581",
    doi            = "10.1140/epjc/s10052-016-4395-z",
    reportNumber   = "CERN-PH-EP-2015-206",
    eprint         = "1510.03823",
    archivePrefix  = "arXiv",
    primaryClass   = "hep-ex",
}

@Article{PERF-2014-07,
    author         = "{ATLAS Collaboration}",
    title          = "{Topological cell clustering in the ATLAS calorimeters and its performance in LHC Run~1}",
    journal        = "Eur. Phys. J. C",
    volume         = "77",
    year           = "2017",
    pages          = "490",
    doi            = "10.1140/epjc/s10052-017-5004-5",
    reportNumber   = "CERN-PH-EP-2015-304",
    eprint         = "1603.02934",
    archivePrefix  = "arXiv",
    primaryClass   = "hep-ex",
}

@Article{PERF-2015-01,
    author         = "{ATLAS Collaboration}",
    title          = "{Reconstruction of primary vertices at the ATLAS experiment in Run~1 proton--proton collisions at the LHC}",
    journal        = "Eur. Phys. J. C",
    volume         = "77",
    year           = "2017",
    pages          = "332",
    doi            = "10.1140/epjc/s10052-017-4887-5",
    reportNumber   = "CERN-EP-2016-150",
    eprint         = "1611.10235",
    archivePrefix  = "arXiv",
    primaryClass   = "hep-ex",
}

@Article{EXOT-2016-10,
    author         = "{ATLAS Collaboration}",
    title          = "{Search for heavy resonances decaying into a \(W\) or \(Z\) boson and a Higgs boson in final states with leptons and \(b\)-jets in \(36\,\text{fb}^{-1}\) of \(\sqrt{s} = 13\,\text{TeV}\) \(pp\) collisions with the ATLAS detector}",
    journal        = "JHEP",
    volume         = "03",
    year           = "2018",
    pages          = "174",
    doi            = "10.1007/JHEP03(2018)174",
    reportNumber   = "CERN-EP-2017-250",
    eprint         = "1712.06518",
    archivePrefix  = "arXiv",
    primaryClass   = "hep-ex",
    related        = "EXOT-2016-10-err",
    relatedstring  = "Erratum:",
}

@Article{HIGG-2016-29,
    author         = "{ATLAS Collaboration}",
    title          = "{Evidence for the \(H \rightarrow b\bar{b}\) decay with the ATLAS detector}",
    journal        = "JHEP",
    volume         = "12",
    year           = "2017",
    pages          = "024",
    doi            = "10.1007/JHEP12(2017)024",
    reportNumber   = "CERN-EP-2017-175",
    eprint         = "1708.03299",
    archivePrefix  = "arXiv",
    primaryClass   = "hep-ex",
}

@Article{HIGG-2016-32,
    author         = "{ATLAS Collaboration}",
    title          = "{Search for heavy neutral Higgs bosons produced in association with \(b\)-quarks and decaying into \(b\)-quarks at \(\sqrt{s} = 13\,\text{TeV}\) with the ATLAS detector}",
    year           = "2019",
    reportNumber   = "CERN-EP-2019-092",
    eprint         = "1907.02749",
    archivePrefix  = "arXiv",
    primaryClass   = "hep-ex",
}

@Article{PERF-2016-04,
    author         = "{ATLAS Collaboration}",
    title          = "{Jet energy scale measurements and their systematic uncertainties in proton--proton collisions at \(\sqrt{s} = 13\,\text{TeV}\) with the ATLAS detector}",
    journal        = "Phys. Rev. D",
    volume         = "96",
    year           = "2017",
    pages          = "072002",
    doi            = "10.1103/PhysRevD.96.072002",
    reportNumber   = "CERN-EP-2017-038",
    eprint         = "1703.09665",
    archivePrefix  = "arXiv",
    primaryClass   = "hep-ex",
}

@Article{TRIG-2016-01,
    author         = "{ATLAS Collaboration}",
    title          = "{Performance of the ATLAS trigger system in 2015}",
    journal        = "Eur. Phys. J. C",
    volume         = "77",
    year           = "2017",
    pages          = "317",
    doi            = "10.1140/epjc/s10052-017-4852-3",
    reportNumber   = "CERN-EP-2016-241",
    eprint         = "1611.09661",
    archivePrefix  = "arXiv",
    primaryClass   = "hep-ex",
}

@Booklet{ATL-SOFT-PUB-2020-001,
    author         = "{ATLAS Collaboration}",
    title          = "{ATLAS Computing Acknowledgements}",
    howpublished   = "{ATL-SOFT-PUB-2020-001}",
    url            = "https://cds.cern.ch/record/2717821",
}

@Article{CMS-HIG-12-028,
    author         = "{CMS Collaboration}",
    title          = "{Observation of a new boson at a mass of 125 GeV with the CMS experiment at the LHC}",
    journal        = "Phys. Lett. B",
    volume         = "716",
    year           = "2012",
    pages          = "30",
    doi            = "10.1016/j.physletb.2012.08.021",
    reportNumber   = "CERN-PH-EP-2012-220",
    eprint         = "1207.7235",
    archivePrefix  = "arXiv",
    primaryClass   = "hep-ex",
}

@Booklet{ATLAS-CONF-2015-029,
    author         = "{ATLAS Collaboration}",
    title          = "{Selection of jets produced in \(13~\text{TeV}\) proton--proton collisions with the ATLAS detector}",
    howpublished   = "{ATLAS-CONF-2015-029}",
    url            = "https://cds.cern.ch/record/2037702",
    year           = "2015",
}

@Booklet{ATLAS-CONF-2018-001,
    author         = "{ATLAS Collaboration}",
    title          = "{Measurement of \(b\)-tagging efficiency of \(c\)-jets in \(t\bar{t}\) events using a likelihood approach with the ATLAS detector}",
    howpublished   = "{ATLAS-CONF-2018-001}",
    url            = "https://cds.cern.ch/record/2306649",
    year           = "2018",
}

@Booklet{ATLAS-CONF-2018-006,
    author         = "{ATLAS Collaboration}",
    title          = "{Calibration of light-flavour \(b\)-jet mistagging rates using ATLAS proton--proton collision data at \(\sqrt{s} = 13~\text{TeV}\)}",
    howpublished   = "{ATLAS-CONF-2018-006}",
    url            = "https://cds.cern.ch/record/2314418",
    year           = "2018",
}

@Booklet{ATL-PHYS-PUB-2010-013,
    author         = "{ATLAS Collaboration}",
    title          = "{The simulation principle and performance of the ATLAS fast calorimeter simulation FastCaloSim}",
    howpublished   = "{ATL-PHYS-PUB-2010-013}",
    url            = "https://cds.cern.ch/record/1300517",
    year           = "2010",
}

@Booklet{ATL-PHYS-PUB-2012-003,
    author         = "{ATLAS Collaboration}",
    title          = "{Summary of ATLAS Pythia 8 tunes}",
    howpublished   = "{ATL-PHYS-PUB-2012-003}",
    url            = "https://cds.cern.ch/record/1474107",
    year           = "2012",
}

\clearpage 
 
\begin{flushleft}
\hypersetup{urlcolor=black}
{\Large The ATLAS Collaboration}

\bigskip

\AtlasOrcid[0000-0002-6665-4934]{G.~Aad}$^\textrm{\scriptsize 102}$,    
\AtlasOrcid[0000-0002-5888-2734]{B.~Abbott}$^\textrm{\scriptsize 129}$,    
\AtlasOrcid{D.C.~Abbott}$^\textrm{\scriptsize 103}$,    
\AtlasOrcid{O.~Abdinov}$^\textrm{\scriptsize 13,*}$,    
\AtlasOrcid[0000-0002-2788-3822]{A.~Abed~Abud}$^\textrm{\scriptsize 71a,71b}$,    
\AtlasOrcid[0000-0002-1002-1652]{K.~Abeling}$^\textrm{\scriptsize 53}$,    
\AtlasOrcid[0000-0002-2987-4006]{D.K.~Abhayasinghe}$^\textrm{\scriptsize 94}$,    
\AtlasOrcid[0000-0002-8496-9294]{S.H.~Abidi}$^\textrm{\scriptsize 167}$,    
\AtlasOrcid[0000-0002-8279-9324]{O.S.~AbouZeid}$^\textrm{\scriptsize 40}$,    
\AtlasOrcid{N.L.~Abraham}$^\textrm{\scriptsize 156}$
% ...
\bigskip
\\

% ...
$^{13}$Institute of Physics, Azerbaijan Academy of Sciences, Baku; Azerbaijan.\\
% ...
$^{40}$Niels Bohr Institute, University of Copenhagen, Copenhagen; Denmark.\\
% ...
$^{53}$II. Physikalisches Institut, Georg-August-Universit\"{a}t G\"ottingen, G\"ottingen; Germany.\\
% ...
$^{71}$$^{(a)}$INFN Sezione di Pavia;$^{(b)}$Dipartimento di Fisica, Universit\`a di Pavia, Pavia; Italy.\\
% ...
$^{94}$Department of Physics, Royal Holloway University of London, Egham; United Kingdom.\\
% ...
$^{102}$CPPM, Aix-Marseille Universit\'e, CNRS/IN2P3, Marseille; France.\\
$^{103}$Department of Physics, University of Massachusetts, Amherst MA; United States of America.\\
% ...
$^{129}$Homer L. Dodge Department of Physics and Astronomy, University of Oklahoma, Norman OK; United States of America.\\
% ...
$^{156}$Department of Physics and Astronomy, University of Sussex, Brighton; United Kingdom.\\
% ...
$^{167}$Department of Physics, University of Toronto, Toronto ON; Canada.\\
% ...
$^{a}$ Also at Borough of Manhattan Community College, City University of New York, New York NY; United States of America.\\
$^{b}$ Also at Centre for High Performance Computing, CSIR Campus, Rosebank, Cape Town; South Africa.\\
% ...
$^{*}$ Deceased

\end{flushleft}


\end{document}